\documentclass[11pt]{article}
\usepackage{eurosym}
\usepackage{amsfonts}
\usepackage{amssymb}
\usepackage{graphicx}
\usepackage{amsmath}
\usepackage{makeidx}
\usepackage{indentfirst}
\usepackage[T1]{fontenc}
\usepackage[utf8]{inputenc}

\newcounter{resultnum}[section]
\newcounter{conclusionnum}[section]
\newcounter{conditionnum}[section]
\newcounter{conjecturenum}[section]
\newcounter{examplenum}[section]
\newcounter{exercisenum}[section]
\newcounter{lemmanum}[section]
\newcounter{notationnum}[section]
\newcounter{theoremnum}[section]
\newcounter{definitionnum}[section]
\newcounter{corollarynum}[section]
\newcounter{remarknum}[section]
\newcounter{propositionnum}[section]
\newcounter{acknowledgementnum}[section]
\newcounter{algorithmnum}[section]
\newcounter{axiomnum}[section]
\newcounter{casenum}[section]
\newcounter{claimnum}[section]
\newcounter{summarynum}[section]
\newcounter{problemnum}[section]
\begin{document}

\title{ Dark energy and dark matter configurations for wormholes and
solitionic hierarchies of nonmetric Ricci flows and $F(R,T,Q,T_{m})$ gravity}
\date{accepted by EPJC:\ February 6, 2024}
\author{ {\textbf{Lauren\c{t}iu Bubuianu}\thanks{%
email: laurentiu.bubuianu@tvr.ro and laurfb@gmail.com}} \\
{\small \textit{SRTV - Studioul TVR Ia\c{s}i} and \textit{University
Appolonia}, 2 Muzicii street, Ia\c{s}i, 700399, Romania} \vspace{.1 in} \and 
\textbf{Sergiu I. Vacaru} \thanks{%
emails: sergiu.vacaru@fulbrightmail.org ; sergiu.vacaru@gmail.com ;  \textit{%
Address for correspondence in 2024 as a visiting fellow at CAS LMU, Munich, Germany and   YF CNU, Chernitvsi, Ukraine:\ } Vokzalna  street, 37-3, Chernivtsi,
Ukraine, 58008} \and {\small \textit{Department of Physics, California State
University at Fresno, Fresno, CA 93740, USA; }} \and {\small \textit{%
Institute of Applied-Physics and Computer Sciences, Yu. Fedkovych
University, Chernivtsi, 58012, Ukraine;}} 
\and {\small \textit{CAS Center for Advanced Studies, Ludwig-Maximilians-Universit\"{a}t, Seestrasse 13, M\"{u}nchen, 80802, Germany}}
 \vspace{.1 in} \and {\textbf{El\c{s}en Veli Veliev}} 
\thanks{%
email: elsen@kocaeli.edu.tr and elsenveli@hotmail.com} \\
{\small \textit{\ Department of Physics,\ Kocaeli University, 41380, Izmit,
Turkey }} \vspace{.1 in} \and {\textbf{Assel Zhamysheva}} \thanks{%
email: assel.zhamysheva1@gmail.com} \\
{\small \textit{Department of General and Theoretical Physics, Eurasian
National University, Astana, 010000, Kazakhstan}} }
\maketitle

\begin{abstract}
We extend the anholonomic frame and connection deformation method, AFCDM, for constructing exact and parametric solutions in general relativity, GR, to geometric flow models and modified gravity theories, MGTs, with nontrivial torsion and nonmetricity fields. Following abstract geometric or variational methods, we can derive corresponding systems of nonmetric gravitational and matter field equations which consist of very sophisticate systems of coupled nonlinear PDEs. Using nonholonomic frames with dyadic spacetime splitting and applying the AFCDM, we prove that such systems of PDEs can be decoupled and integrated in general forms for generic off-diagonal metric structures and generalized affine connections. We generate new classes of quasi-stationary solutions (which do not depend on time like coordinates) and study the physical properties of some physically important examples. Such exact or parametric solutions are determined by nonmetric solitonic distributions and/or ellipsoidal deformations of wormhole hole configurations. It is not possible to describe the thermodynamic properties of such solutions in the framework of the Bekenstein-Hawking paradigm because such metrics do not involve, in general, certain horizons, duality, or holographic configurations. Nevertheless, we can always elaborate on associated Grigori Perelman thermodynamic models elaborated for nonmetric geometric flows. In explicit form, applying the AFCDM, we construct and study the physical implications of new classes of traversable wormhole solutions describing solitonic deformation and dissipation of non-Riemannian geometric objects. Such models with nontrivial gravitational off-diagonal vacuum are important for elaborating models of dark energy and dark matter involving wormhole configurations and solitonic- type structure formation.

\vskip3pt

\textbf{Keywords:}\ exact and parametric solutions; modified gravity; nonholonomic geometric flows; dark energy; dark matter; wormholes
\end{abstract}

\tableofcontents

\section{Introduction, motivations and objectives}

The standard approach to the gravity theory and general relativity, GR (i.e. Einstein's gravity theory) is formulated in the framework of pseudo-Riemannian geometry, see   \cite{misner,hawking73,wald82,kramer03} as typical monographs and reviews of physically important exact solutions. In GR, a four dimensional, 4-d, curved spacetime $V$ is modelled as a Lorentz manifold endowed with metric structure, $g=\{g_{\alpha \beta }\},$ when the Levi-Civita, LC-connection $\nabla =\{\ ^{\nabla }\Gamma _{\ \beta \gamma}^{\alpha }\}$ is uniquely determined by coefficients $g_{\alpha \beta }$ following two conditions: 1) zero nonmetricity (i. e. metric compatibility),  $Q:=\nabla g=0$ and 2) zero torsion, $\ ^{\nabla}T=0.$\footnote{In our approach, we use the signature $(+++-)$, when coordinate indices $\alpha ,\beta ,....$ run values 1,2,3,4; in general, indices may be abstract ones; we use also certain left/up labels for geometric/physical objects;
local coordinates are labelled $u=\{u^{\alpha }\}$ and the Einstein summation rule on up-low indices is used. All necessary notation conventions and definitions of geometric/physical objects are defined in the next sections and appendices. We assume that readers are familiar with standard results and methods outlined in the mentioned monographs and understand basic ideas and motivations for how to extend the geometric constructions to modern modified gravity theories, MGTs, and accelerating cosmology.} The Einstein relativity theory has a remarkable success and deep influence both in physics and mathematics. Nevertheless, various alternatives and modifications of GR were elaborated when instead of standard Lorentz manifolds there are considered metric-affine spaces determined by some general metric and (independent) affine/ linear connection structures, $(g,D=\{\Gamma _{\ \beta \gamma}^{\alpha }\}).$ Such non-Riemannian spaces can be characterized by nonzero torsion, $T=\{T_{\ \beta \gamma }^{\alpha}\},$ and/or nontrivial nonmetricity, $Q:=Dg=\{Q_{\alpha \beta \gamma}:=
D_{\alpha }g_{\beta \gamma }\},$ fields. We cite \cite{hehl95} as an early review of metric-affine gravity theories. For applications in modern cosmology, various generalizations of such geometric and gravity models are formulated as modified gravity theories, MGTs, when the gravitational and matter field Lagrangians in GR are changed into some functionals $F(R,T,Q,\ ^{tr}T),$ where $R$ is the Ricci scalar for $D,$ and $\ ^{tr}T$ is the trace of the energy-momentum tensor for matter \cite{harko21,iosifidis22}.\footnote{Readers may find in the cited works many references with chronological
historical remarks, criticism, and applications in modern cosmology and astrophysics. In this paper, we do not provide a comprehensive review on nonmetricity and MGTs but concentrate on elaborating new geometric methods for constructing exact and parametric solutions in nonmetric geometric flow and gravity models. In our works, we write $F(R,T,...)$ instead of $f(,R,T,...)$ used in papers by other authors.}

\vskip5pt Generalizations of the Einstein equations with geometric distortions of linear connections, $\nabla \rightarrow D$, can be written in some effective forms:%
\begin{equation}
E_{\mu \nu }=\varkappa \lbrack T_{\mu \nu }^{[br]}+T_{\mu \nu }^{[DM]}(\phi
,\psi ,...)+T_{\mu \nu }^{[DE]}(\phi ,\psi ,...)+T_{\mu \nu
}^{[geom]}(g,R,T,Q,\mathcal{L}^{[m]},T^{[m]},\square R,\square T,...)+...].
\label{effectngteq}
\end{equation}%
In such formulas, $\varkappa $ is defined by the gravitational constant and the energy-momentum tensors contain respective labels $[br],$ for barionic matter (which can be written also $T_{\mu \nu }^{[m]});$ [DM] is for dark matter models with some scalar, $\phi ,$ spinor, $\psi ,$ and other type fields. The label [DE] is used for respective dark energy terms. The
tensor $T_{\mu \nu}^{[geom]}$ is a functional of different geometric and physical values including classical and quasi-classical and/or extra dimension contributions, string terms contributions, distortions of mater fields Lagrange densities, $\mathcal{L}^{[m]},$ and corresponding traces of energy-momentum tensors, $T^{[m]}$. In some MGTs, there are considered nonlocal terms of type $\square R$ and/or $\square T,$ where $\square $ is a corresponding variant of d'Alambert (wave operator) for respective curved spacetime etc. The modified Einstein tensor $E_{\mu \nu }$ and respective
(effective) barionic matter terms $T_{\mu \nu }\approx T_{\mu \nu}^{[br]}+...$ can be derived from a corresponding action 
\begin{equation*}
S=\int [\varkappa ^{-1} F(R,...)+\mathcal{L}^{[m]}+...]\sqrt{|g|}d^{4}u.
\end{equation*}%
Typically, such formulas define a four-dimensional, 4-d, spacetime model which is distorted in the symbolic form $D=\nabla +Z[g,T,Q,...],$ where $Z$ is the distortion tensor. Physical motivations and details on such MGTs and dark gravity/ matter / energy formalism are presented in \cite{harko21,iosifidis22,khyllep23} and references therein ( there are studied
also certain applications in modern acceleration cosmology and DM and DE physics).

\vskip5pt
Modified Einstein equations of type (\ref{effectngteq}) consist of very sophisticate systems of coupled nonlinear partial differential equations, PDEs. It is very difficult to find exact/ parametric solutions for such dynamical equations using standard methods elaborated in GR \cite{kramer03}, when for some higher symmetry and diagonal ansatz form metrics, nonlinear PDEs transform into certain nonlinear systems of ordinary differential equations, ODEs. For instance, it is not clear how to construct black hole, BH, like solutions for nontrivial $Q$-terms, for generic off-diagonal $g_{\alpha \beta };$ and how to define extensions of the Einstein-Dirac equations. Such problems were discussed in detail in \cite{vplb10,vmon05},
(in a general form for Finsler modifications of gravity theories). Similar problems exist for metric-affine distortions of physical models elaborated on Lorentz manifolds. The DM and DE coupling theories and various MGTs involve constructions with nonminimal coupling between geometry and matter. The equations (\ref{effectngteq}) lead to the "nonconservation" of matter energy-momentum tensor which makes more sophisticate the physical interpretation of such models and solutions of dynamical or evolution equations. Nevertheless, we can elaborate on a nonholonomic deformation formalism with adapted distortions $\nabla \rightarrow D,$ when "nonconservation" is related to certain (equivalently, anholonomic, i.e. nonintegrable) constraints like in nonholonomic mechanics. By introducing integration constants, then solving the constraint equations and redefining the effective Lagrangians, we can formulate some types of modified conservation laws.

\vskip5pt The main goal of this work is to prove that modified Einstein equations of type (\ref{effectngteq}) can be decoupled and integrated in some general forms for $Q\neq 0$. We shall provide explicit examples for generic off-diagonal solutions defining $Q$-deformations of gravitational solitonic hierarchies and wormhole configurations. Such methods for
generating exact/ parametric solutions of physically important systems of nonlinear PDEs are not contained in standard monographs on GR and various MGTs reviews \cite{misner,hawking73,wald82,kramer03,hehl95,harko21,iosifidis22,khyllep23}. During the last 30 years, it was elaborated the anholonomic frame and connection deformation method, AFCDM, for constructing exact and parametric off-diagonal solutions of physically important systems of nonlinear PDEs. We cite \cite{vplb10,vmon05,bubuianu17,vacaru18} for a  review of main results and methods, examples and applications.  Here we note that the AFCDM involves nonholonomic distributions of geometric objects and nonholonomic frames, with respective types 2+2+2+..., 3+1 and
(3+1)+(2+2), 3+2+2+... (correspondingly, for 4-d, and extra dimension spacetimes). Then, it is important to construct an auxiliary connection $\widehat{\mathbf{D}}=\nabla + \widehat{\mathbf{Z}}$ when equations of type (\ref{effectngteq}) with redefined linear connections, $\nabla \rightarrow  \widehat{\mathbf{D}}$ can be decoupled and integrated in certain general forms. In the next section, we shall provide the necessary definitions and explicit formulas.

\vskip5pt Our nonholonomic geometric approach to generating solutions in gravity theories should not be confused with the well-known Cartan moving frame method, Neuman-Penrose formalism or other nontrivial string torsion generalizations and/or equivalent constructions involving various types of tetradic, dyadic, and Arnowit-Deser-Wheeler, ADM, formalisms. Main geometric and analytic constructions related to GR are summarized in \cite{kramer03}). The main difference of the AFCDM from other ones is that it involves deformations both of the frame and linear connection structures which are adapted to certain canonical nonholonomic distributions. In such nonholonomic adapted variables, physically important systems of nonlinear PDEs can be decoupled and integrated in certain general forms, when the technique of constructing solutions is not restricted only to special diagonal ansatz transforming systems of nonlinear PDEs into systems of nonlinear ODEs. For generic off-diagonal ansatz, we can prescribe some special symmetries with Killing vectors and/or Lie algebra structure, for
spherical, cylindrical, toroid or other type configurations. Such solutions also posses certain nonlinear symmetries. We can prescribe necessary symmetries of solutions, and compute certain deformations (for instance, by nonmetricity fields) for some physically important classes of solutions. Usually, it is possible to extract LC configurations if there are imposed
additional nonholonomic constraints on the nonlinear and linear connection structures, and respective generating and integration functions. For some models, the off-diagonal terms of metrics and the nontrivial nonholonomic structure may encode nonmetricity contributions even for constraints to LC models.

\vskip5pt This paper is a generalization to nonmetric geometric flows  providing a metric-affine development of the methods reviewed in \cite{vmon05,bubuianu17,vacaru18}. Such geometric constructions can be performed for nonassociative and noncommutative geometric flow and gravity theories \cite{partner02}, when the constructions are performed for star product R-flux deformations in metric compatible forms but including nonsymmetric metrics. In this work, we consider associative and commutative metric-affine structures with symmetric metrics when the nonholonomic deformations and nonmetricity fields are related both to nonmetricity-induced torsion fields and canonically induced torsions. For this work, we state three general objectives:

\vskip5pt The \textbf{first objective} (Obj1, in section \ref{sec2}) is to provide an introduction to theories of geometric flow and gravity including nonmetricity. The 4-d metric--affine geometry is formulated in an N-connection adapted form with (dyadic) nonholonomic (2+2)-splitting. We define $Q$-modified (for Riemannian metrics introduced by G. Perelman 
\cite{perelman1}) F- and W-functionals and sketch how they can be derived from respective Hamilton-Friedan geometric flow equations with nonmetricity. Nonmetric Ricci solitons and related modified Einstein equations are derived as self-similar nonholonomic geometric flow configurations for a fixed flow parameter.

\vskip5pt The \textbf{second objective}, Obj2, stated for section \ref{sec3}, is to construct and analyse solitonic deformations of exact/parametric quasi-stationary geometric flow solutions encoding nonmetricity fields, which for fixed flow parameters and LC-configurations define solutions of modified Einstein equations (\ref{effectngteq}) and their nonholonomic deformations. Such systems of nonlinear PDEs are extended in certain forms encoding nonmetric geometric flow data. We provide necessary examples of solitonic distributions defined in quasi-stationary geometric form and study models with locally anisotropic wormhole solutions encoding nonmetricity. Nonmetric quasi-stationary deformations of 4-d wormhole metrics in GR are constructed for general and small parametric off-diagonal and nonmetric gravitational polarizations.

\vskip5pt The \textbf{third objective}, Obj3, is stated for section \ref{sec4}. It consists of a study of $Q$-modified Grigori Perelman thermodynamics and its applications for quasi-stationary configurations. We show how to define and compute (using integration functions and nonlinear symmetries to some flow-running cosmological constants) respective volume elements. This allows us to compute thermodynamic variables with running cosmological constants and nonmetricity. For nonmetric modified wormhole configurations and their solitonic deformations, we show that, in principle, we can construct two thermodynamic models: a 'l'{a} Perelman' and/ or do not follow the Bekenstein-Hawking approach because the last one is applicable only for solutions with conventional horizons, holographic models and similar. Modified Perelman thermodynamic models can be formulated for all classes of nonholonomic geometric flow theories,  including theories with $Q$-deformations.

\vskip5pt
In this article and a series of further partner works on nonmetric geometric flow and gravity theories we follow the \textbf{Hypothesis:} \emph{Metric-affine geometric flow models can be exploited as alternatives for describing DE and DM effects and elaborating new physical theories. Such approaches can be elaborated in self-consistent and solvable forms using
nonholonomic variables with conventional 2(3)+2+... splitting which allows to decouple and integrate of physically important systems of nonlinear PDEs for such theories. The solutions with conventional $\tau $ -running effective cosmological constants physics can be used for modelling DE physics when nonlinear symmetries relate such configurations to models of DM physics. Generic off-diagonal metrics are determined by respective generating functions and effective matter sources which encode nonmetric $Q$-deformations and describe DE and DM off-diagonal interactions or geometric evolution scenarios. Corresponding systems of $Q$--deformed geometric flow/ gravitational and (effective) matter field equations admit exact and parametric solutions describing certain quasi-stationary (BH, wormhole etc.) configurations and (locally anisotropic and inhomogeneous) cosmological scenarios. For well-defined nonholonomic geometric constraints, such models
can be defined almost equivalently in canonical metric compatible backgrounds with effective N-connection structure, when generalized conservation laws and nonlinear symmetries are well-defined. In such cases, we can formulate self-consistent nonmetric modifications of classical gravity and quantum gravity theories, quantum mechanical models and quantum
field theories, which can be unified as thermodynamic information theories in the framework of respective geometric and quantum information models.}

\vskip5pt In Appendix \ref{appendixa}, we revise in a nonmetric quasi-stationary form all formulas which are necessary for general decoupling and integration of $Q$-modified nonholonomic Ricci flow/ soliton equations. Details and proofs for general metric compatible canonical d-connections are provided in \cite{vmon05,bubuianu17,vacaru18,partner02}.
In this work, abstract and N-adapted coefficient formulas are re-defined by $Q$-deformations and $Q$-generating sources. A summary of basic concepts and formulas for generating solitonic hierarchies via d-metrics and nonmetricity effective sources is presented in Appendix \ref{appendixb}.

\section{Metric noncompatible geometric flows and MGTs}

\label{sec2} This section contains an introduction to the geometry of four dimensional, 4-d, metric-affine spaces with nontrivial torsion and nonmetricity fields. The approach is formulated in canonical nonholonomic variables with (2+2)-splitting defined by a nonlinear connection, N-connection, structure stating dyadic frame decompositions. We consider
N-adapted distortions of linear connections and fundamental geometric objects. The constructions are performed in such forms that physically important systems of nonlinear PDEs (such as nonmetric geometric flow evolution and modified Einstein equations) can be decoupled and integrated in certain general off-diagonal forms. Necessary concepts and additional
technical formulas for the nonmetric anholonomic frame and connection deformation method, AFCDM, are outlined in appendix \ref{appendixa}. Such a nonholonomic geometric formalism was elaborated for (co) tangent bundles, see reviews of results and methods in \cite{vmon05,vacaru18}. In this section, we develop the approach in a form which allows to construct exact and parametric solutions with nonmetricity for gravity theories of type in 
\cite{hehl95,harko21,iosifidis22,khyllep23}. The Grigori Perelman functionals \cite{perelman1} (see \cite{ibsvevv22} for recent developments related to GR, MGTs and quantum information flow theories) are modified in nonmetric forms. For self-similar configurations (i.e. for nonholonomic Ricci solitons) such models encode the action functionals for gravity
theories with nontrivial $Q$-fields \cite{harko21}.

\subsection{Geometric preliminaries on metric-affine spaces and nonholonomic deformations}

\label{ss21}In this work, the background geometric arena consists from a
Lorentz spacetime manifold $V$ enabled with standard geometric data $(V,%
\mathbf{g,}\nabla ).$ Such a (primary) spacetime is defined as a 4-d
pseudo-Riemannian manifold of necessary smooth/ differentiability class,
when the symmetric metric tensor $\mathbf{g}$ is of signature $(+++-)$ and
can be written in the form 
\begin{equation}
\mathbf{g}=g_{\alpha ^{\prime }\beta ^{\prime }}(u)e^{\alpha ^{\prime
}}\otimes e^{\beta ^{\prime }},  \label{mst}
\end{equation}%
using the tensor product $\otimes $ of general co-frames $e^{\alpha ^{\prime
}}$, which are dual to frame bases $e_{\alpha ^{\prime }}.$ In general form,
the geometric and physical constructions are performed for metric-affine
spaces (target ones) determined by geometric data $(V,\mathbf{g,}D),$ when $%
\nabla \rightarrow D$ and, in general, the nonmetricity field is nontrivial, 
$Q:=D\mathbf{g}\neq 0.$ To elaborate theories of geometric flows \cite%
{perelman1,ibsvevv22} one considers families of metrics $\mathbf{g}(\tau
)=\{g_{\alpha \beta }(\tau ,u)\},$ where $\tau $ is a temperature like
parameter considered for an interval $0\leq \tau \leq \tau _{1},$ or $\tau
=\tau _{0}$ for a fixed value. Frame vectors can be prescribed to depend, or
not, on $\tau $-parameter, i.e. $e_{\alpha ^{\prime }}(\tau ),$ or $%
e_{\alpha ^{\prime }}.$ For simplicity, we shall write only $g_{\alpha \beta
}(\tau )$ instead of $g_{\alpha \beta }(\tau ,u)$ if that will not result in
ambiguities.

\subsubsection{N-adapted metric-affine structures with nonholonomic (2+2)
splitting}

We introduce a nonlinear connection, N-connection, structure as a Whitney
sum: 
\begin{equation}
\mathbf{N}:\ TV=hV\oplus vV,  \label{ncon}
\end{equation}
which is globally defined on $V$ and its tangent bundle $TV.$ A $\mathbf{N}$
defines a conventional horizontal and vertical splitting ( h- and
v--decomposition) into respective 2-d and 2-d subspaces, $hV$ and $vV.$ In
local coordinate form, a N-connection is defined by a set of coefficients $%
N_{i}^{a}(u)$ when $\mathbf{N}=N_{i}^{a}(x,y)dx^{i}\otimes \partial
/\partial y^{a}.$\footnote{%
We can always define local coordinates $u=\{u^{\alpha }=(x^{i},y^{a})\}$
involving a conventional $2+2$ splitting into h-coordinates, $x=(x^{i}),$
and v-coordinates, $y=(y^{a}),$ for indices $j,k,...=1,2$ and $%
a,b,c,...=3,4, $ when $\alpha ,\beta ,...=1,2,3,4.$ Using partial
derivatives, local coordinate basis and a co-base are computed respectively
as $e_{\alpha }=\partial _{\alpha }=\partial /\partial u^{\beta }$ and $%
e^{\beta }=du^{\beta }.$ Transforms to arbitrary frames (tetrads /
vierbeinds) are defined as $e_{\alpha ^{\prime }}=e_{\ \alpha
^{\prime}}^{\alpha }(u)e_{\alpha }$ and $e^{\alpha ^{\prime }}=e_{\alpha \
}^{\ \alpha ^{\prime }}(u)e^{\alpha }.$ Usually, such (co) bases are
orthonormalized by the conditions $e_{\alpha \ }^{\ \alpha ^{\prime }}e_{\
\alpha ^{\prime }}^{\beta }=\delta _{\alpha }^{\beta },$ where $\delta
_{\alpha }^{\beta }$ is the Kronecker symbol.
\par
On Lorentz manifolds, a N-connection (\ref{ncon}) states a nonholonomic
distribution defining a fibred 2+2 structure. We use the term nonholonomic
Lorentz / pseudo-Riemannian manifold when a conventional h-v-splitting is
considered. Typically, "boldface" symbols are used to emphasize that certain
spaces or geometric objects are enabled (adapted) with (to) a N-connection
structure.}

N--elongated/adapted local bases, $\mathbf{e}_{\nu },$ and co-bases
(N--differentials), $\mathbf{e}^{\mu },$ are defined 
\begin{eqnarray}
\mathbf{e}_{\nu } &=&(\mathbf{e}_{i},e_{a})=(\mathbf{e}_{i}=\partial
/\partial x^{i}-\ N_{i}^{a}(u)\partial /\partial y^{a},\ e_{a}=\partial
_{a}=\partial /\partial y^{a}),\mbox{ and  }  \label{nader} \\
\mathbf{e}^{\mu } &=&(e^{i},\mathbf{e}^{a})=(e^{i}=dx^{i},\ \mathbf{e}%
^{a}=dy^{a}+\ N_{i}^{a}(u)dx^{i}),  \label{nadif}
\end{eqnarray}%
to be linear on $N_{i}^{a}.$ The term nonholonomic (equivalently,
anholonomic) is used because, for instance, a N-elongated basis (\ref{nader}%
) satisfies certain nonholonomy relations 
\begin{equation}
\lbrack \mathbf{e}_{\alpha },\mathbf{e}_{\beta }]=\mathbf{e}_{\alpha } 
\mathbf{e}_{\beta }-\mathbf{e}_{\beta }\mathbf{e}_{\alpha }= W_{\alpha
\beta}^{\gamma }\mathbf{e}_{\gamma },  \label{nonholr}
\end{equation}%
with nontrivial anholonomy coefficients 
\begin{equation}
W_{ia}^{b}=\partial _{a}N_{i}^{b},W_{ji}^{a}=\Omega _{ij}^{a}=\mathbf{e}%
_{j}\left( N_{i}^{a}\right) -\mathbf{e}_{i}(N_{j}^{a}).  \label{anhcoef}
\end{equation}%
In these formulas, $\Omega _{ij}^{a}$ define the coefficients of
N-connection curvature. If all $W_{ia}^{b}$ (\ref{anhcoef}) are zero for a $%
\mathbf{e}_{\alpha },$ such a N-adapted base is holonomic and we can write
it as a partial derivative $\partial _{\alpha }$ with $N_{i}^{a}=0.$ In
curved local coordinates, the coefficients $N_{j}^{a}$ may be nontrivial
even all $W_{\alpha \beta }^{\gamma }=0$ and we may chose a holonomic base.%
\footnote{%
For instance, we define and write a d--vector in N-adapted form as $\mathbf{X%
}=(hX,vX)$. The geometric objects on a nonholonomic manifold $\mathbf{V}$
enabled with a N-connection structure $\mathbf{N}$ are called distinguished
(i.e d-objects, d-vectors, d-tensors etc) if they are adapted to the
N--connection structure via corresponding decompositions with respect to
frames of type (\ref{nader}) and (\ref{nadif}).}

The geometric objects on a nonholonomic manifold $\mathbf{V}$ enabled with a
N-connection structure $\mathbf{N}$ (and on extensions to tangent, $T\mathbf{%
V,}$ and cotangent, $T^{\ast }\mathbf{V}$, bundles; and their tensor
products, for instance, $T\mathbf{V\otimes }T^{\ast }\mathbf{V}$) are called
distinguished (in brief, d-objects, d-vectors, d-tensors etc) if they are
adapted to the N--connection structure via corresponding decompositions with
respect to frames of type (\ref{nader}) and (\ref{nadif}). For instance, we
write a d--vector as $\mathbf{X}=(hX,vX)$.

Any spacetime metric $\mathbf{g}=(hg,vg)$ (\ref{mst}) can be represented
equivalently as a d--metric\footnote{%
Introducing coefficients of (\ref{nadif}) into (\ref{dm}) and regrouping
with respect to the coordinate dual basis, we obtain the formulas for the
coefficients in (\ref{cm}), 
\begin{equation*}
\underline{g}_{\alpha \beta }=\left[ 
\begin{array}{cc}
g_{ij}+N_{i}^{a}N_{j}^{b}g_{ab} & N_{j}^{e}g_{ae} \\ 
N_{i}^{e}g_{be} & g_{ab}%
\end{array}%
\right] .
\end{equation*}%
A metric $\mathbf{g}=\{\underline{g}_{\alpha \beta }\}$ is generic
off--diagonal if the anholonomy coefficients $W_{\alpha \beta }^{\gamma }$
are not identical to zero. For 4-d spacetimes, such a matrix can't be
diagonalized via coordinate transforms.}, when 
\begin{eqnarray}
\ \mathbf{g} &=&\ g_{ij}(x,y)\ e^{i}\otimes e^{j}+\ g_{ab}(x,y)\ \mathbf{e}%
^{a}\otimes \mathbf{e}^{b},\mbox{ in N-adapted form with }hg=\{\
g_{ij}\},vg=\{g_{ab}\};  \label{dm} \\
&=&\underline{g}_{\alpha \beta }(u)du^{\alpha }\otimes du^{\beta }.
\label{cm}
\end{eqnarray}

A\textbf{\ d--connection} $\mathbf{D}=(hD,vD)$ is defined as a linear
connection preserving under parallelism the N--connection splitting (\ref%
{ncon}). In N-adapted coefficient form with respect to frames (\ref{nader})
and (\ref{nadif}), we can write decompositions of $\mathbf{D}$ in terms of
h- and v-indices, 
\begin{equation}
\mathbf{D}=\{\mathbf{\Gamma }_{\ \alpha \beta }^{\gamma }=(L_{jk}^{i},\acute{%
L}_{bk}^{a};\acute{C}_{jc}^{i},C_{bc}^{a})\},\mbox{ where }hD=(L_{jk}^{i},%
\acute{L}_{bk}^{a})\mbox{ and }vD=(\acute{C}_{jc}^{i},C_{bc}^{a}).
\label{hvdcon}
\end{equation}

We define a nonholonomic metric-affine space by geometric data $(\mathbf{%
V,N,g,D}).$

\subsubsection{Geometric objects adapted to a N-connection structure and
nonmetricity}

The fundamental geometric d-objects of nonholonomic metric-affine space are
defined: 
\begin{eqnarray}
\mathcal{T}(\mathbf{X,Y}) &:=&\mathbf{D}_{\mathbf{X}}\mathbf{Y}-\mathbf{D}_{%
\mathbf{Y}}\mathbf{X}-[\mathbf{X,Y}],\mbox{ torsion d-tensor,  d-torsion};
\label{fundgeom} \\
\mathcal{R}(\mathbf{X,Y})&:=&\mathbf{D}_{\mathbf{X}}\mathbf{D}_{\mathbf{Y}}-%
\mathbf{D}_{\mathbf{Y}}\mathbf{D}_{\mathbf{X}}-\mathbf{D}_{\mathbf{[X,Y]}},%
\mbox{ curvature d-tensor, d-curvature};  \notag \\
\mathcal{Q}(\mathbf{X}) &:=&\mathbf{D}_{\mathbf{X}}\mathbf{g,}%
\mbox{\
nonmetricity d-fiels, d-nonmetricity}.  \notag
\end{eqnarray}%
The N-adapted coefficient formulas involving (\ref{dm}), (\ref{cm}) and (\ref%
{hvdcon}) are provided and computed in \cite%
{vmon05,bubuianu17,vacaru18}. Here we only
present, respectively, their 4-d N-adapted coefficient representations, 
\begin{equation*}
\mathcal{R}=\mathbf{\{R}_{\ \beta \gamma \delta }^{\alpha }\},\mathcal{T}=\{%
\mathbf{T}_{\ \alpha \beta }^{\gamma }\},\mathcal{Q}=\mathbf{\{Q}_{\gamma
\alpha \beta }\}.
\end{equation*}

In geometric flow and gravity theories, there are also another important
geometric d-objects: 
\begin{eqnarray*}
\mathbf{R}ic &:=&\{\mathbf{R}_{\ \beta \gamma }:=\mathbf{R}_{\ \beta \gamma
\alpha }^{\alpha }\},\mbox{ the Ricci d-tensor }; \\
\mathbf{R}sc &:=&\mathbf{g}^{\alpha \beta }\mathbf{R}_{\ \alpha \beta }, %
\mbox{ the scalar curvature },
\end{eqnarray*}%
where we uses the inverse d-tensor $\{\mathbf{g}^{\alpha \beta }\}$ of a
d-metric (\ref{dm}).

Using a d-metric $\mathbf{g}$ (\ref{dm}), we can define two important linear
connection structures: 
\begin{equation}
(\mathbf{g,N})\rightarrow \left\{ 
\begin{array}{cc}
\mathbf{\nabla :} & \mathbf{\nabla g}=0;\ _{\nabla }\mathcal{T}=0,\ 
\mbox{\
LC--connection }; \\ 
\widehat{\mathbf{D}}: & \widehat{\mathbf{Q}}=0;\ h\widehat{\mathcal{T}}=0,v%
\widehat{\mathcal{T}}=0,\ hv\widehat{\mathcal{T}}\neq 0,%
\mbox{ the canonical
d-connection}.%
\end{array}%
\right.  \label{twocon}
\end{equation}%
In this paper, "hat" labels are used for geometric d-objects defined by a $%
\widehat{\mathbf{D}}.$ Such an auxiliary d-connection defines a canonical
distortion relation 
\begin{equation}
\widehat{\mathbf{D}}[\mathbf{g}]=\nabla \lbrack \mathbf{g}]+\widehat{\mathbf{%
Z}}[\mathbf{g}],  \label{cdist}
\end{equation}%
when the canonical distortion d-tensor, $\widehat{\mathbf{Z}}$, and $\nabla
\lbrack \mathbf{g}]$ are determined by the same metric structure $\mathbf{g.}
$\footnote{%
A canonical d-connection (\ref{twocon}) is defined by N-adapted coefficients 
$\widehat{\mathbf{D}}=\{\widehat{\mathbf{\Gamma }}_{\ \alpha \beta
}^{\gamma} =(\widehat{L}_{jk}^{i},\widehat{L}_{bk}^{a},\widehat{C}_{jc}^{i}, 
\widehat{C}_{bc}^{a})\},$ for 
\begin{eqnarray*}
\widehat{L}_{jk}^{i} &=&\frac{1}{2}g^{ir}(\mathbf{e}_{k}g_{jr}+\mathbf{e}%
_{j}g_{kr}-\mathbf{e}_{r}g_{jk}),\widehat{L}_{bk}^{a}=e_{b}(N_{k}^{a})+\frac{%
1}{2}g^{ac}(\mathbf{e}_{k}g_{bc}-g_{dc}\ e_{b}N_{k}^{d}-g_{db}\
e_{c}N_{k}^{d}), \\
\widehat{C}_{jc}^{i} &=&\frac{1}{2}g^{ik}e_{c}g_{jk},\ \widehat{C}_{bc}^{a}=%
\frac{1}{2}g^{ad}(e_{c}g_{bd}+e_{b}g_{cd}-e_{d}g_{bc}).
\end{eqnarray*}%
}

The coefficients of the canonical fundamental geometric d-objects (\ref%
{fundgeom}) are labeled by "hat" symbols, for instance, $\widehat{\mathcal{R}%
}=\{\widehat{\mathbf{R}}_{\ \beta \gamma \delta }^{\alpha }\}.$ Similar
fundamental geometric objects can be defined and computed for $\nabla ,$ for
instance, $\ _{\nabla }\mathcal{R}=\{\ _{\nabla }R_{\ \beta\gamma \delta
}^{\alpha }\}$ (in such cases, boldface indices are not used). Considering
the canonical distortion relation for linear connections (\ref{cdist}), we
can compute respective canonical distortions of fundamental geometric
d-objects (\ref{fundgeom}). Such formulas relate, for instance, two
different curvature tensors, $\ _{\nabla }\mathcal{R}=\{\ _{\nabla }R_{\
\beta \gamma \delta }^{\alpha }\}$ and $\widehat{\mathcal{R}}=\{\widehat{%
\mathbf{R}}_{\ \beta \gamma \delta }^{\alpha }\}$ etc.

An arbitrary d-connection $\mathbf{D=\{\mathbf{\Gamma }}_{\ \alpha \beta
}^{\gamma }\}$ with nontrivial nonmetric d-tensor $\mathbf{Q}_{\gamma \alpha
\beta }$ can be expressed  via distortion d-tensors with respect to $%
\nabla =\{\breve{\Gamma}_{\ \alpha \beta }^{\gamma }\}$ or $\widehat{\mathbf{%
D}}=\{\widehat{\mathbf{\mathbf{\Gamma }}}_{\ \alpha \beta }^{\gamma }\},$%
\begin{equation}
\mathbf{\mathbf{\Gamma }}_{\ \alpha \beta }^{\gamma }=\breve{\Gamma}_{\
\alpha \beta }^{\gamma }+K_{\ \alpha \beta }^{\gamma }+\ ^{q}Z_{\ \alpha
\beta }^{\gamma }\mbox{ and/or  }\mathbf{\mathbf{\Gamma }}_{\ \alpha \beta
}^{\gamma }=\widehat{\mathbf{\mathbf{\Gamma }}}_{\ \alpha \beta }^{\gamma }+%
\widehat{\mathbf{K}}_{\ \alpha \beta }^{\gamma }+\ ^{q}\widehat{\mathbf{Z}}%
_{\ \alpha \beta }^{\gamma }.  \label{cdist1}
\end{equation}%
In these formulas, for instance, for $(T,Q)$-deformations of
LC-configurations, we use and construct such distortion tensors: 
\begin{equation*}
K_{\alpha \beta \gamma }=\frac{1}{2}(T_{\alpha \beta \gamma }+T_{\beta
\gamma \alpha }-T_{\gamma \alpha \beta }),\ S_{\gamma \ }^{\ \alpha \beta }=%
\frac{1}{2}(K_{\ \ \ \gamma }^{\alpha \beta }+\delta _{\gamma }^{\alpha
}T_{\ \ \ \tau }^{\tau \beta }-\delta _{\gamma }^{\beta }T_{\ \ \ \tau
}^{\tau \alpha }),\ ^{q}Z_{\ \alpha \beta \gamma }=\frac{1}{2}(Q_{\alpha
\beta \gamma }-Q_{\beta \gamma \alpha }-Q_{\gamma \beta \alpha }),
\end{equation*}%
where $Q_{\alpha \beta \gamma }:=D_{\alpha }g_{\beta \gamma }$ and $T_{\
\beta \gamma }^{\alpha }$ is computed for any $D=\{\Gamma _{\ \alpha \beta
}^{\gamma }\}$ which can an arbitrary affine connection (with coefficients
in coordinate or arbitrary frames). The torsion d-tensor $\mathbf{T}_{\
\beta \gamma }^{\alpha }$ can be computed for an arbitrary d-connection $%
\mathbf{D}=\{{\Gamma }_{\ \alpha \beta }^{\gamma }\},$ using N-adapted
bases. Such objects are defined with respect to coordinate or N-adapted
frames and used for introducing three scalar values considered in the
Weyl--Cartan geometry:%
\begin{equation*}
Rsc[D]=R=g^{\beta \gamma }\ R_{\beta \gamma },\ ^{s}T=\ S_{\gamma \ }^{\
\alpha \beta }T_{\ \alpha \beta }^{\gamma },Q=\ ^{q}Z_{\ \beta \alpha
}^{\alpha }\ ^{q}Z_{\quad \mu }^{\beta \mu }-\ ^{q}Z_{\ \beta \mu }^{\alpha
}\ ^{q}Z_{\quad \alpha }^{\beta \mu }.
\end{equation*}%
For the nonholonomic Weyl-Cartan geometry, corresponding scalar values are
defined for $\mathbf{D}=\{{\Gamma }_{\ \alpha \beta }^{\gamma }\}$ and $%
\mathbf{Q}_{\alpha \beta \gamma }:=\mathbf{D}_{\alpha }\mathbf{g}_{\beta
\gamma }$ as distortions (\ref{cdist}) of $\widehat{\mathbf{D}}=\{\widehat{%
\Gamma }_{\ \alpha \beta }^{\gamma }\}$ and can be written%
\begin{equation*}
\mathbf{R}sc[\mathbf{D}]=\ ^{s}R=\mathbf{g}^{\beta \gamma }\ \mathbf{R}%
_{\beta \gamma },\ ^{s}\mathbf{T}=\ \mathbf{S}_{\gamma \ }^{\ \alpha \beta }%
\mathbf{T}_{\ \alpha \beta }^{\gamma },\ ^{q}\mathbf{Q}=\ ^{q}\mathbf{Z}_{\
\beta \alpha }^{\alpha }\ ^{q}\mathbf{Z}_{\quad \mu }^{\beta \mu }-\ ^{q}%
\mathbf{Z}_{\ \beta \mu }^{\alpha }\ ^{q}\mathbf{Z}_{\quad \alpha }^{\beta
\mu }.
\end{equation*}

MGTs with torsion and nonmetricity are modelled for actions of type 
\begin{equation}
S(g,D,\phi )=\int \sqrt{|g|}d^{4}u[\ ^{g}\mathcal{L}+\ ^{m}\mathcal{L}],
\label{actmafmgt}
\end{equation}%
where the gravitational Lagrange density is a functional $\ ^{g}\mathcal{L}%
=F(R,\ ^{s}T,Q,T^{[m]})$ and $\ ^{m}\mathcal{L}[\phi ]$ is the Lagrange
density for conventional matter fields $\phi .$ Such a model is studied in 
\cite{harko21} but, in this work, we follow a different system of notations
considered, for instance, in \cite{bubuianu17} (for instance, we write $%
F(...)$ instead of $f(...)$). The gravitational and matter field equations
of type (\ref{effectngteq}) derived variationally from an action (\ref%
{actmafmgt}) consist sophisticate coupled systems of nonlinear PDEs. It is
very difficult to find exact/parametric solutions in such MGTs even, for
instance, certain cosmological and DE and DM models were studied in \cite%
{harko21,iosifidis22,khyllep23}.

In this work, we shall consider MGTs of type (\ref{actmafmgt}) written in
canonical d-variables $(\mathbf{g,D}=\widehat{\mathbf{D}}+\widehat{\mathbf{K}%
}+\ ^{q}\widehat{\mathbf{Z}}),$ see distortions (\ref{cdist1}), and
following a N-adapted variational calculus for actions of type 
\begin{equation}
S(g,\mathbf{D},\phi )=\int \sqrt{|\mathbf{g}|}\delta ^{4}u[\ ^{g}\widehat{%
\mathcal{L}}\mathcal{+}\ ^{e}\widehat{\mathcal{L}}+\ ^{m}\widehat{\mathcal{L}%
}],  \label{actmafmgt1}
\end{equation}%
where $\delta ^{4}u$ is the volume element defined with N-elongated
differentials (\ref{nadif}), $\ ^{g}\widehat{\mathcal{L}}=F(\widehat{\mathbf{%
R}}sc,\ ^{s}\widehat{T},T^{[m]})$ is computed as $\ ^{g}\mathcal{L}$ from (%
\ref{actmafmgt}) when $\nabla \rightarrow \widehat{\mathbf{D}},$ for $%
\widehat{\mathbf{Q}}_{\alpha \beta \gamma }=0$; $\ ^{e}\widehat{\mathcal{L}}(%
\mathbf{Q}_{\alpha \beta \gamma },\widehat{\mathbf{D}},\phi )$ includes
distortions of geometric d-objects and Lagrangians deformations for $%
\widehat{\mathbf{D}}\rightarrow \mathbf{D}$ is nontrivial $\mathbf{Q}%
_{\alpha \beta \gamma };$ and $\ ^{m}\widehat{\mathcal{L}}(\mathbf{g}%
_{\alpha \beta },\phi ,\widehat{\mathbf{D}})$ if $\ ^{m}\mathcal{L}%
(g_{\alpha \beta },\phi ,\nabla ),$ or (we can consider some simplified
models) when $\ ^{m}\widehat{\mathcal{L}}(\mathbf{g}_{\alpha \beta },\phi
)=\ ^{m}\mathcal{L}(\mathbf{g}_{\alpha \beta },\phi ).$

In a series of works \cite{bubuianu17,vacaru18,partner02,ibsvevv22}, we
proved that geometric flow and gravitational field equations in MGTs with $%
\mathbf{D}=\widehat{\mathbf{D}}$ or $\mathbf{D\rightarrow }\widehat{\mathbf{D%
}}\rightarrow \nabla $ can be decoupled and integrated in certain general
off-diagonal forms using the AFCDM. The general goal of this article is to
show how those methods can be generalized for nontrivial $\mathbf{Q}%
_{\alpha\beta \gamma }(\tau )$ and applied for research of the relativistic
Ricci flows of nonholonomic metric-affine structures, or corresponding
nonholonomic Ricci soliton equations for any $\tau _{0}.$

\subsection{Relativistic geometric flows encoding nonmetricity fields}

\label{ss22}The theory of Ricci flows has a high scientific impact in modern mathematics and physics after Grigori Perelman proved \cite{perelman1} the famous Poincar\'{e}--Thurston conjecture, see original works \cite{hamilt1} and \cite{friedan2,friedan3} and monographs \cite{monogrrf1,monogrrf2,monogrrf3} for reviews of mathematical results and
methods. A crucial step in elaborating such theories consisted in definition of the so-called $\mathcal{F}$- and $\mathcal{W}$-functionals from which the geometric flow equations (called also as R. Hamilton or Hamilton-Friedan
equations) can be proved in variational form. It is not clear how mathematically can be formulated and proved relativistic variants of such conjectures and generalizations for nonmetric/ supersymmetric/ nonassociative / noncommutative / Finsler like geometries. Nevertheless, generalizations of $\mathcal{F}$- and $\mathcal{W}$-functionals allow to prove modified versions of geometric flow equations  and solve such systems of nonlinear PDEs using the AFCDM. The results of a series of recent papers \cite{ibsvev20,ibsvevv22} demonstrate that Perelman like information thermodynamics may play an important role in the theory of quantum geometric and information, QGIF, flows. In this work, we show how such constructions
can be performed for nonmetric geometric flows for families $\tau $-evolving metric-affine data $(\mathbf{g}(\tau ),\mathbf{D}(\tau ))$ and Lagrange densities $\ ^{g}\widehat{\mathcal{L}}(\tau )+\ ^{e}\mathcal{L}(\tau )+\ ^{m}%
\widehat{\mathcal{L}}(\tau ).$

\subsubsection{$Q$-modified Perelman's F- and W-functionals in canonical nonholonomic variables}

\label{sss221}The modified Perelman's functionals for nonmetric geometric
flows are postulated in the form 
\begin{eqnarray}
\mathcal{F}(\tau ) &=&\int_{t_{1}}^{t_{2}}\int_{\Xi _{t}}e^{-f(\tau )}\sqrt{|%
\mathbf{g}(\tau )|}\delta ^{4}u[F(\widehat{\mathbf{R}}sc(\tau ),\ ^{s}%
\widehat{T}(\tau ),T^{[m]}(\tau ))+\ ^{e}\mathcal{L}(\tau )+\ ^{m}\widehat{%
\mathcal{L}}(\tau )+|\mathbf{D}(\tau )f(\tau )|^{2}],  \label{fperelm4matter}
\\
\mathcal{W}(\tau ) &=&\int_{t_{1}}^{t_{2}}\int_{\Xi _{t}}\left( 4\pi \tau
\right) ^{-2}e^{-f(\tau )}\sqrt{|\mathbf{g}(\tau )|}\delta ^{4}u[\tau (F(%
\widehat{\mathbf{R}}sc(\tau ),\ ^{s}\widehat{T}(\tau ),T^{[m]}(\tau ))+
\label{wfperelm4matt} \\
&&\ ^{e}\mathcal{L}(\tau )+\ ^{m}\widehat{\mathcal{L}}(\tau )+|\mathbf{D}%
(\tau )f(\tau )|^{2}+f(\tau )-4],  \notag
\end{eqnarray}%
where the condition $\int_{t_{1}}^{t_{2}}\int_{\Xi _{t}}\left( 4\pi \tau
\right) ^{-2}e^{-f(\tau )}\sqrt{|\mathbf{g}|}d^{4}u=1$ is imposed on the
normalizing function $f(\tau )=f(\tau ,u).$ The difference from the original
F- and W-functionals \cite{perelman1} introduced for 3-d Riemannian $\tau $%
-flows $(g(\tau ),\nabla (\tau ))$ is that in this work we study geometric
flows of canonical geometric data $(\mathbf{g}(\tau ),$ $\mathbf{N}(\tau ),%
\mathbf{D}(\tau ))$ for $Q$-deformations of nonholonomic Lorentz manifolds.

We can compute relativistic effective functionals (\ref{fperelm4matter}) and
(\ref{wfperelm4matt}) for any 3+1 splitting with 3-d closed hypersurface
fibrations $\widehat{\Xi }_{t}$ and considering nonholonomic canonical
d-connections and respective geometric variables. In general, it is possible
to work with any class of normalizing functions $f(\tau )$. Such a function
can be fixed by some constant values or some parametrization conditions
simplifying corresponding systems of nonlinear PDEs. Such $f(\tau )$ define
respective integration measures which may be important, or not, for
elaborating topological and/or geometric models. The $\mathcal{W}$%
-functional possess the properties of "minus" entropy. This can be stated by
choosing corresponding nonholonomic configurations along some causal curves
taking values $\mathcal{W}(\tau )$ on $\widehat{\Xi }_{t}.$ Using N-adapted
variations, we can derive nonmetric geometric flow evolution equations which
can be solved using the AFCDM for metrics with pseudo-Euclidean signature
even analogs of Poincar\'{e}--Thurston conjecture have not been formulated
and proven in modern mathematics.

\subsubsection{Hamilton-Friedan geometric flow equations with nonmetricity}

There are two possibilities to derive geometric flow equations from
functionals $\mathcal{F}(\tau )$ (\ref{fperelm4matter}) and $\mathcal{W}%
(\tau )$ (\ref{wfperelm4matt}). In the first case, we can use $\mathbf{D}%
(\tau )$ instead of $\nabla (\tau )$ and reproduce in N-adapted form all
covariant differential and integral calculus from \cite%
{perelman1,monogrrf1,monogrrf2,monogrrf3}. This would consist proofs on some
hundred of pages.

We can follow geometric abstract principles \cite{misner} when all geometric
and physically important objects and fundamental physical equations are
derived by corresponding generalizations of Riemannian geometry to certain
nonholonomic metric-affine geometries with $\nabla(\tau)\rightarrow \mathbf{D%
}(\tau )=\widehat{\mathbf{D}}(\tau )+\widehat{\mathbf{K}}(\tau )+\ ^{q}%
\widehat{\mathbf{Z}}(\tau )$ respective generalizations of Ricci, torsion,
and energy-momentum d-tensors. Such an abstract geometric calculus allows to
prove for some primary data $(\mathbf{g}=\{\mathbf{g}_{\mu \nu
}=[g_{ij},g_{ab}]\}, \mathbf{N}= \{ N_{i}^{a}\},\mathbf{D},\ ^{tot}\widehat{%
\mathcal{L}}= \ ^{e}\mathcal{L} + \ ^{m}\widehat{\mathcal{L}}),$ see
definitions related to formulas (\ref{actmafmgt1}) the nonholonomic
geometric flow evolution equations: 
\begin{eqnarray}
\partial _{\tau }g_{ij}(\tau ) &=&-2[\widehat{\mathbf{R}}_{ij}(\tau )-\
^{tot}\widehat{\Upsilon }_{ij}(\tau )];\ \partial _{\tau }g_{ab}(\tau )=-2[%
\widehat{\mathbf{R}}_{ab}(\tau )-\ ^{tot}\widehat{\Upsilon }_{ab}(\tau )];
\label{ricciflowr2} \\
\widehat{\mathbf{R}}_{ia}(\tau ) &=&\widehat{\mathbf{R}}_{ai}(\tau )=0;%
\widehat{\mathbf{R}}_{ij}(\tau )=\widehat{\mathbf{R}}_{ji}(\tau );\widehat{%
\mathbf{R}}_{ab}(\tau )=\widehat{\mathbf{R}}_{ba}(\tau );\   \notag \\
\partial _{\tau }\widehat{f}(\tau ) &=&-\widehat{\square }(\tau )[\widehat{f}%
(\tau )]+\left\vert \widehat{\mathbf{D}}(\tau )[\widehat{f}(\tau
)]\right\vert ^{2}-\ ^{s}\widehat{R}(\tau )+\ ^{tot}\widehat{\Upsilon }_{\
\alpha }^{\alpha }(\tau ).  \notag
\end{eqnarray}%
In these formulas, there are used such geometric d-objects and N-adapted
operators: $\widehat{\square }(\tau )=\widehat{\mathbf{D}}^{\alpha }(\tau )%
\widehat{\mathbf{D}}_{\alpha }(\tau )$ when the conditions $\widehat{\mathbf{%
R}}_{ia}=\widehat{\mathbf{R}}_{ai}=0$ for the Ricci tensor $\widehat{R}ic[%
\widehat{\mathbf{D}}]=\{\widehat{\mathbf{R}}_{\alpha \beta} = [\widehat{R}%
_{ij},\widehat{R}_{ia},\widehat{R}_{ai},\widehat{R}_{ab,}]\}$ are necessary
if we want to keep the metric $\mathbf{g}(\tau )$ to be symmetric under
nonholonomic Ricci flow evolution. Such constraints are not obligatory, for
instance, in nonassociative geometric flow theory with nonsymmetric metrics 
\cite{partner02}.

The definition of $\ ^{tot}\widehat{\Upsilon }_{ab}(\tau )$ from (\ref%
{ricciflowr2}) will be discussed in section (\ref{ssriccisolit}) for $%
\tau=\tau _{0}.$ Here we note that such equations describe nonmetric
geometric flow evolution of d-metrics $\mathbf{g}_{\mu \nu }(\tau )$
described in nonholonomic canonical variables. Alternatively, such equations
can be introduced as relativistic generalizations and nonholonomic canonical
deformations of the R. Hamilton \cite{hamilt1} and D. Friedan \cite%
{friedan2,friedan3} Ricci flow equations for $\nabla (\tau ).$

The normalizing function $f(\tau )$ can be re-defined in such a way that it
compensates certain $Q$-deforms, or other type nonholonomic distortions,
when $\widehat{f}(\tau )\rightarrow f(\tau )$ for 
\begin{equation*}
\partial _{\tau }f=-\square f+\left\vert \mathbf{D}f\right\vert ^{2}-\
^{s}R+\ ^{tot}\Upsilon _{\ \alpha }^{\alpha }.
\end{equation*}%
Such an equation involves nonlinear partial differential operators and
usually it is not possible to solve it in an explicit form and define the
evolution of topological configurations determined, for instance, by
nontrivial nonmetric structures. Nevertheless, we can fix a variant of $%
\widehat{f}(\tau )$ which together with some off-diagonal ansatz for metrics
we can solve the nonholonomic system of nonlinear PDEs (\ref{ricciflowr2})
in certain general/ parametric forms and then to re-define the constructions
in for arbitrary systems of reference, other types of distortions of
connections and normalizing functions. The formulas for nonholonomic
frame/coordinate/ normalizing transforms could be found in certain series/
recurrent form when the solutions of geometric flow equations are generated
in explicit form.

\subsection{Nonmetric Ricci solitons and modified Einstein equations}

\label{ssriccisolit}

A nonholonomic and nonmetric Ricci soliton configuration is a self-similar
one for the geometric flow equations (\ref{ricciflowr2}). For Riemannian
metrics, such configurations homothetically strink, remain steady or expand
under geometric flow evolution, see details in \cite%
{perelman1,hamilt1,monogrrf1,monogrrf2,monogrrf3}, and can be respectively
studied for a fixed point $\tau =\tau _{0}$. Considering relativistic and
torsion and nonmetricity modified nonlinear systems with $\partial _{\tau }%
\mathbf{g}_{\mu \nu }=0$ and for a specific choice of the normalizing
geometric flow function $f,$ the equations (\ref{ricciflowr2}) transform
into nonholonomic Ricci soliton equations encoding $Q$-distortions into
effective sources. Such systems of nonlinear PDEs are equivalent to modified
Einstein equations in nonholonomic metric-affine gravity for corresponding
definitions of effective sources $\ ^{tot}\widehat{\Upsilon }_{\alpha
\beta}(\tau _{0})=[\ ^{tot}\widehat{\Upsilon }_{ij},\ ^{tot}\widehat{%
\Upsilon }_{ab}],$ 
\begin{eqnarray}
\widehat{\mathbf{R}}_{ij} &=&\ ^{tot}\widehat{\Upsilon }_{ij},\ \widehat{%
\mathbf{R}}_{ab} =\ ^{tot}\widehat{\Upsilon }_{ab},  \label{nonheinst} \\
\widehat{\mathbf{R}}_{ia} &=&\widehat{\mathbf{R}}_{ai}=0;\widehat{\mathbf{R}}%
_{ij}=\widehat{\mathbf{R}}_{ji};\widehat{\mathbf{R}}_{ab}=\widehat{\mathbf{R}%
}_{ba}.  \notag
\end{eqnarray}

The effective sources in (\ref{nonheinst}) are can be parameterized as for
the effective Lagrangians (\ref{actmafmgt1}) 
\begin{equation}
\ ^{tot}\widehat{\mathbf{\Upsilon }}_{\mu \nu }:=\varkappa \left( \ ^{tot}%
\widehat{\mathbf{T}}_{\mu \nu }-\frac{1}{2}\mathbf{g}_{\mu \nu }\ ^{tot}%
\widehat{\mathbf{T}}\right) =(\frac{\partial F}{\partial \ ^{s}R})^{-1}\ ^{m}%
\widehat{\mathbf{\Upsilon }}_{\mu \nu }+\ ^{F}\widehat{\mathbf{\Upsilon }}%
_{\mu \nu }+\ ^{e}\widehat{\mathbf{\Upsilon }}_{\mu \nu },  \label{totsourc}
\end{equation}%
where $\varkappa $ is determined in standard form by the Newton
gravitational constant $G,$%
\begin{equation*}
\ ^{m}\widehat{\mathbf{T}}_{\mu \nu }:=-\frac{2}{\sqrt{|\mathbf{g}|}}\frac{%
\delta (\sqrt{|\mathbf{g}|}\ ^{m}\widehat{\mathcal{L}}\mathcal{)}}{\delta 
\mathbf{g}^{\mu \nu }}=2\frac{\delta \ ^{m}\widehat{\mathcal{L}}}{\delta 
\mathbf{g}^{\mu \nu }}+\mathbf{g}_{\mu \nu } \ ^{m}\widehat{\mathcal{L}}.
\end{equation*}%
For the full system relating nonholonomic Ricci solitons to modified
gravity, the effective energy-momentum d-tensor is computed for 
\begin{equation*}
\ ^{F}\widehat{\mathbf{T}}_{\beta \gamma }=[\frac{1}{2}(F-\frac{\partial F}{%
\partial \ ^{s}R})\mathbf{g}_{\beta \gamma }-(\mathbf{g}_{\beta \gamma }%
\widehat{\mathbf{D}}_{\alpha }\widehat{\mathbf{D}}^{\alpha }-\widehat{%
\mathbf{D}}_{\beta }\widehat{\mathbf{D}}_{\gamma })\frac{\partial F}{%
\partial \ ^{s}R}](\frac{\partial F}{\partial \ ^{s}R})^{-1},
\end{equation*}%
when 
\begin{equation}
\ ^{tot}\widehat{\mathbf{T}}_{\mu \nu }=(\frac{\partial F}{\partial \ ^{s}R}%
)^{-1}\ ^{m}\widehat{\mathbf{T}}_{\mu \nu }+\ ^{F}\widehat{\mathbf{T}}_{\mu
\nu }+\ ^{e}\widehat{\mathbf{T}}_{\mu \nu }.  \label{totem}
\end{equation}%
Choosing $F(\ ^{s}R)=\ ^{s}R$ and the Levi-Civita connection $\mathbf{D}%
\rightarrow \nabla ,$ we can relate above formulas to GR.

\subsubsection{Connecting nonholonomic solitons to nonmetric modified
gravitational equations}

The gravitational field equations in Weyl-Cartan MGT can be constructed by
considering variations of the action on a metric-affine manifold determined
by geometric objects using such values constructed for the affine connection 
$\mathbf{D}$ expressed as a canonical distortion (\ref{cdist1}) from $%
\widehat{\mathbf{D}}$ involving $\ ^{e}\widehat{\mathbf{T}}_{\mu \nu }$ as
an effective source containing nontrivial contributions from $\mathbf{Q}%
_{\alpha \beta \gamma }:$ 
\begin{eqnarray*}
\widehat{\mathbf{H}}_{\lambda }^{\ \mu \nu } &=&-\frac{2}{\sqrt{|\mathbf{g}|}%
}\frac{\delta (\sqrt{|\mathbf{g}|}\ \delta \ ^{m}\widehat{\mathcal{L}}%
\mathcal{)}}{\delta \widehat{\mathbf{\mathbf{\Gamma }}}_{\ \mu \nu }^{\alpha
}}\mbox{  the canonical hyper-momentum d-tensor }, \\
&\mbox{and}& \mathbf{A}_{\mu \alpha \nu } = \frac{\partial F}{\partial Q}(%
\mathbf{g}_{\mu \nu }\ ^{q}\mathbf{Z}_{\ \ \alpha \nu }^{\alpha }+\mathbf{g}%
_{\alpha \nu }\ ^{q}\mathbf{Z}_{\mu \beta }^{\quad \beta }-\ ^{q}\mathbf{Z}%
_{\ \mu \alpha \nu }-\ ^{q}\mathbf{Z}_{\ \nu \alpha \mu }).
\end{eqnarray*}
Such values can be constructed in terms of $\widehat{\mathbf{\mathbf{\Gamma }%
}}_{\ \mu \nu }^{\alpha }$ and/or $\mathbf{\mathbf{\Gamma }}_{\ \mu
\nu}^{\alpha }.$ In coordinate bases and in non N-adapted form, such results
are presented, for instance by formulas (23) - (26) in \cite{harko21}.

In this work, we consider a model of nonholonomic nonmetric Ricci solitons
with a Weyl d-vector $\mathbf{q}_{\alpha },$ when $\mathbf{Q}_{\alpha
\beta\gamma }=\mathbf{q}_{\alpha }\mathbf{g}_{\beta \gamma }$ and nontrivial
d-torsion $\mathbf{T}_{\mu \nu \alpha }=\mathbf{A}_{\nu }\mathbf{g}_{\mu
\alpha }-\mathbf{A}_{\alpha }\mathbf{g}_{\mu \nu },$ for $\mathbf{A}_{\mu}=q 
\mathbf{q}_{\mu }, q=const.$ For such approximations considering $\frac{%
\partial F}{\partial T}=1/2q\frac{\partial F}{\partial Q},$ the variational
N-adapted gravitational field equations with $\widehat{\mathbf{D}}$ can be
written in the form:%
\begin{eqnarray}
\widehat{\mathbf{D}}_{\beta }\frac{\partial F}{\partial \ ^{s}R}+[\frac{%
\partial F}{\partial Q}+ (1-q)\frac{\partial F}{\partial \ ^{s}R}]\mathbf{q}%
_{\beta } &=&0;  \label{modifcomp} \\
\widehat{\mathbf{R}}_{\ \beta \gamma } &=&\ ^{q}\widehat{\Upsilon }_{\alpha
\beta }(\tau _{0})  \label{riccsolq}
\end{eqnarray}%
for effective $Q$-source%
\begin{eqnarray}
\ ^{q}\widehat{\Upsilon }_{\alpha \beta } &=&(\frac{\partial F}{\partial \
^{s}R})^{-1}[\frac{F}{2}\mathbf{g}_{\alpha \beta }+\frac{\partial F}{%
\partial \ ^{m}T}(\widehat{\mathbf{T}}_{\alpha \beta }-\mathbf{g}_{\alpha
\beta }\ ^{m}\widehat{\mathcal{L}})-\mathbf{g}_{\alpha \beta }\widehat{%
\square }\frac{\partial F}{\partial \ ^{s}R}-\widehat{\mathbf{D}}_{\alpha }%
\widehat{\mathbf{D}}_{\beta }\frac{\partial F}{\partial \ ^{s}R}  \notag \\
&&-(q-\frac{1}{2}+\frac{3}{q})(\widehat{\mathbf{D}}_{\alpha }\mathbf{q}%
_{\beta }-\widehat{\mathbf{D}}_{\beta }\mathbf{q}_{\alpha })(\frac{\partial F%
}{\partial \ ^{s}R})-\frac{1}{2}\widehat{\mathbf{T}}_{\alpha \beta }]
\label{qsourc} \\
&&+(2+\frac{1}{2}-\frac{3}{q})\mathbf{q}_{\alpha }\mathbf{q}_{\beta }-3(%
\frac{1}{2}-\frac{1}{q})(\widehat{\mathbf{D}}_{\alpha }\mathbf{q}_{\beta }-%
\widehat{\mathbf{D}}_{\beta }\mathbf{q}_{\alpha })+\frac{1}{2}(1+q)(\mathbf{q%
}_{\tau }\mathbf{q}^{\tau })~\mathbf{g}_{\alpha \beta }.  \notag
\end{eqnarray}

We defined the system of constraints and nonlinear PDEs (\ref{modifcomp})-(%
\ref{qsourc}) in a form that for $\widehat{\mathbf{D}}\rightarrow \nabla $
it transforms into respective equations (37) and (38) in \cite{harko21}. For
such nonholonomic Ricci soliton equations, we can decouple and integrate in
certain general forms the modified Einstein equations (\ref{riccsolq}) if
the $Q$-source (\ref{qsourc}) is generated by two effective sources (see
below). It is not possible to decouple such equations for generic
off-diagonal $\mathbf{g}_{\beta \gamma }$ if it is considered only the
LC-connection $\nabla $ and/or general nonmetricity fields.

\subsubsection{Generating sources for $\protect\tau $-running
quasi-stationary effective matter fields and nonmetricity deformations}

In this work, we shall construct and study physical implications of
quasi-stationary solutions of nonmetric geometric flow equations (\ref%
{ricciflowr2}) when the metric (\ref{mst}) (in equivalent form, the d-metric
(\ref{dm})) is determined by N-adapted coefficients $\mathbf{g}%
(\tau)=[g_{i}(\tau ),g_{a}(\tau ),N_{i}^{a}(\tau )],$ when such coefficients
do not depend on variable $y^{4}=t$ and can be parameterized in the form 
\begin{equation}
g_{i}(\tau )=e^{\psi {(\tau ,x^{j})}},g_{a}(\tau )=h_{a}(\tau
,x^{k},y^{3}),\ N_{i}^{3}=w_{i}(\tau
,x^{k},y^{3}),\,\,\,\,N_{i}^{4}=n_{i}(\tau ,x^{k},y^{3}).
\label{stationarydm}
\end{equation}

Let us consider effective sources $\ ^{tot}\widehat{\Upsilon }_{ab}(\tau )$
from (\ref{ricciflowr2}) which via N--adapted frames can be parameterized in
the form 
\begin{equation}
\mathbf{\Upsilon }_{\ \nu }^{\mu }(\tau )=\mathbf{e}_{\ \mu }^{\mu ^{\prime
}}(\tau )\mathbf{e}_{\nu }^{\ \nu ^{\prime }}(\tau )[~\ ^{tot}\mathbf{%
\Upsilon }_{\mu ^{\prime }\nu ^{\prime }}(\tau )-\frac{1}{2}~\partial _{\tau
}\mathbf{g}_{\mu ^{\prime }\nu ^{\prime }}(\tau )]=[~\ _{h}^{tot}\Upsilon
(\tau ,{x}^{k})\delta _{j}^{i},~\ ^{tot}\Upsilon (\tau ,x^{k},y^{3})\delta
_{b}^{a}].  \label{dsourcparam}
\end{equation}%
In these formulas, there are considered $\tau $-families of vierbein
transforms $\mathbf{e}_{\ \mu ^{\prime }}^{\mu }(\tau )=\mathbf{e}_{\ \mu
^{\prime }}^{\mu }(\tau ,u^{\gamma })$ and their dual $\mathbf{e}_{\nu }^{\
\nu ^{\prime }}(\tau ,u^{\gamma })$, when $\mathbf{e}_{\ }^{\mu }=\mathbf{e}%
_{\ \mu ^{\prime }}^{\mu }du^{\mu ^{\prime }}.$ The values $[\
_{h}^{tot}\Upsilon (\tau ,{x}^{k}),\ ^{tot}\Upsilon (\tau ,x^{k},y^{3})]$
can be fixed as generating functions for (effective) matter sources imposing
nonholonomic frame constraints on quasi-stationary distributions of
(effective) matter fields. In particular, we can change $\ ^{tot}\Upsilon
_{\mu ^{\prime }\nu ^{\prime }}(\tau )\rightarrow \ ^{q}\widehat{\Upsilon }%
_{\alpha \beta }(\tau )$ for modeling nonholonomic flow evolution of a $Q$%
-source (\ref{qsourc}), when the generating sources are written in the form 
\begin{equation}
\ ^{q}\mathbf{\Upsilon }_{\ \nu }^{\mu }(\tau )=[~\ _{h}^{q}\Upsilon (\tau ,{%
x}^{k})\delta _{j}^{i},~\ ^{q}\Upsilon (\tau ,x^{k},y^{3})\delta _{b}^{a}].
\label{qgenersourc}
\end{equation}

In Appendix \ref{appendixa}, we show how nonlinear systems of PDEs (\ref%
{ricciflowr2}) can be decoupled in general forms for any quasi-stationary
ansatz (\ref{stationarydm}) and any variant of generating sources (\ref%
{dsourcparam}) or (\ref{qgenersourc}). For certain classes of nonholonomic
constraints and small parametric deformations, we are able to change
symbolic data $[\ _{h}^{q}\Upsilon ,\ ^{q}\Upsilon ]$ into some
approximations of (\ref{qsourc}) and study in explicit form for
contributions from any $\frac{\partial F}{\partial \ ^{s}R},\frac{\partial F%
}{\partial \ ^{m}T}$ and/or $q$-term. In general, we can consider that $[\
_{h}^{tot}\Upsilon ,~\ ^{tot}\Upsilon ]$ or $[\ _{h}^{q}\Upsilon ,\
^{q}\Upsilon ]$ impose certain nonholonomic constraints on respective
geometric evolution / dynamical field generating sources which allow to
generate solutions with nontrivial canonical d-torsion $\widehat{\mathbf{T}}%
_{\ \alpha \beta }^{\gamma }(\tau ,x^{k},y^{3}).$ Such nonholnomic values
can be eliminated by additional nonholonomic constraints $\widehat{\mathbf{D}%
}[\mathbf{g}]\rightarrow \nabla \lbrack \mathbf{g}]$ even the d-torsion $%
\mathbf{T}_{\mu \nu \alpha }=\mathbf{A}_{\nu }\mathbf{g}_{\mu \alpha }- 
\mathbf{A}_{\alpha }\mathbf{g}_{\mu \nu },$ for $\mathbf{A}_{\mu }=q\mathbf{q%
}_{\mu },q=const,$ can be nonzero because of nonmetricity.

\section{Off-diagonal quasi-stationary solutions encoding nonmetricity}

\label{sec3} In this section, we construct and analyze physical properties
of two classes of respective nonholonomic geometric flow and Ricci soliton
equations encoding quasi-stationary nonmetricity effects. First, we consider
examples of nonmetric solitionic hierarchies. Then, we generate wormhole
solutions determined by nonmetric fields and study generic off-diagonal
deformations, ellipsoidal deformations and embedding into nonmetric
backgrounds determined by solitonic hierarchies with general or small
parametric polarizations. In Appendix \ref{appendixb}, we provide necessary
formulas for bi-Hamilton structures and solitonic hierarchies.

\subsection{Effective nonmetric and $\protect\tau $-running Einstein
equations}

Let us consider a system of nonlinear PDEs (\ref{ricciflowr2}), for
nonmetric Ricci flows, or (\ref{nonheinst}), for nonmetric Ricci solitons,
with generating sources of nonmetric type $[\ _{h}^{q}\Upsilon (\tau), \
^{q}\Upsilon (\tau )]$ (\ref{qgenersourc}). Such effective sources can be
substituted by formulas of type $[\ _{h}^{tot}\Upsilon (\tau ),\
^{tot}\Upsilon (\tau )]$ (\ref{dsourcparam}) involving additional effective
sources $\frac{1}{2}\partial _{\tau }\mathbf{g}_{\mu \nu }(\tau ).$ Such
conditions involve a more special class of nonholonomic constraints on the
geometric evolution and dynamics of effective sources which allows to
decouple the nonlinear systems of PDEs in general form. For elaborating
evolution scenarios in explicit forms, we can consider product
parameterizations of type $\mathbf{g}_{\mu \nu }(\tau ,x^{k},y^{3})= \ ^{1}%
\mathbf{g}_{\mu \nu}(\tau )\times \ ^{2}\mathbf{g}_{\mu \nu }(x^{k},y^{3})$.
In this work, we shall write the general form only parametric solutions in
terms of generating functions and generating sources without discussing
particular details on how we may apply methods with separation of variables.
In abstract geometric form, any quasi-stationary d-metric generated by a $\
^{q}\mathbf{\Upsilon }_{\ \nu }^{\mu }(\tau )$ and respective generating
functions and nonlinear symmetries to effective cosmological constants, see
details in appendix \ref{assnonlinearsym}, can be transformed by frame and
connection deformations into a more general $\mathbf{\Upsilon }_{\ \nu
}^{\mu }(\tau ).$ Different types of nonholonomic deformations determined by
a $\ ^{q}\mathbf{\Upsilon }_{\ \nu }^{\mu }$ or a general type $\mathbf{%
\Upsilon }_{\ \nu }^{\mu }$, and their physical properties and implications
in physical theories can be studied using Perelman thermodynamic variables
as we shall consider in section \ref{sec4}.

With respect to N-adapted frames (\ref{nader}) and (\ref{nadif}), we can
write the modified Einstein equations (\ref{nonheinst}) in $\tau $%
-parametric form for nonmetric sources (\ref{qgenersourc}) and using the
canonical d-connection $\widehat{\mathbf{D}},$ 
\begin{eqnarray}
\widehat{\mathbf{R}}_{\ \ \beta }^{\alpha }(\tau ) &=&\ ^{q}\mathbf{\Upsilon 
}_{\ \ \beta }^{\alpha }(\tau ),  \label{cdeq1} \\
\widehat{\mathbf{T}}_{\ \alpha \beta }^{\gamma }(\tau ) &=&0,\mbox{ for }%
\nabla .  \label{lccond1}
\end{eqnarray}%
The equations (\ref{lccond1}) do not involve zero conditions for another
types of torsion which may be present in theory, for instance, of type $%
\mathbf{T}_{\mu \nu \alpha }=\mathbf{A}_{\nu }\mathbf{g}_{\mu \alpha }-%
\mathbf{A}_{\alpha }\mathbf{g}_{\mu \nu }.$ Such a torsion is induced by a
nontrivial nonmetricity d-vector $\mathbf{A}_{\nu }$ for a d-metric $\mathbf{%
g}_{\mu \alpha }$ defined as a solution of (\ref{cdeq1}).

We note that, in general, for above considered systems of nonmetric
gravitational and matter fields, 
\begin{equation*}
\widehat{\mathbf{D}}^{\beta }\widehat{\mathbf{E}}_{\ \ \beta }^{\alpha }= 
\widehat{\mathbf{D}}(\widehat{\mathbf{R}}_{\ \ \beta }^{\alpha }- \frac{1}{2}%
\ ^{s}\widehat{R}\delta _{\ \ \beta }^{\alpha })\neq 0 \mbox{ and } \widehat{%
\mathbf{D}}^{\beta }\widehat{\mathbf{\Upsilon }}_{\ \ \beta }^{\alpha }\neq
0,
\end{equation*}
which is typical for nonholonomic systems. In some similar forms, such
nonholonomic configurations are modeled in nonholonomic mechanics when the
conservation laws are not formulated in a standard form. For mechanical
systems, there are introduced the so-called Lagrange multiples associated to
certain classes of nonholonomic constraints. Solving the constraint
equations, it is possible to re-define the variables. Such nonholonomic
variables allow us to introduce new effective Lagrangians and, finally, to
define conservation laws in certain standard form if $\mathbf{Q}_{\alpha
\beta \gamma }=0.$ In explicit general forms, such constructions can be
performed only for some "toy" models. Using distortions of connections, we
can rewrite (\ref{cdeq1}) in terms of $\nabla ,$ when $\nabla ^{\beta }E_{\
\ \beta }^{\alpha }=\nabla ^{\beta }T_{\ \ \beta }^{\alpha }=0$ for $\mathbf{%
Q}_{\alpha \beta \gamma}\rightarrow 0.$

In Appendix \ref{appendixb}, we show how using the AFCDM the equations (\ref%
{cdeq1}) and (\ref{lccond1}) can be decoupled and integrated in general
quasi-stationary forms for certain prescribed nonmetric effective sources (%
\ref{qgenersourc}).

\subsection{Nonmetric solitonic hierarchies}

\label{ss32} Nonholonomic geometric flow models with solitonic hierarchies,
in metric compatible form, are studied in sections 4 and 5 of \cite%
{ibsvevv22} for constructing theories of geometric information flows of
nonholonomic Einstein systems. Similar geometric models can be elaborated
for solutions of\ $\tau $-modified Einstein equations (\ref{cdeq1}), when
the effective source $\ ^{q}\mathbf{\Upsilon }_{\ \ \beta }^{\alpha }(\tau )$
is determined by nonmetricity fields as we considered in previous
subsection. The physical interpretation of such off-diagonal solutions
involving solitonic wave/ distributions and describing geometric evolution
flow processes is different. In this work, the nonmetricity is encoded into
effective generating sources. We present a brief summary on the theory of
quasi-stationary solitonic hierarchies and their nonmetric $\tau $-evolution
in Appendix \ref{appendixb}. The goal of this subsection is to provide
explicit formulas for general off-diagonal solutions defining nonmetric
geometric flow equations determined by solitonic distributions for
generating functions and/or generating sources written in solitonic
functional form, for instance, as $[\ _{1}^{q}\Upsilon \lbrack \mathbf{\wp }%
],\ _{2}^{q}\Upsilon \lbrack \mathbf{\wp }]]$, see formulas (\ref%
{qgenersourc1}).

\subsubsection{$\protect\tau $-running quasi-stationary generating functions
with solitonic hierarchies}

We show how $\tau $-evolution of quasi-stationary solitonic configurations
can be defined by respective classes of generating functions when the
nonmetricity generating source $\ ^{q}\mathbf{\Upsilon }_{\ \ \beta}^{\alpha
}(\tau )$ is an arbitrary one (i.e. it is not obligatory of solitonic
nature). For instance, we consider that $\Phi \lbrack \mathbf{\wp }]$ is any
functional on a solitonic hierarchy $\wp (\tau ,x^{i},y^{3})$ as we stated
for d-metrics $\mathbf{g}(\tau )=(g_{i}[\mathbf{\wp }],g_{a}[\mathbf{\wp }])$
(\ref{solitondm}). Using nonlinear symmetries (\ref{nonlinsymrex}), we can
consider as a generating function any coefficient 
\begin{equation*}
h_{4}[\tau ,x^{i},\mathbf{\wp }]=h_{4}^{[0]}(\tau ,x^{i})-\Phi ^{2}[\mathbf{%
\wp }]/4\ _{2}^{q}\Lambda (\tau ),\mbox{ for }h_{4}^{\ast }(\tau )\neq 0.
\end{equation*}%
We can also express 
\begin{equation*}
\ (\Psi ^{2})^{\ast }=\ \left\vert \ _{2}^{q}\Upsilon (\tau )\right\vert
(h_{4})^{\ast }\mbox{ and }\Psi ^{2}[\tau ,x^{i},\mathbf{\wp }]=\left\vert \
\ _{2}^{q}\Upsilon (\tau )\right\vert h_{4}-\int dy^{3}\left\vert \
_{2}^{q}\Upsilon (\tau )\right\vert ^{\ast }h_{4}
\end{equation*}%
and use such a $Q$-deformed (by a generating source $\ _{2}^{q}\Upsilon
(\tau)$) solitonic configuration $\Psi (\tau )$ as a new generating
function. For simplicity, the integration function $h_{4}^{[0]}(\tau ,x^{i})$
can be approximated to some $\tau $-running values or constants when the
generating functions are some functionals of type $h_{4}[\mathbf{\wp }],\Psi
\lbrack \mathbf{\wp }]$ and $\Phi \lbrack \mathbf{\wp }].$

We can express above quadratic element in three equivalent forms: 
\begin{eqnarray}
ds^{2}(\tau ) &=&e^{\ \psi \lbrack \mathbf{\wp }%
]}[(dx^{1})^{2}+(dx^{2})^{2}]+  \label{gensolstat} \\
&&\left\{ 
\begin{array}{c}
\begin{array}{c}
-\frac{[h_{4}^{\ast }]^{2}}{|\int dy^{3}[\ _{2}^{q}\mathbf{\Upsilon }(\tau
)h_{4}]^{\ast }|\ h_{4}}[dy^{3}+\frac{\partial _{i}[\int dy^{3}(\ \ _{2}^{q}%
\mathbf{\Upsilon }(\tau ))\ h_{4}^{\ast }]}{\ \ _{2}^{q}\mathbf{\Upsilon }%
(\tau )\ h_{4}^{\ast }}dx^{i}]- \\ 
h_{4}[dt+(\ _{1}n_{k}(\tau )+\ _{2}n_{k}(\tau )\int dy^{3}\frac{[h_{4}^{\ast
}]^{2}}{|\int dy^{3}[\ \ _{2}^{q}\mathbf{\Upsilon }(\tau )h_{4}]^{\ast }|\
[h_{4}]^{5/2}}\ )dx^{k}]; \\ 
\mbox{ or }%
\end{array}
\\ 
\begin{array}{c}
-\frac{[\Psi ^{\ast }]^{2}}{4(\ _{2}^{q}\mathbf{\Upsilon }%
)^{2}\{h_{4}^{[0]}-\int dy^{3}[\Psi ^{2}]^{\ast }/4(\ \ _{2}^{q}\mathbf{%
\Upsilon })\}}[dy^{3}+\frac{\partial _{i}\ \Psi }{\ \partial _{3}\Psi }%
dx^{i}]-(h_{4}^{[0]}(\tau )-\int dy^{3}\frac{[\Psi ^{2}]^{\ast }}{4(\
_{2}^{q}\mathbf{\Upsilon })}) \\ 
\lbrack dt+(_{1}n_{k}(\tau )+\ _{2}n_{k}(\tau )\int dy^{3}\frac{[(\Psi
)^{2}]^{\ast }}{4(\ _{2}^{q}\mathbf{\Upsilon }(\tau ))^{2}|h_{4}^{[0]}-\int
dy^{3}[\Psi ^{2}]^{\ast }/4(\ _{2}^{q}\mathbf{\Upsilon }(\tau ))|^{5/2}}%
)dx^{k}]; \\ 
\mbox{ or }%
\end{array}
\\ 
\begin{array}{c}
-\frac{\Phi ^{2}[\Phi ^{\ast }]^{2}}{|\ _{2}^{q}\Lambda (\tau )\int dy^{3}\
_{2}^{q}\mathbf{\Upsilon }(\tau )[\Phi ^{2}]^{\ast }|[g_{4}^{[0]}(\tau
)-\Phi ^{2}/4\ _{2}^{q}\Lambda (\tau )]}[dy^{3}+\frac{\partial _{i}\ \int
dy^{3}\ \ _{2}^{q}\mathbf{\Upsilon }(\tau )\ [\Phi ^{2}]^{\ast }}{\ \
_{2}^{q}\mathbf{\Upsilon }(\tau )\ [\Phi ^{2}]^{\ast }}dx^{i}]-(h_{4}^{[0]}-%
\frac{\Phi ^{2}}{4\ _{2}^{q}\Lambda (\tau )}) \\ 
\lbrack dt+(_{1}n_{k}(\tau )+\ _{2}n_{k}(\tau )\int dy^{3}\frac{\Phi
^{2}[\Phi ^{\ast }]^{2}}{|\ _{2}^{q}\Lambda (\tau )\int dy^{3}\ \ \ _{2}^{q}%
\mathbf{\Upsilon }(\tau )[\Phi ^{2}]^{\ast }|[g_{4}^{[0]}(\tau )-\Phi
^{2}/4\ _{2}^{q}\Lambda (\tau )]^{5/2}})dx^{k}].%
\end{array}%
\end{array}%
\right.  \notag
\end{eqnarray}%
In these formulas, there are used respective generating data: $\{{\small h}%
_{4}{\small [\wp ],\ }_{2}^{q}{\small \Upsilon (\tau ),\ }_{2}^{q}{\small %
\Lambda (\tau )}\}$ from (\ref{offdsolgenfgcosmc}); $\{{\small \Psi \lbrack
\wp ],\ }_{2}^{q}{\small \Upsilon (\tau )}\}$ from (\ref{qeltors}); and $\{%
{\small \Phi \lbrack \wp ],\ }_{2}^{q}{\small \Lambda (\tau ),\ }_{2}^{q}%
{\small \Upsilon (\tau )}\}$ from (\ref{offdiagcosmcsh}).

The solutions (\ref{gensolstat}) can be re-defined for $\eta (\tau )$%
-polarizations in a form (\ref{offdiagpolfr}) (for instance, with
functionals $\eta _{4}[\mathbf{\wp }]$) or considering $\kappa $-parametric
deformations to $\tau $--families of quasi-stationary d-metrics with $\chi $%
-generating functions for solutions of type (\ref{offdncelepsilon}) (with
functionals $\chi _{4}[\mathbf{\wp }]$). The prime metrics can be taken of
any nature (being or not solutions of some modified gravitational equations,
or some quasi-periodic/ solitonic configurations) and than subjected to
nonmetric quasi-stationary solitonic deformations via generating/
gravitational polarization functions. In such cases, the target solutions
will define a mixture of solitonic distributions under $\tau $-evolution and
respective prescribed geometric data for primary d-metrics and effective
nonmetric generating sources.

\subsubsection{Nonmetric quasi-stationary generating sources with solitonic
hierarchies}

We can generate a different class of generic off-diagonal solitonic
solutions of nonmetric geometric flow equations (\ref{cdeq1}) when the
generating source is of type $\ ^{q}\mathbf{\Upsilon }_{\ \nu }^{\mu}(\tau)=
[\ _{h}^{q}\Upsilon \lbrack \mathbf{\wp }], \ ^{q}\Upsilon \lbrack \mathbf{%
\wp }]]$ (\ref{qgenersourc1}) but the d-metric coefficients of $\mathbf{g}%
(\tau )=(g_{i}(\tau ),g_{a}(\tau ))$ are computed via corresponding
nonholonomic deformations of nonmetric sources. The respective quadratic
elements are written in the form, 
\begin{eqnarray}
ds^{2}(\tau ) &=&e^{\ \psi \lbrack ~\ _{h}^{q}\Upsilon \lbrack \mathbf{\wp }%
]]}[(dx^{1})^{2}+(dx^{2})^{2}]+  \label{gensolsourc} \\
&&\left\{ 
\begin{array}{c}
\begin{array}{c}
-\frac{[h_{4}^{\ast }(\tau )]^{2}}{|\int dy^{3}[\ ^{q}\mathbf{\Upsilon }[%
\mathbf{\wp }]\ h_{4}(\tau )]^{\ast }|\ h_{4}(\tau )}[dy^{3}+\frac{\partial
_{i}[\int dy^{3}(\ \ ^{q}\mathbf{\Upsilon }[\mathbf{\wp }])\ h_{4}^{\ast
}(\tau )]}{\ \ ^{q}\mathbf{\Upsilon }[\mathbf{\wp }]\ \ h_{4}^{\ast }(\tau )}%
dx^{i}]- \\ 
h_{4}(\tau )[dt+(\ _{1}n_{k}(\tau )+\ _{2}n_{k}(\tau )\int dy^{3}\frac{%
[h_{4}^{\ast }(\tau )]^{2}}{|\int dy^{3}[\ ^{q}\mathbf{\Upsilon }[\mathbf{%
\wp }]\ h_{4}(\tau )]^{\ast }|\ [h_{4}(\tau )]^{5/2}}\ )dx^{k}]; \\ 
\mbox{ or }%
\end{array}
\\ 
\begin{array}{c}
-\frac{[\Psi ^{\ast }(\tau )]^{2}}{4(\ _{2}^{q}\mathbf{\Upsilon }%
)^{2}\{h_{4}^{[0]}(\tau )-\int dy^{3}[\Psi ^{2}(\tau )]^{\ast }/4\ \ ^{q}%
\mathbf{\Upsilon }[\mathbf{\wp }]\ \}}[dy^{3}+\frac{\partial _{i}\ \Psi }{\
\partial _{3}\Psi }dx^{i}]-(h_{4}^{[0]}(\tau )-\int dy^{3}\frac{[\Psi
^{2}(\tau )]^{\ast }}{4\ ^{q}\mathbf{\Upsilon }[\mathbf{\wp }]\ }) \\ 
\lbrack dt+(_{1}n_{k}(\tau )+\ _{2}n_{k}(\tau )\int dy^{3}\frac{[(\Psi (\tau
))^{2}]^{\ast }}{4(\ \ ^{q}\mathbf{\Upsilon }[\mathbf{\wp }]\
)^{2}|h_{4}^{[0]}(\tau )-\int dy^{3}[\Psi ^{2}(\tau )]^{\ast }/4(\ ^{q}%
\mathbf{\Upsilon }[\mathbf{\wp }]\ )|^{5/2}})dx^{k}]; \\ 
\mbox{ or }%
\end{array}
\\ 
\begin{array}{c}
-\frac{\Phi ^{2}(\tau )[\Phi ^{\ast }(\tau )]^{2}}{|\ _{2}^{q}\Lambda (\tau
)\int dy^{3}\ \ ^{q}\mathbf{\Upsilon }[\mathbf{\wp }]\ [\Phi ^{2}(\tau
)]^{\ast }|[g_{4}^{[0]}(\tau )-\Phi ^{2}(\tau )/4\ _{2}^{q}\Lambda (\tau )]}%
[dy^{3}+\frac{\partial _{i}\ \int dy^{3}\ \ ^{q}\mathbf{\Upsilon }[\mathbf{%
\wp }]\ \ [\Phi ^{2}]^{\ast }}{\ \ ^{q}\mathbf{\Upsilon }[\mathbf{\wp }]\ \
[(\ _{2}\Phi )^{2}]^{\ast }}dx^{i}]-(h_{4}^{[0]}(\tau ) \\ 
-\frac{\Phi ^{2}(\tau )}{4\ _{2}^{q}\Lambda (\tau )})[dt+(_{1}n_{k}(\tau )+\
_{2}n_{k}(\tau )\int dy^{3}\frac{\Phi ^{2}[\Phi ^{\ast }]^{2}}{|\
_{2}^{q}\Lambda (\tau )\int dy^{3}\ \ ^{q}\mathbf{\Upsilon }[\mathbf{\wp }]\
\ [\Phi ^{2}]^{\ast }|[g_{4}^{[0]}(\tau )-\Phi ^{2}/4\ _{2}^{q}\Lambda (\tau
)]^{5/2}})dx^{k}].%
\end{array}%
\end{array}%
\right.  \notag
\end{eqnarray}%
In d-metrics (\ref{gensolsourc}), the corresponding generating data for (\ref%
{offdsolgenfgcosmc}), (\ref{qeltors}), \ and (\ref{offdiagcosmcsh}) are
stated in such forms: \newline
$\{{\small h}_{4}(\tau )={\small h}_{4}(\tau,x^{i},y^{3}){\small ,\ ^{q}%
\mathbf{\Upsilon } [\mathbf{\wp }]\ ,\ }_{2}^{q}{\small \Lambda (\tau )}\};$ 
$\{{\small \Psi (\tau ),\ \ ^{q}\mathbf{\Upsilon }[\mathbf{\wp }]\ }\}$ from
(\ref{qeltors}); and $\{{\small \Phi (\tau ),\ }_{2}^{q}{\small \Lambda
(\tau ),\ \ ^{q}\mathbf{\Upsilon } [\mathbf{\wp }]\ }\}.$

\subsubsection{Interacting $\protect\tau $-running solitonic hierarchies for
generating functions with respective solitonic hierarchies for nonmetric
sources}

The generic off-diagonal solutions (\ref{gensolstat}) or (\ref{gensolsourc})
can be generalized for $\tau $-running configurations when a set of
solitionic hierarchies $\ _{1}{\small \mathbf{\wp }}$ is prescribed for
generating functions and another set $\ _{2}{\small \mathbf{\wp }}$ is
generated for generating functions of nonmetric origin. For a prime d-metric 
$\mathbf{\mathring{g}}$ (\ref{offdiagpm}) of arbitrary nature, we generate a
target quasi-stationary d-metric double generating function/source solitonic
hierarchies, 
\begin{eqnarray}
d\widehat{s}^{2}(\tau ) &=&\widehat{g}_{\alpha \beta }(\tau ,x^{k},y^{3};%
\mathring{g}_{\alpha };\psi (\tau ),\eta _{4}[\ _{1}\mathbf{\wp }];\ \ ^{q}%
\mathbf{\Upsilon }[\ _{2}\mathbf{\wp }])du^{\alpha }du^{\beta }=e^{\psi
(\tau )}[(dx^{1})^{2}+(dx^{2})^{2}]  \label{gensolmixed} \\
&&-\frac{[\partial _{3}(\eta _{4}[\ _{1}\mathbf{\wp }]\ \mathring{g}%
_{4})]^{2}}{|\int dy^{3}\ _{2}^{q}\mathbf{\Upsilon }[\ _{2}\mathbf{\wp }%
]\partial _{3}(\eta _{4}[\ _{1}\mathbf{\wp }]\ \mathring{g}_{4})|\ \eta
_{4}[\ _{1}\mathbf{\wp }]\mathring{g}_{4}}\{dy^{3}+\frac{\partial _{i}[\int
dy^{3}\ _{2}^{q}\mathbf{\Upsilon }[\ _{2}\mathbf{\wp }]\ \partial _{3}(\eta
_{4}[\ _{1}\mathbf{\wp }]\mathring{g}_{4})]}{\ _{2}^{q}\mathbf{\Upsilon }[\
_{2}\mathbf{\wp }]\partial _{3}(\eta _{4}[\ _{1}\mathbf{\wp }]\mathring{g}%
_{4})}dx^{i}\}^{2}  \notag \\
&&+\eta _{4}[\ _{1}\mathbf{\wp }]\mathring{g}_{4}\{dt+[\ _{1}n_{k}(\tau )+\
_{2}n_{k}(\tau )\int dy^{3}\frac{[\partial _{3}(\eta _{4}[\ _{1}\mathbf{\wp }%
]\mathring{g}_{4})]^{2}}{|\int dy^{3}\ \ _{2}^{q}\mathbf{\Upsilon }[\ _{2}%
\mathbf{\wp }]\partial _{3}(\eta _{4}[\ _{1}\mathbf{\wp }]\mathring{g}%
_{4})|\ (\eta _{4}[\ _{1}\mathbf{\wp }]\mathring{g}_{4})^{5/2}}]dx^{k}\}^{2}.
\notag
\end{eqnarray}%
In such nonlinear quadratic elements, we can prescribe $\mathbf{\mathring{g}}
$ to define, for instance, a black hole, BH, solution in GR or a MGT like in 
\cite{bubuianu17,ibsvevv22}. Nonmetric soltionic deformations of type
(\ref{gensolmixed}) do not preserve, in general, the primary BH character.
Nevertheless, for small parametric deformations like in (\ref%
{offdncelepsilon}), we can generate $\tau $-families of quasi-stationary
d-metrics with $\chi $-generating functions for solutions with generating
functionals $\chi _{4}[\ _{1}\mathbf{\wp }]$ and/or $\ _{2}^{q}\mathbf{%
\Upsilon }[\ _{2}\mathbf{\wp }].$ Various variants with ellipsoid and
solitonic deformations, for instance, for black ellipsoids, BE, with
additional solitonic modifications (of physical constants, or embedding into
certain solitonic gravitational vacuum) can be modelled. We do not study in
this work solutions for BHs, or BEs, solitonic $\tau $-running but
concentrate only on nonmetric wormhole-soliton configurations.

\subsection{Nonmetric wormhole solutions and their solitonic deformations}

\label{ss33} The goal of this subsection is to construct analyze physical
properties of wormhole solutions and their solitonic deformations in
nonmetric geometric flow and gravity theories with $\tau $-modified Einstein
equations (\ref{cdeq1}). As prime d-metric configurations, we shell consider
certain curve coordinate transforms of the Morris--Thorne and generalized
Ellis--Bronnikov wormholes to certain trivial off-diagonal configurations
which allow to apply the AFCDM. We cite \cite%
{morris88,kar94,roy20,souza22,wormh19a,wormh21a} for details and a recent
review of results on wormhole solutions. Then, we shall construct new
classes of quasi-stationary solutions as target metrics, for certain
gravitational polarizations determined by additional nonmetricity source,
and study their possible traversable nonmetric properties. Necessary
technical results are summarized in appendix \ref{appendixa}, containing a
brief summary of the AFCDM adapted to nonmetric gravity, and appendix \ref%
{appendixb}, for necessary formulas on solitoinic hierarchies.

\subsubsection{Prime metrics for 4-d metric compatible wormhole
configurations}

Let us consider a prime d-metric 
\begin{equation}
d\mathring{s}^{2}=\check{g}_{\alpha }(l,\theta ,\varphi )[\mathbf{\check{e}}%
^{\alpha }(l,\theta ,\varphi )]^{2},  \label{pmwh}
\end{equation}%
where the (tortoise) coordinates $u^{\alpha }=(r,\theta ,\varphi ,t)$ are
defined for $r(l)=(l^{2k}+\ _{0}b^{2k})^{1/2k}$ and the cylindrical angular
coordinate $\phi \in \lbrack 0,2\pi )$ is called parallel. In such
coordinates, $-\infty <l<\infty $ which is different from the standard
cylindrical radial coordinate $\rho ,$ when $0\leq \rho <\infty .$ We can
fix $\check{g}_{1}=1,\check{g}_{2}=r^{2}(l),\check{g}_{3}=r^{2}(l)\sin
^{2}\theta $ and $\check{g}_{4}=-1$ and can consider frame transforms to a
parametrization with trivial N-connection coefficients $\check{N}_{i}^{a}=%
\check{N}_{i}^{a}(u^{\alpha }(l,\theta ,\varphi ,t))$ and $\check{g}%
_{\beta}(u^{j}(l,\theta ,\varphi ),u^{3}(l,\theta ,\varphi )),$ when new
coordinates are $u^{1}=x^{1}=l,u^{2}=\theta ,$ and $u^{3}=y^{3}=\varphi +\
^{3}B(l,\theta ),u^{4}=y^{4}=t+\ ^{4}B(l,\theta ),$ for 
\begin{eqnarray*}
\mathbf{\check{e}}^{3} &=&d\varphi =du^{3}+\check{N}_{i}^{3}(l,\theta
)dx^{i}=du^{3}+\check{N}_{1}^{3}(l,\theta )dl+\check{N}_{2}^{3}(l,\theta
)d\theta , \\
\mathbf{\check{e}}^{4} &=&dt=du^{4}+\check{N}_{i}^{4}(l,\theta
)dx^{i}=du^{4}+\check{N}_{1}^{4}(l,\theta )dl+\check{N}_{2}^{4}(l,\theta
)d\theta ,
\end{eqnarray*}%
for $\mathring{N}_{i}^{3}=-\partial \ ^{3}B/\partial x^{i}$ and $\mathring{N}%
_{i}^{4}=-\partial \ ^{4}B/\partial x^{i}.$

We consider a prime d-metric (\ref{pmwh}) which is related via coordinate
transforms to the generic Morris-Thorne wormhole solution \cite{morris88}, 
\begin{equation*}
d\mathring{s}^{2}=(1-\frac{b(r)}{r})^{-1}dr^{2}+r^{2}d\theta ^{2}+r^{2}\sin
^{2}\theta d\varphi ^{2}-e^{2\Phi (r)}dt^{2},
\end{equation*}%
where $e^{2\Phi (r)}$ is a red-shift function and $b(r)$ as a shape function
defined in spherically polar coordinates $u^{\alpha }=(r,\theta ,\varphi,t). 
$ We can also parameterize this metric to get usual Ellis-Bronnikov, EB,
wormholes which are defined for $\Phi (r)=0$ and $b(r)=\ _{0}b^{2}/r$
characterizing a zero tidal wormhole with $\ _{0}b$ the throat radius. A
generalized EB is characterized additionally by even integers $2k$ (with $%
k=1,2,...$) This allows us to define a prime metric 
\begin{equation*}
d\mathring{s}^{2}=dl^{2}+r^{2}(l)d\theta ^{2}+r^{2}(l)\sin ^{2}\theta
d\varphi ^{2}-dt^{2},
\end{equation*}%
when 
\begin{equation*}
dl^{2}=(1-\frac{b(r)}{r})^{-1}dr^{2}\mbox{ and }b(r)=r-r^{3(1-k)}(r^{2k}-\
_{0}b^{2k})^{(2-1/k))}.
\end{equation*}

Generic off-diagonal nonholonomic deformations of prime metrics (\ref{pmwh})
can be analyzed for effective sources of type (\ref{qsourc}%
) when the $Q$-deformations are stated to be zero. For such quasi-stationary
diagonalizable solutions, the effective sources are of type $\ ^{tot}%
\widehat{\mathbf{\Upsilon }}_{\mu \nu }$ (\ref{totsourc}) when $\ ^{m}%
\widehat{\mathbf{T}}_{\mu \nu }$ in (\ref{totem}) is taken for a
energy-momentum tensor for matter fields. For generating off-diagonal
deformations of wormhole solutions, we shall consider generating sources of
type (\ref{dsourcparam}) or (\ref{qgenersourc}). Corresponding classes of
generating and integration functions are related via nonlinear symmetries (%
\ref{ntransf1}) or (\ref{ntransf2}), when a class of solutions for $(\
_{2}^{tot}\mathbf{\Upsilon }(\tau )\leftrightarrow \ _{2}^{tot}\Lambda
(\tau))$ can be distinguished from another class of solutions for $(\
_{2}^{q}\mathbf{\Upsilon }(\tau )\leftrightarrow \ _{2}^{q}\Lambda (\tau )).$
In general, such $\tau $-running or nonholonomic Ricci flow configurations
are characterized by different types of thermodynamic variables as we shall
prove in section \ref{sec4}.

\subsubsection{Nonholonomic quasi-stationary gravitational polarizations of
wormholes}

Nonmetric off-diagonal quasi-stationary deformations of wormhole d-metrics (%
\ref{pmwh}) can be generated by introducing $\left( \check{g}_{\beta },%
\check{N}_{i}^{a}\right) $ instead of primary geometric data $\mathbf{%
\mathring{g}}=[\mathring{g}_{\alpha },\ \mathring{N}_{i}^{a}]$ (\ref%
{offdiagpm}) in (\ref{offdiagpolfr}) using some generating sources $\
_{2}^{q}\mathbf{\Upsilon }(\tau )$ and/or $_{2}^{q}\Lambda (\tau ).$ We
generate new classes of a target $\tau $-family of d-metrics $\mathbf{g}%
(\tau )$ determined by nonmetric geometric flows and respective $\eta $%
-polarization functions, when $\mathbf{\check{g}}\rightarrow \mathbf{g}%
(\tau)= [g_{\alpha }(\tau )=\eta _{\alpha }(\tau )\check{g}%
_{\alpha},N_{i}^{a}(\tau )= \eta _{i}^{a}\ (\tau )\check{g}_{i}^{a}].$
Corresponding quadratic linear elements 
\begin{eqnarray}
d\widehat{s}^{2}(\tau ) &=&\widehat{g}_{\alpha \beta }(l,\theta ,\varphi
;\psi ,\eta _{4};\ _{2}^{q}\Lambda (\tau )=\ _{2}^{q}\Lambda (\tau ),\ \
_{2}^{q}\mathbf{\Upsilon }(\tau ),\ \check{g}_{\alpha })du^{\alpha
}du^{\beta }  \notag \\
&=&e^{\psi (\tau l,\theta )}[(dx^{1}(l,\theta ))^{2}+(dx^{2}(l,\theta ))^{2}]
\label{whpolf} \\
&&-\frac{[\partial _{\varphi }(\eta _{4}(\tau )\ \check{g}_{4})]^{2}}{|\int
d\varphi \ \ _{2}^{q}\mathbf{\Upsilon }(\tau )\partial _{\varphi }(\eta
_{4}(\tau )\ \breve{g}_{4})|\ \eta _{4}(\tau )\ \check{g}_{4}}\{dy^{3}+\frac{%
\partial _{i}[\int d\varphi \ \ _{2}^{q}\mathbf{\Upsilon }(\tau )\ \partial
_{\varphi }(\eta _{4}(\tau )\ \check{g}_{4})]}{\ _{2}^{q}\mathbf{\Upsilon }%
(\tau )\partial _{\varphi }(\eta _{4}(\tau )\ \check{g}_{4})}dx^{i}\}^{2} 
\notag \\
&&+\eta _{4}(\tau )\breve{g}_{4}\{dt+[\ _{1}n_{k}+\ _{2}n_{k}\int d\varphi 
\frac{\lbrack \partial _{\varphi }(\eta _{4}(\tau )\ \breve{g}_{4})]^{2}}{%
|\int d\varphi \ \ _{2}^{q}\mathbf{\Upsilon }(\tau )\partial _{\varphi
}(\eta _{4}(\tau )\ \breve{g}_{4})|\ (\eta _{4}(\tau )\ \breve{g}_{4})^{5/2}}%
]dx^{k}\}.  \notag
\end{eqnarray}%
This class of solutions are determined by respective generating function $%
\eta _{4}(\tau )=\eta _{4}(\tau ,l,\theta ,\varphi )$ and integration
functions $\ _{1}n_{k}(\tau ,l,\theta )$ and $\ _{2}n_{k}(\tau ,l,\theta ).$
The $\tau $-family of functions $\psi (\tau ,l,\theta )$ are defined as a
solution of a respective family of 2-d Poisson equations,  
 $\partial _{11}^{2}\psi (\tau )+\partial _{22}^{2}\psi (\tau )=2\ \ _{1}^{q}%
\mathbf{\Upsilon }(\tau ,l,\theta )$, 
when the horizontal generating effective sources, $\ _{1}^{q}\mathbf{%
\Upsilon }(\tau ,l,\theta ),$ can be different from the vertical ones which
may depend on a vertical coordinate, $\ _{2}^{q}\mathbf{\Upsilon }%
(\tau,l,\theta ,z).$

For families of off-diagonal solutions (\ref{whpolf}), we can fix $\
_{1}^{q}\Lambda (\tau )=\ _{2}^{q}\Lambda (\tau )$ and analyze
quasi-stationary configurations with running cosmological constants. We
suppose that such effective $\tau $-families of cosmological constants can
be expressed in additive form 
\begin{equation}
\ _{2}^{q}\Lambda (\tau )=\ _{2}^{tot}\Lambda (\tau )+\ _{2}^{1}\Lambda
(\tau )=\ _{2}^{m}\Lambda (\tau )+\ _{2}^{F}\Lambda (\tau )+\
_{2}^{e}\Lambda (\tau ),  \label{additcosm}
\end{equation}%
where $\ _{2}^{tot}\Lambda (\tau )$ model metric compatible configurations
and $\ _{2}^{1}\Lambda (\tau )$ describe possible additional nonmetric
contributions. The three terms with left labels $m,F,e$ in these formulas
correspond to (effective) energy-momentum tensors (\ref{totem}). We can
consider nonlinear symmetries of type (\ref{nonlinsymrex}) relating $\mathbf{%
\Upsilon }_{\ \nu }^{\mu }(\tau )\simeq $ $[~\ _{h}^{tot}\Upsilon ,~\
^{tot}\Upsilon ]$ (\ref{dsourcparam}) to a $\tau $-family $\
_{s}^{tot}\Lambda (\tau ),$ for $s=1,2,$ and $\ ^{q}\mathbf{\Upsilon }_{\
\nu }^{\mu }(\tau )\simeq \lbrack \ _{h}^{q}\Upsilon ,~\ ^{q}\Upsilon ]$ (%
\ref{qgenersourc}) to another family $\ _{2}^{q}\Lambda (\tau )=\
_{2}^{q}\Lambda \lbrack \tau ,\ _{2}^{tot}\Lambda (\tau )]$ which may be a
nonlinear functional on $\ _{2}^{tot}\Lambda (\tau ).$ In this work, we
elaborate on models with nontrivial nonmetricity effective sources and
additional type functionals (\ref{additcosm}) for cosmological constants.

The class of quasi-stationary nonmetric deformed wormholes (\ref{whpolf})
can be written in terms of generating data $\left[ \Phi (\tau ),\
_{2}^{tot}\Lambda (\tau )+\ _{2}^{1}\Lambda (\tau )\right] $ as in (\ref%
{gensolsourc}),%
\begin{eqnarray}
d\widehat{s}^{2}(\tau ) &=&e^{\ \psi \lbrack ~\ _{h}^{q}\Upsilon (\tau
)]}[(dx^{1})^{2}+(dx^{2})^{2}]-  \label{whcosm1} \\
&&\frac{\Phi ^{2}(\tau )[\Phi ^{\ast }(\tau )]^{2}}{|\ [\ _{2}^{tot}\Lambda
(\tau )+\ _{2}^{1}\Lambda (\tau )]\int d\varphi \ ^{q}\mathbf{\Upsilon }%
(\tau )\ [\Phi ^{2}(\tau )]^{\ast }|[g_{4}^{[0]}(\tau )-\frac{\Phi ^{2}(\tau
)}{4\ [\ _{2}^{tot}\Lambda (\tau )+\ _{2}^{1}\Lambda (\tau )]}]}  \notag \\
&&[dy^{3}+\frac{\partial _{i}\ \int d\varphi \ \ ^{q}\mathbf{\Upsilon }(\tau
)\ \ [\Phi ^{2}]^{\ast }}{\ \ ^{q}\mathbf{\Upsilon }(\tau )\ [(\ _{2}\Phi
)^{2}]^{\ast }}dx^{i}]-[h_{4}^{[0]}(\tau )-\frac{\Phi ^{2}(\tau )}{4\ [\
_{2}^{tot}\Lambda (\tau )+\ _{2}^{1}\Lambda (\tau )]}][dt+(_{1}n_{k}(\tau )+
\notag \\
&&\ _{2}n_{k}(\tau )\int d\varphi \frac{\Phi ^{2}(\tau )[\Phi ^{\ast }(\tau
)]^{2}}{|\ [\ _{2}^{tot}\Lambda (\tau )+\ _{2}^{1}\Lambda (\tau )]\int
d\varphi \ ^{q}\mathbf{\Upsilon }(\tau )\ [\Phi ^{2}(\tau )]^{\ast
}|[g_{4}^{[0]}(\tau )-\frac{\Phi ^{2}(\tau )}{4\ [\ _{2}^{tot}\Lambda (\tau
)+\ _{2}^{1}\Lambda (\tau )]}]^{5/2}})dx^{k}].  \notag
\end{eqnarray}%
In such a form, the data for a prime wormhole metric are "hidden" into
generating functions and the equations for nonlinear symmetries.

The target d-metrics (\ref{whpolf}) or (\ref{whcosm1}) do not describe
wormhole configurations for general classes of generating and integrating
data. There are necessary additional assumptions on polarization functions
and integration functions which allows us to provide certain physical
interpretation of such generic off-diagonal solutions. Typically, we can
prescribe some generating data for solitonic waves, or some small
deformations of wormhole configurations. For instance, to elaborate
cosmological scenarios with acceleration and quasi-periodic structure, or to
transform quasi-stationary d-metrics into locally anisotropic cosmological
ones, we have to consider other types of gravitational polarization and
generating data.

\subsubsection{Off-diagonal quasi-stationary solitonic deformations of
wormhole d-metrics}

We can generate a class of d-metrics (\ref{gensolmixed}) when the prime
d-metric $\mathbf{\mathring{g}}$ (\ref{offdiagpm}) changed into a wormhole
one (\ref{pmwh}) with space coordinates $(l,\theta ,\varphi ).$ Such a
target d-metric is of type (\ref{whpolf}) when off-diagonal deformations are
determined by a generating function $\eta _{4}(\tau )=\eta _{4}[\wp ],$
where $\wp =\wp (\tau ,x^{i}),$ $=\wp (\tau ,x^{1},y^{3}),$ or $=\wp (\tau
,x^{2},y^{3})$ is any solution for $\tau $-running solitonic configurations (%
\ref{swaves}).\footnote{%
We can consider $\wp $ as any solitonic hierarchy considered in appendix \ref%
{ssb1} and re-defined in coordinates $(\tau,l,\theta ,\varphi ).$} The
quadratic linear element for $\wp (\tau,x^{1},y^{3})$--solitonic
deformations of wormhole metrics is constructed 
\begin{eqnarray*}
d\widehat{s}^{2}(\tau ) &=&\widehat{g}_{\alpha \beta }(l,\theta ,y^{3};\psi
,\eta _{4};\ _{2}^{m}\Lambda (\tau )+\ _{2}^{F}\Lambda (\tau )+\
_{2}^{e}\Lambda (\tau ),\ \ _{2}^{q}\mathbf{\Upsilon }(\tau ),\ \check{g}%
_{\alpha })du^{\alpha }du^{\beta } \\
&=&e^{\psi (\tau ,l,\theta )}[(dx^{1}(l,\theta ))^{2}+(dx^{2}(l,\theta
))^{2}] \\
&&-\frac{[\partial _{3}(\eta _{4}[\wp ]\ \check{g}_{4})]^{2}}{|\int dy^{3}\
\ _{2}^{q}\mathbf{\Upsilon }(\tau )\partial _{3}(\eta _{4}[\wp ]\ \breve{g}%
_{4})|\ \eta _{4}[\wp ]\ \check{g}_{4}}\{dy^{3}+\frac{\partial _{i}[\int
dy^{3}\ \ _{2}^{q}\mathbf{\Upsilon }(\tau )\ \partial _{3}(\eta _{4}[\wp ]\ 
\check{g}_{4})]}{\ _{2}^{q}\mathbf{\Upsilon }(\tau )\partial _{3}(\eta
_{4}[\wp ]\ \check{g}_{4})}dx^{i}\}^{2} \\
&&+\eta _{4}[\wp ]\breve{g}_{4}\{dt+[\ _{1}n_{k}+\ _{2}n_{k}\int dy^{3}\frac{%
[\partial _{3}(\eta _{4}[\wp ]\ \breve{g}_{4})]^{2}}{|\int dy^{3}\ \ _{2}^{q}%
\mathbf{\Upsilon }(\tau )\partial _{3}(\eta _{4}[\wp ]\ \breve{g}_{4})|\
(\eta _{4}[\wp ]\ \breve{g}_{4})^{5/2}}]dx^{k}\}.
\end{eqnarray*}%
Such solutions describe nonmetric nonholonomic dissipations of a prime
wormhole metric into certain $\tau $-running generic off-diagonal solitonic
distributions. Nonlinear symmetries of type (\ref{ntransf1}) or (\ref%
{ntransf2}) involve additive decompositions of running cosmological
constants, $\ _{2}^{m}\Lambda (\tau )+\ _{2}^{F}\Lambda (\tau )+\
_{2}^{e}\Lambda (\tau )$, and a nonmetric source $\ _{2}^{q}\mathbf{\Upsilon 
}(\tau )$ which can be of non-solitonic character.

\subsubsection{Small parametric off-diagonal quasi-stationary deformations
of wormhole d-metrics}

We can generate new classes of solutions which preserve wormhole character
under nonmetric geometric flow evolution scenarios of a prime d-metric $%
\mathbf{\check{g}}$ (\ref{pmwh}) if we consider small $\varepsilon $%
-parametric deformations of type (\ref{offdncelepsilon}). The generating
functions can be linearized on $\varepsilon $ as in (\ref{epsilongenfdecomp}%
) when in terms of $\chi $-polarization functions, the quadratic linear
elements (\ref{whpolf}) can be expressed 
\begin{eqnarray*}
d\ \widehat{s}^{2}(\tau ) &=&\widehat{g}_{\alpha \beta }(l,\theta ,\varphi
;\psi ,\chi _{4};\ _{2}^{q}\Lambda (\tau )=\ _{2}^{q}\Lambda (\tau ),\ \
_{2}^{q}\mathbf{\Upsilon }(\tau ),\ \check{g}_{\alpha },\check{N}%
_{i}^{3})du^{\alpha }du^{\beta } \\
&=&e^{\psi _{0}(l,\theta )}[1+\varepsilon \ ^{\psi (l,\theta )}\chi
(l,\theta )][(dx^{1}(l,\theta ))^{2}+(dx^{2}(l,\theta ))^{2}] \\
&&-\{\frac{4[\partial _{\varphi }(|\zeta _{4}\ \breve{g}_{4}|^{1/2})]^{2}}{%
\breve{g}_{3}|\int d\varphi \{\ \ _{2}^{q}\mathbf{\Upsilon }(\tau )\partial
_{\varphi }(\zeta _{4}\ \breve{g}_{4})\}|}-\varepsilon \lbrack \frac{%
\partial _{\varphi }(\chi _{4}|\zeta _{4}\breve{g}_{4}|^{1/2})}{4\partial
_{\varphi }(|\zeta _{4}\ \breve{g}_{4}|^{1/2})}-\frac{\int d\varphi \{\
_{2}^{q}\mathbf{\Upsilon }(\tau )\partial _{\varphi }[(\zeta _{4}\ \breve{g}%
_{4})\chi _{4}]\}}{\int d\varphi \{\ _{2}^{q}\mathbf{\Upsilon }(\tau
)\partial _{\varphi }(\zeta _{4}\ \breve{g}_{4})\}}]\}\ \breve{g}_{3} \\
&&\{d\varphi +[\frac{\partial _{i}\ \int d\varphi \ \ _{2}^{q}\mathbf{%
\Upsilon }(\tau )\ \partial _{\varphi }\zeta _{4}}{(\check{N}_{i}^{3})\
_{2}^{q}\mathbf{\Upsilon }(\tau )\partial _{\varphi }\zeta _{4}}+\varepsilon
(\frac{\partial _{i}[\int d\varphi \ _{2}^{q}\mathbf{\Upsilon }(\tau )\
\partial _{\varphi }(\zeta _{4}\chi _{4})]}{\partial _{i}\ [\int d\varphi \
_{2}^{q}\mathbf{\Upsilon }(\tau )\partial _{\varphi }\zeta _{4}]}-\frac{%
\partial _{\varphi }(\zeta _{4}\chi _{4})}{\partial _{\varphi }\zeta _{4}})]%
\check{N}_{i}^{3}dx^{i}\}^{2}
\end{eqnarray*}%
\begin{eqnarray}
&&+\zeta _{4}(1+\varepsilon \ \chi _{4})\ \breve{g}_{4}\{dt+[(\check{N}%
_{k}^{4})^{-1}[\ _{1}n_{k}+16\ _{2}n_{k}[\int d\varphi \frac{\left( \partial
_{\varphi }[(\zeta _{4}\ \breve{g}_{4})^{-1/4}]\right) ^{2}}{|\int d\varphi
\partial _{\varphi }[\ \ _{2}^{q}\mathbf{\Upsilon }(\tau )(\zeta _{4}\ 
\breve{g}_{4})]|}]  \notag \\
&&+\varepsilon \frac{16\ _{2}n_{k}\int d\varphi \frac{\left( \partial
_{\varphi }[(\zeta _{4}\ \breve{g}_{4})^{-1/4}]\right) ^{2}}{|\int d\varphi
\partial _{\varphi }[\ \ _{2}^{q}\mathbf{\Upsilon }(\tau )(\zeta _{4}\ 
\breve{g}_{4})]|}(\frac{\partial _{\varphi }[(\zeta _{4}\ \breve{g}%
_{4})^{-1/4}\chi _{4})]}{2\partial _{\varphi }[(\zeta _{4}\ \breve{g}%
_{4})^{-1/4}]}+\frac{\int d\varphi \partial _{\varphi }[\breve{g}_{4}(\zeta
_{4}\chi _{4}\ \breve{g}_{4})]}{\int d\varphi \partial _{\varphi }[\breve{g}%
_{4}(\zeta _{4}\ \breve{g}_{4})]})}{\ _{1}n_{k}+16\ _{2}n_{k}[\int d\varphi 
\frac{\left( \partial _{\varphi }[(\zeta _{4}\ \breve{g}_{4})^{-1/4}]\right)
^{2}}{|\int d\varphi \partial _{\varphi }[\breve{g}_{4}(\zeta _{4}\ \breve{g}%
_{4})]|}]}]\check{N}_{k}^{4}dx^{k}\}^{2}.  \label{whpolf1}
\end{eqnarray}

We can model elliptic deformations of the wormhole throat as a particular
case of d-metrics of type (\ref{whpolf1}) if we chose a generating function
of type%
\begin{equation}
\chi _{4}(\tau ,l,\theta ,\varphi )=\underline{\chi }(\tau ,l,\theta )\sin
(\omega _{0}\varphi +\varphi _{0})  \label{ellipsoid}
\end{equation}%
as for cylindric configurations with $\varphi $-anisotropic deformations and 
$\tau $-running small deformations. Such classes of solutions describe
nonmetric $\tau $-evolution of some ellipsoidal wormholes.

If in (\ref{whpolf1}) we take a vacuum profile with $\zeta _{4}=[\ _{1}\wp ]$
i.e. as a functional of a solitonic distribution $\ _{1}\wp =\ _{1}\wp
(l,\theta ,\varphi )$ as in (\ref{solitdistr}) but with correspondingly
re-defined coordinates, the wormhole ellipsoidal configuration is modelled
as a $\tau $-evolution in such nonmetric gravitational vacuum. Instead of
ellipsoidal configurations (\ref{ellipsoid}), we can consider small
solitonic $\chi $-deformations. For such classes of generic off-diagonal
solutions, we can consider any $\chi _{4}(\tau ,l,\theta ,\varphi )$ defined
as functional $\chi _{4}[\wp ]$ of a $\wp =\wp (\tau ,x^{i}),$ $=\wp
(\tau,x^{1},y^{3}),$ or $=\wp (\tau ,x^{2},y^{3}),$ of some solitonic $\tau $%
-waves (\ref{swaves}).

The class of nonmetric locally anisotropic wormhole solutions (\ref{whpolf1}%
) may possess a multiple solitonic wave or solitionic distribution character
even the generating source $\ _{2}^{q}\mathbf{\Upsilon }(\tau )$ can be
non-solitonic. Nonlinear symmetries allow to associate to such
configurations certain effective $\tau $-running cosmological constants of
type (\ref{additcosm}). Considering, for instance, zero values of $\
_{2}^{e}\Lambda (\tau ),$ we can model some metric compatible locally
anisotropic wormholes and then to extend the constructions for nonmetric
configurations with nontrivial $\ _{2}^{e}\Lambda (\tau ).$

\subsubsection{On traversable nonmetric wormholes}

Wormhole solutions in GR and MGTs are considered as hypothetical geometric
structures that link two distinct regions of the same spacetime. References 
\cite{morris88,kar94,roy20,souza22,warpwormh} contain reviews of
results and methods of constructing wormhole solutions. Standard wormhole
solutions in GR are not traversable because for various classes of such
solutions it is not possible to send causal light signals through it throat
faster than we can send it through outside. The first wormhole model of the
so-called Einstein-Rosen bridge was elaborated using as a vacuum solution of
gravitational field equations \cite{er35}. That solution was derived as a
modification of the Schwarzschild BH when the corresponding wormhole model
is not traversable because of singularity of such solutions. Latter, a
static and spherically symmetric wormhole configuration with a traversable
throat at the center was constructed in \cite{morris88}. Various classes of
wormhole metrics were found in the framework of GR and MGTs \cite%
{wormh21a,roy20,souza22}. The existence of such solutions demands the
presence of some exotic matter and additional geometric distortions for
which the null energy condition, NEC, is violated in order to achieve a
stable and traversable structure.

Let us remember in brief how traversable wormholes require a violation of
the so-called \textit{average null energy condition,} ANEC, and how such
conditions are modified in the framework of MGTs with nonmetricity. The ANEC
states that the energy-momentum tensor for matter fields, $T_{\mu \nu },$
for a local quantum field theory, QFT, along a complete achronal null
geodesic, there are satisfied the conditions $\int T_{\mu \nu }k^{\mu}k^{\nu
}d\lambda \geq 0,$ where $k^{\mu }$ is a tangent d-vector and $\lambda $ is
an affine parameter. For elaborating quantum gravity, QG, models, such
conditions have to be considered for certain effective matter field and
distortion of geometric objects. Here we note that in classical theories the
violation of the ANEC is prevented by the null energy conditions, NEC, $%
T_{\mu \nu }k^{\mu }k^{\nu }d\lambda \geq 0.$ It is considered that such
conditions must be valid for any physically reasonable theory at least for
classical and semi-classical configurations. Additionally, there are
considered other important criteria on causality, topological censorship,
absence of singularities etc.

Due to problems with transversability, many authors excluded wormholes to be
considered as reliable astrophysical objects. In another turn, a number of
authors concluded that there are possibilities to realize wormholes without
considering exotic matter but modifying GR. In various cosmological models,
there are used solutions with NECs. One supposes that such theories are not
experimentally prohibited in the condition when MGTs are involved and due to
ideas on existence of particles beyond the standard model.

However, for elaborating explicit physical models and applications in modern
cosmology and astrophysics, it is admitted that QM and distortions of
geometric effects may induce negative null energy, leading to violations of
some NECs and/or ANECs. For nonholonomic systems, the variational and
conservation laws are different from those stated for unconstrained ones.
For diagonal traversable systems, to sustain a traversable wormhole there
are introduced certain negative null energy and various nonlocal /
nonachronal constructions. The matter fields are considered as quantum ones,
but the gravitational field is treated classically. In such models, there
are solved some semi-classical Einstein's equations with an effective source 
$<T_{\mu \nu }>$ computed as the expectation value of the stress-energy
tensor in a given quantum state. For certain classes of wormhole solutions,
the 1-loop expectation value of the stress-energy tensor satisfy in some
spacetime regions the conditions $\int <T_{\mu \nu }>k^{\mu }k^{\nu}d\lambda
<0$. Such configurations allows us to construct traversable Einstein-Rosen
bridges with certain interesting physical properties.

Quantum effects and wormhole solutions are studied in various MGTs and
quantum information theories. For elaborating quantum computing models, this
provides the possibility to transfer information between the two asymptotic
spacetime boundaries. Such a process can be viewed as a teleportation
protocol, see details and references in \cite{wormh19}. Here we note that
wormholes seem to be traversable for qubits \cite{wormh19a,wormh21,wormh22}.

In this work, we follow an approach to constructing wormhole solutions with off-diagonal deformations of some prime wormhole metrics in 4-d theories (see also generalizations for higher dimensions and MGTs \cite{vwh1,vwh2,vwh3,harko,vwh4,sv14a}). Such solutions can be constructed using the AFCDM and extended to nonmetric geometric flow and gravity theories as we have shown in previous subsections. The generating sources $\
^{q}\mathbf{\Upsilon }_{\ \ \beta }^{\alpha }(\tau )\simeq [\
_{h}^{q}\Upsilon (\tau ), \ ^{q}\Upsilon (\tau )]$ (\ref{qgenersourc}) for $%
Q $-modified Einstein equations$\ $(\ref{cdeq1}) can be prescribed in such a
form that $\int \ ^{q}\mathbf{\Upsilon }_{\mu \nu }k^{\mu }k^{\nu
}d\lambda<0 $ even using the matter energy-momentum d-tensor from (\ref%
{totem}), we have $\ ^{m}\widehat{\mathbf{T}}_{\mu \nu }k^{\mu }k^{\nu
}d\lambda \geq 0.$ This means that certain types of nonmetric geometric flow
and off-diagonal deformations resulting in certain locally anisotropic
wormhole solutions result in traversable conditions even at the classical
level. Such conditions can be valid even for locally anisotropic wormhole
solutions in GR because of additional generic off-diagonal terms. This is
also a result of nonholonomic modifications of the spacetime structure.

\section{Perelman thermodynamics for nonmetric quasi-stationary
configurati\-ons}

\label{sec4} The parametric solutions constructed in previous section describe nonmetric geometric flow and solitonic deformations of wormhole metrics. They are characterized by respective Perelman statistical/
geometric thermodynamic variables \cite{perelman1}. The Bekenstein-Hawking
thermodynamic paradigm is not applicable to such solutions because, in
general, they do not involve certain hypersurface configurations. We cite 
\cite{ibsvev20,ibsvevv22} for details on
relativistic generalizations and MGTs. For geometric flows and metric and
nonmetric gravity theories, the W-functional (\ref{wfperelm4matt}) can be
treated as a "minus" entropy. The goal of this section is to show how
nonmetric geometric flow thermodynamic variables can be defined and computed
for quasi-stationary off-diagonal solutions of type (\ref{whcosm1}).

\subsection{Statistical thermodynamic variables for $Q$-deformed
relativistic geometric flows}

Let us consider nonmetric geometric flow equations (\ref{cdeq1}). They can
be derived in geometric or variational form, following the methods outlined
in section 3.1 and 5 of \cite{perelman1},\footnote{%
in this work, we consider 4-d Lorentz manifols which correspond to dimension 
$n=4$, which a conventional changing to Riemannian signature; Perelman's
formulas for deriving geometric flows and thermodynamic variables can be
used for arbitrary dimension $n\geq 3$ and signature but such formulas can't
be used for formulations/ proofs of analogs of Poincare--Thorston
conjectures in non-Riemannian cases} from the W-functional, 
\begin{equation}
\widehat{\mathcal{W}}(\tau )=\int_{t_{1}}^{t_{2}}\int_{\Xi _{t}}\left( 4\pi
\tau \right) ^{-2}e^{-\widehat{f}(\tau )}\sqrt{|\mathbf{g}(\tau )|}\delta
^{4}u[\tau (\widehat{\mathbf{R}}sc(\tau )+|\widehat{\mathbf{D}}(\tau )%
\widehat{f}(\tau )|^{2}+\widehat{f}(\tau )-4].  \label{wf1}
\end{equation}%
In this formula, the normalizing function $\widehat{f}(\tau ,u)$ and the
parameter $\tau $ are such way re-defined that corresponding geometric flow
equations and normalizing conditions formulas contributions from possible
matter fields, $F$- and $Q$-distortions of the geometric data $%
(g(\tau),\nabla (\tau ))$ to canonical ones, $(\mathbf{g}(\tau ), \mathbf{N}%
(\tau),\widehat{\mathbf{D}}(\tau )).$ We use different integration measures
and nonlinear symmetries for (\ref{wf1}) comparing to $\mathcal{W}(\tau )$ (%
\ref{wfperelm4matt}). The effects of nonmetricity will be emphasized below
when there will be used solutions of (\ref{cdeq1}) determined by respective $%
\tau$--running generating sources (\ref{qgenersourc}) or effective
cosmological constants (\ref{additcosm}).

On a metric-affine space $\mathcal{M}$ endowed with canonical geometric data
and an additional nonholonomic (3+1) splitting,\footnote{%
such a conventional splitting is necessary for introducing thermodynamic
variables, when the nonholonomic 2+2 decompositions is important for
generating off-diagonal solutions} we introduce the statistical partition
function 
\begin{equation}
\ ^{q}\widehat{Z}(\tau )=\exp [\int_{\widehat{\Xi }}[-\widehat{f}+2]\ \left(
4\pi \tau \right) ^{-2}e^{-\widehat{f}}\delta \ ^{q}\mathcal{V}(\tau )],
\label{spf}
\end{equation}%
where the volume element is defined and computed as 
\begin{equation}
\delta \ ^{q}V(\tau )=\sqrt{|\mathbf{g}(\tau )|}dx^{1}dx^{2}\delta
y^{3}\delta y^{4}\ .  \label{volume}
\end{equation}%
We use a left label $q$ because nonmetric $Q$-contributions can be encoded
in $\mathbf{g}(\tau ).$ Such a label will be omitted in formulas below if
that will not result in ambiguities. Here we note that a a statistical
thermodynamic model can be constructed using a partition function $Z=\int
\exp (-\beta E)d\omega (E)$ for a canonical ensemble at temperature $\beta
^{-1}=\tau $ and when the measure is stated as the density of states $\omega
(E).$ The thermodynamical variables are computed as the average energy, $%
\left\langle E\right\rangle :=-\partial \log Z/\partial \beta ,$ the entropy 
$S:=\beta \left\langle E\right\rangle +\log Z$ and the fluctuation parameter 
$\sigma :=\left\langle \left( E-\left\langle E\right\rangle
\right)^{2}\right\rangle = \partial ^{2}\log Z/\partial \beta ^{2}.$

Using $\widehat{Z}$ (\ref{spf}) and $\widehat{\mathcal{W}}(\tau )$ (\ref{wf1}%
) and following for canonical variables a variational procedure on a closed
region of $\mathcal{M}$ as in section 5 of \cite{perelman1}, we can define
and compute respective thermodynamic variables: 
\begin{eqnarray}
\ ^{q}\widehat{\mathcal{E}}\ (\tau ) &=&-\tau ^{2}\int_{\widehat{\Xi }}\
\left( 4\pi \tau \right) ^{-2}\left( \widehat{\mathbf{R}}sc+|\ \widehat{%
\mathbf{D}}\ \widehat{f}|^{2}-\frac{2}{\tau }\right) e^{-\ \widehat{f}}\
\delta \ ^{q}\mathcal{V}(\tau ),  \label{qthermvar} \\
\ \ ^{q}\widehat{S}(\tau ) &=&-\int_{\widehat{\Xi }}\left( 4\pi \tau \right)
^{-2}\left( \tau (\widehat{\mathbf{R}}sc+|\widehat{\mathbf{D}}\ \widehat{f}%
|^{2})+\widehat{f}-4\right) e^{-\ \widehat{f}}\delta \ ^{q}\mathcal{V}(\tau
),  \notag \\
\ \ ^{q}\widehat{\sigma }(\tau ) &=&2\ \tau ^{4}\int_{\widehat{\Xi }}\left(
4\pi \tau \right) ^{-2}|\ \widehat{\mathbf{R}}_{\alpha \beta }+\widehat{%
\mathbf{D}}_{\alpha }\ \widehat{\mathbf{D}}_{\beta }\ \widehat{f}-\frac{1}{%
2\tau }\mathbf{g}_{\alpha \beta }|^{2}e^{-\widehat{f}}\delta \ ^{q}\mathcal{V%
}(\tau ).  \notag
\end{eqnarray}%
We note that such a thermodynamic systems can be associated to solution of
the nonholonomic nonmetric geometric flow equations (\ref{ricciflowr2}). In
particular, we can fix conventionally the temperature and consider such data
for nonmetric Ricci solitons characterized by $\left[ \ ^{q}\widehat{%
\mathcal{E}}(\tau _{0}),\ ^{q}\widehat{\mathcal{S}}(\tau _{0}),\ ^{q}%
\widehat{\sigma }(\tau _{0})\right] .$ Certain classes of solutions can be
not well-defined in the framework of such a statistical and geometric
thermodynamic approach, for instance, if $\ ^{q}\widehat{\mathcal{S}}(\tau
_{0})<0.$ We have to restrict certain classes of nonholonomic frames/
distributions/distortions in order to generate physically viable solutions.
The nonmetric $Q$-deformations may have different sign contributions
comparing to certain metric compatible classes of solutions determined by
corresponding $\left[ \widehat{\mathcal{E}}(\tau ),\widehat{\mathcal{S}}%
(\tau ),\widehat{\sigma }(\tau )\right].$

\subsection{Thermodynamic variables for nonmetric deformations of wormhole
solutions}

We compute in explicit form the variables $\widehat{Z}$ (\ref{spf}), and $\
^{q}\widehat{\mathcal{E}}\ (\tau ), \ ^{q}\widehat{S}(\tau )$ from (\ref%
{qthermvar}) for quasi-stationary off-diagonal solutions (\ref{whcosm1}).%
\footnote{%
We omit more cumbersome calculations for $\ ^{q}\widehat{\sigma }(\tau ).$}
The simplest way is to consider that 
\begin{equation*}
\widehat{\mathbf{R}}sc=2[\ _{1}^{tot}\Lambda (\tau )+\ _{1}^{1}\Lambda (\tau
)+\ _{2}^{tot}\Lambda (\tau )+\ _{2}^{1}\Lambda (\tau )]
\end{equation*}
choosing such a normalizing function when $\widehat{\mathbf{D}}_{\alpha }\ 
\widehat{f}=0$ and approximating $\widehat{f}\approx 0.$ Such conditions can
be considered for a frame/coordinate system and then the results can be
redefined for arbitrary bases and normalizing functions. Correspondingly, we
obtain%
\begin{eqnarray}
\ ^{q}\widehat{Z}(\tau ) &=&\exp [\int_{\widehat{\Xi }}\frac{1}{8\left( \pi
\tau \right) ^{2}}\ \delta \ ^{q}\mathcal{V}(\tau )],  \label{thermvar1} \\
\ ^{q}\widehat{\mathcal{E}}\ (\tau ) &=&-\tau ^{2}\int_{\widehat{\Xi }}\ 
\frac{1}{8\left( \pi \tau \right) ^{2}}[\ _{1}^{tot}\Lambda (\tau )+\
_{1}^{1}\Lambda (\tau )+\ _{2}^{tot}\Lambda (\tau )+\ _{2}^{1}\Lambda (\tau
)-\frac{1}{\tau }]\ \delta \ ^{q}\mathcal{V}(\tau ),  \notag \\
\ \ ^{q}\widehat{S}(\tau ) &=&-\ ^{q}\widehat{W}(\tau )=-\int_{\widehat{\Xi }%
}\frac{1}{8\left( \pi \tau \right) ^{2}}[\tau (\ _{1}^{tot}\Lambda (\tau )+\
_{1}^{1}\Lambda (\tau )+\ _{2}^{tot}\Lambda (\tau )+\ _{2}^{1}\Lambda (\tau
)-2]\delta \ ^{q}\mathcal{V}(\tau ).  \notag
\end{eqnarray}

To compute the volume form $\delta \ ^{q}\mathcal{V}(\tau )$ (\ref{volume})
is better to consider the equivalent d-metric (\ref{whpolf}) with $\eta $%
--polarization functions, or (\ref{whpolf1}) for $\eta $--polarization
functions, and including data for nonmetric generating sources.
Respectively, we can write 
\begin{eqnarray}
\ _{2}\Phi (\tau ) &=&2\sqrt{|[\ _{2}^{tot}\Lambda (\tau )+\ _{2}^{1}\Lambda
(\tau )]\ g_{4}(\tau )|}=\ 2\sqrt{|\ [\ _{2}^{tot}\Lambda (\tau )+\
_{2}^{1}\Lambda (\tau )]\ \eta _{4}(\tau )\ \breve{g}_{4}(\tau )|}  \notag \\
&\simeq &2\sqrt{|\ [\ _{2}^{tot}\Lambda (\tau )+\ _{2}^{1}\Lambda (\tau )]\
\zeta _{4}(\tau )\ \breve{g}_{4}|}[1-\frac{\varepsilon }{2}\chi _{4}(\tau )].
\label{genf1}
\end{eqnarray}%
For simplicity, we shall elaborate on nonholonomic evolution models with
trivial integration functions $\ _{1}n_{k}=0$ and $\ _{2}n_{k}=0.$
Introducing formulas (\ref{genf1}) in (\ref{volume}), then separating terms
with shell $\tau $-running cosmological constants, we express: 
\begin{eqnarray*}
\ \delta \ ^{q}\mathcal{V} &=&\delta \mathcal{V}[\tau ,\ \ _{1}^{tot}\Lambda
(\tau )+\ _{1}^{1}\Lambda (\tau ),\ _{2}^{tot}\Lambda (\tau )+\
_{2}^{1}\Lambda (\tau );\ _{h}^{q}\Upsilon (\tau ),\ ^{q}\Upsilon (\tau
);\psi (\tau ),\ g_{4}(\tau )] \\
&=&\delta \mathcal{V}(\ \ _{h}^{q}\Upsilon (\tau ),\ ^{q}\Upsilon (\tau ),\
\ _{1}^{tot}\Lambda (\tau )+\ _{1}^{1}\Lambda (\tau ),\ _{2}^{tot}\Lambda
(\tau )+\ _{2}^{1}\Lambda (\tau ),\eta _{4}(\tau )\breve{g}_{4}) \\
&=&\frac{1}{\sqrt{|[\ \ _{1}^{tot}\Lambda (\tau )+\ _{1}^{1}\Lambda (\tau
)][\ _{2}^{tot}\Lambda (\tau )+\ _{2}^{1}\Lambda (\tau )]|}}\ \delta \
_{\eta }\mathcal{V},\mbox{ where }\ \delta \ _{\eta }\mathcal{V}=\ \delta \
_{\eta }^{1}\mathcal{V}\times \delta \ _{\eta }^{2}\mathcal{V}.
\end{eqnarray*}%
In these formulas, we use the functionals:%
\begin{eqnarray}
\delta \ _{\eta }^{1}\mathcal{V} &=&\delta \ _{\eta }^{1}\mathcal{V}[\
_{1}^{tot}\Lambda (\tau )+\ _{1}^{1}\Lambda (\tau ),\eta _{1}(\tau )\ \breve{%
g}_{1}]  \label{volumfuncts} \\
&=&e^{\widetilde{\psi }(\tau )}dx^{1}dx^{2}=\sqrt{|\ \ \ _{1}^{tot}\Lambda
(\tau )+\ _{1}^{1}\Lambda (\tau )|}e^{\psi (\tau )}dx^{1}dx^{2},\mbox{ for }%
\psi (\tau )\mbox{ being a solution of  }(\ref{eq1}),  \notag \\
\delta \ _{\eta }^{2}\mathcal{V} &=&\delta \ _{\eta }^{2}\mathcal{V}[\
^{q}\Upsilon (\tau ),\eta _{4}(\tau )\ \breve{g}_{4}]  \notag \\
&=&\frac{\partial _{3}|\ \eta _{4}(\tau )\ \breve{g}_{4}|^{3/2}}{\ \sqrt{%
|\int dy^{3}\ \ ^{q}\Upsilon (\tau )\{\partial _{3}|\ \eta _{4}(\tau )\ 
\breve{g}_{4}|\}^{2}|}}[dy^{3}+\frac{\partial _{i}\left( \int dy^{3}\ \
^{q}\Upsilon (\tau )\partial _{3}|\ \eta _{4}(\tau )\ \breve{g}_{4}|\right)
dx^{i}}{\ \ ^{q}\Upsilon (\tau )\partial _{3}|\ \eta _{4}(\tau )\ \breve{g}%
_{4}|}]dt,  \notag
\end{eqnarray}%
where numeric coefficients were used for re-defining the generating
functions. We note that we can define $\widetilde{\psi }(\tau )$ as a $\tau $%
--family of solutions of 2-d Poisson equations with effective source $\
_{1}^{tot}\Lambda (\tau )+\ _{1}^{1}\Lambda (\tau ),$ or use $\psi (\tau )$
for a respective source $\ _{h}^{q}\Upsilon (\tau ).$ Integrating on a
closed hypersurface $\widehat{\Xi }$ such products of $h$- and $v$-forms, we
obtain a running phase space volume functional 
\begin{equation*}
\ _{\eta }^{\shortmid }\mathcal{\mathring{V}}(\tau )=\int_{\ \widehat{\Xi }%
}\delta \ _{\eta }\mathcal{V}(\ _{h}^{q}\Upsilon (\tau ),\ ^{q}\Upsilon
(\tau ),\ \breve{g}_{\alpha })  \label{volumfpsp}
\end{equation*}%
determined by prescribed classes of generating $\eta $-functions, effective
generating sources $\left[ _{h}^{q}\Upsilon (\tau ),\ ^{q}\Upsilon (\tau )%
\right] ,$ coefficients of a prime s-metric $\ \mathring{g}_{\alpha }$ and
nonholonomic distributions defining the hyper-surface $\widehat{\Xi }.$ The
explicit value of $\ _{\eta }^{\shortmid }\mathcal{\mathring{V}}(\tau )$
depends on the data we prescribe for $\widehat{\Xi }$ the type of $Q$%
-deformations (via $\eta $- or $\zeta $-polarizations) we use for deforming
a prime wormhole d-metric into quasi-stationary ones as we considered in
section \ref{sec3}. We emphasize that it is always possible to compute $\
_{\eta }^{\shortmid }\mathcal{\mathring{V}}(\tau )$ for certain nonlinear
solitonic waves/ distributions and some general $Q$-deformations. The
thermodynamic variables depend on the $\tau $-running effective cosmological
constants.

\section{Conclusions and open questions}
\label{sec5} In this work we elaborated on the nonmetric geometric flow theory of metric-affine spaces and 
applied it to modified gravity theories, MGT, as in \cite{hehl95,vmon05,harko21,iosifidis22,khyllep23}. The approach was generalized in nonholonomic form \cite{bubuianu17,partner02,ibsvevv22} with the aim to apply the anholonomic frame and connection deformation method, AFCDM, for constructing physically important exact and parametric solutions in geometric flow and gravity theories with nonmetricity. Such solutions are defined by generic off-diagonal metrics and generalized (non) linear connections and, in general, do not possess hypersurface/ duality / holographic configurations which would allow to treat them in the framework of Bekenstein-Hawking paradigm \cite{bek1,bek2,haw1,haw2}. In another turn, as we have shown in this paper,  the G. Perelman statistical and geometric thermodynamic paradigm  \cite{perelman1} can be applied for all types of solutions in MGTs  including nonmetric geometric flow evolution models as we considered in the previous section. In addition to gaining a more complete understanding of gravity theories with nonmetricity, we also studied in this article certain new classes of wormhole and solitonic solutions encoding nonmetric data. This included such new and original results:
\begin{enumerate}
\item In section \ref{ss21}, the metric-affine geometry was formulated in nonholonomic dyadic variables for nonmetric $Q$-deformed 4-d Lorentz manifolds. Such a formulation allows us to prove general decoupling and integration properties of nonmetric geometric flow equations and modified Einstein equations in MGTs as we outlined in Appendix \ref{appendixa}.

\item The Obj1 of this work was completed in section \ref{ss22} where Lyapunov type F- and W-functionals are defined for nonholonomic variables encoding $Q$-deformations. This allowed us to formulate nonmetric geometric flow models, which for self-similar configurations define nonmetric Ricci solitons containing as particular cases, for instance, nonmetric
gravitational equations studied in \cite{vmon05,harko21,iosifidis22}.

\item In section \ref{sec3}, we solved the goals of Obj2 by constructing in explicit form two classes of physically important quasi-stationary solutions of nonmetric geometric flow equations which for fixed flow parameters define $Q$-deformed Einstein spaces. We proved that such generic off-diagonal solutions can be described in general form by respective solitonic hierarchies and solitonic distributions (see subsection \ref{ss32}, when the necessary concepts and formulas are outlined in Appendix \ref{appendixb}).

\item Wormhole solutions present an important tool for testing MGTs and applications in modern quantum computer science as we show by constructing and analyzing possible nonmetric effects in subsection \ref{ss33}. Such configurations can be with nonholonomic solitonic $Q$-deformations and gravitational polarizations, locally anisotropic, in particular, ellipsoid deformations of throats, when nonmetricity makes such configurations to be transversable.

\item In general, the quasi-stationary solutions encoding nonmetricity do not involve hypersurface / holographic configurations or certain duality conditions when the concept of Bekenstein-Hawking entropy could be applicable. As in GR and other MGTs, general classes of exact/ parametric solutions can be characterized thermodynamically in the framework of corresponding generalization of G. Perelman paradigm with W-entropy. In section \ref{sec4}, we show how such constructions can be performed for $Q$-deformations, which presents a solution of Obj3.
\end{enumerate}

The quasi-stationary solutions constructed in this work are characterized by a $\tau $-family of Weyl d-vectors $\mathbf{q}_{\alpha}(\tau )$ subjected to conditions of type (\ref{modifcomp}). For such nonholonomic configurations,
we can computed respectively induced $\tau $-families of nonmetricity d-tensors, $\mathbf{Q}_{\alpha \beta \gamma }(\tau )=\mathbf{q}_{\alpha}(\tau )\mathbf{g}_{\beta \gamma }(\tau )$, and d-torsion, $\mathbf{T}_{\mu\nu \alpha }(\tau)=\mathbf{A}_{\nu }(\tau )\mathbf{g}_{\mu \alpha}(\tau) -\mathbf{A}_{\alpha }(\tau )\mathbf{g}_{\mu \nu }(\tau ),$ for $\mathbf{A}_{\mu }(\tau )= q\mathbf{q}_{\mu }(\tau ),$ when $q=const$ or depend on $\tau .$ Such a $q$ can be chosen as a small parameter (like $\varepsilon $ in (\ref{epsilongenfdecomp})) for $Q$-deformations which allow to construct parametric deformations of physically important solutions in GR.

\vskip5pt Let us discuss the legacy of using $Q$-deformed Perelman's F- and W-functionals to formulate and prove analogues of Poincar\'{e}--Thorston conjecture \cite{perelman1,monogrrf1,monogrrf2,monogrrf3} for nonmetric geometric flows. For general $Q$-deformations this is an un-defined mathematical problem similar to those for an infinite number of noncommutative/ nonassociative differential and integral calculuses and geometric theories, see discussions and respective variants of solutions in  \cite{ibsvevv22,partner02,ibsvev20}. For metric-affine spaces,  an infinite number of topological and nonmetric geometric models can be formulated because of an infinite number of nonlinear and linear connection structures that can be used. So, it is not possible to formulate a general mathematical framework involving only some fundamental topological theories and nonmetric geometric analysis. Nevertheless, self-consistent generalizations of the statistic and nonmetric geometric thermodynamics are possible if there are used $Q$-deformations as in sections \ref{sss221} and \ref{sec4}. They encode nonmetric geometric data and result in nonholonomic Ricci soliton configurations and $Q$-modified Einstein equations. Such systems of nonlinear PDEs can be solved in some general forms as we show in Appendix  \ref{appendixa} and provide explicit examples in section \ref{sec3}. For instance, we can associate and compute for such generic off-diagonal solutions respective Perelman-like nonmetric geometric thermodynamic variables, see respective formulas (\ref{qthermvar}), (\ref{thermvar1}) and (\ref{thermvar2}). Thus, such nonmetric geometric flow and MGTs and their associated thermodynamic theories can be formulated in a self-consistent form as $\tau $-parametric $Q$-deformations of Lorentz manifolds geometries, and this is possible even if we are not able to formulate in general form a rigorous version of metric-affine Poincar\`{e} hypothesis. Here we also note that the concept of Bekenstein-Hawking entropy is not applicable for the classes of nonmetric solitonic and wormholes solutions considered in section \ref{sec3}. However, the concepts of Perelman's W-entropy and related statistical thermodynamics can be generalized for various classes of nonmetric theories and their solutions. 

\vskip5pt The results of this work support the Hypothesis from the Introduction section in such senses:
\begin{enumerate}
\item We constructed in explicit form certain models of metric-affine geometric flow and MGTs which are exactly/ parametric solvable in certain general off-diagonal forms in nonholonomic dyadic variables.

\item The solutions with $\tau $-running effective cosmological constants can be used for modelling DE physical effects and other type configurations with generating sources for effective matter (all such solutions encoding nonmetricity data) for modelling DM physics.

\item In this paper, we elaborated only on nonmetric quasi-stationary configurations which can be described as solitonic hierarchies or nonmetric wormhole solutions and certain nonlinear $Q$-deformations of such generic off-diagonal solutions subjected to respective nonlinear symmetries.

\item Perelman type nonmetric geometric thermodynamic variables were defined and computed in explicit form for the mentioned classes of quasi-stationary solutions.
\end{enumerate}

Nevertheless, there is a series of important fundamental problems that should be investigated and solved in future works. Here we outline four of the most important open questions on nonmetric geometric and information flow theories and gravity (QNGIFG) and cite some relevant previous works:
\begin{itemize}
\item QNGIFG1: To elaborate full and viable classical and quantum theories on metric-affine spacetimes we have to formulate a theory of spinors and $Q$-deformed Dirac operators, which is not possible in general form for arbitrary nonmetric structures. This problem is discussed in more general forms in \cite{vplb10,vacaru18} for phase or Finsler-Lagrange-Hamilton theories on (co) tangent Lorentz bundles. Corresponding conceptual and technical difficulties exist for metric-affine generalizations of Lorentz manifolds. Certain solutions can be found as $Q$-deformed off-diagonal
Einstein-Dirac systems, see previous results \cite{vsp98}.
\item QNGIFG2: One of the next steps is to study models of $Q$-deformed Einstein-Yang-Mills-Higgs systems. If such systems are derived as star-product R-flux deformations in string theory, the obtain nonholonomic geometric structures with nonsymmetric metrics and $Q$-deformed Einstein-Eisenhart-Moffat theories, see details and references to
    \cite{partner02,vmon05}.
\item QNGIFG3: $Q$-deformed off-diagonal cosmological systems can be considered as certain dual ones to quasi-stationary configurations as stated in \cite{bubuianu17}. Such solutions involve, for instance,
various quasi-periodic (cosmological time quasi-crystals etc.) and time-solitonic hierarchies which can be exploited for modelling DE and DM effects.
\item QNGIFG4: Finally, we point to the possibility to extend the geometric and quantum information flow theory \cite{ibsvevv22,ibsvev20} to certain $Q$-deformed versions with nonmetric qubits, nonmetric entanglement and respective generalizations of conditional entropies with $Q$-modified Perelman's functionals.
\end{itemize}

We shall report on progress to answers for above questions in future works.

\vskip6pt \textbf{Acknowledgments and Historical Remarks:} 

\vskip4pt SV activity on nonmetric geometry and constructing off-diagonal solutions in MGTs was inspired by many discussions during 1986-1988 with Ascar Aringazin, Sergei Ponomarenko and Anatolii Mikhailov who were postgraduate students of Prof. G. A. Asanov at the department of theoretical physics of M. Lomonosov Moscow State University. Those ideas and nonholonomic geometric formalism resulted in elaborating the AFCDM in a series of works by SV and his
students from R. Moldova, for metric affine Finsler-Lagrange-Hamilton geometry and nonmetric locally anisotropic gravity theories \cite{vmon05}, see Introduction, Conclusion and Appendix B to \cite{vacaru18} for historical remarks, main references and reviews of results. This research on nonmetric geometric flows and applications extends SV and AZ visit programs, supported respectively by Fulbright USA-Romania and the Ministry of Education and Science of the Republic of Kazakhstan, at the physics department at California State University at Fresno, USA. The program on the AFCDM for theories with effective sources encoding nonmetricity consists also a generalization of some objectives for a SV's visiting program at CAS LMU in Munich, Germany. Co-affiliations for the universities in Germany, Ukraine and USA are motivated by the mentioned fellowships. Authors are grateful to Prof. D. L\"{u}st, Prof. D. Singleton and Dr. Lena Bouman for respective support of their research activity.

\appendix

\setcounter{equation}{0} \renewcommand{\theequation}
{A.\arabic{equation}} \setcounter{subsection}{0} 
\renewcommand{\thesubsection}
{A.\arabic{subsection}}

\section{Decoupling and integrability of nonmetric quasi-stationary geometric flow equations}

\label{appendixa}In this appendix, we summarize necessary formulas which allow us to prove some general decoupling and integration properties of nonmetric geometric flow equations (\ref{ricciflowr2}) with generating sources (\ref{qgenersourc}). There are studied generic off-diagonal solutions parameterized by a quasi-stationary ansatz (\ref{stationarydm}). Details on nonholonomic geometric methods can be found in generalized Finsler-Lagrange-Hamilton and metric--affine forms in  \cite{vmon05,bubuianu17,vacaru18}. 

\subsection{Decoupling of nonlinear PDEs with nonmetricity fields}

We provide formulas for $\tau $-depending coefficients of fundamental
geometric objects describing nonmetric geometric flows in such a form that
for any fixed $\tau _{0}$ we shall generate solutions for nonmetric Ricci
solitons.

\subsubsection{The coefficients of quasi-stationary ansatz and canonical
Ricci d-tensor}

To generate quasi-stationary $\tau $-evolving solutions of the system of
nonlinear PDEs (\ref{cdeq1}) we consider a generic off-diagonal ansatz with
N-adapted coefficients of type (\ref{stationarydm}), when the nonlinear
quadratic element is 
\begin{eqnarray}
\mathbf{\hat{g}}(\tau ) &=&g_{i}(\tau ,x^{k})dx^{i}\otimes dx^{i}+h_{3}(\tau
,x^{k},y^{3})\mathbf{e}^{3}(\tau )\otimes \mathbf{e}^{3}(\tau )+h_{4}(\tau
,x^{k},y^{3})\mathbf{e}^{4}(\tau )\otimes \mathbf{e}^{4}(\tau ),  \notag \\
&&\mathbf{e}^{3}(\tau )=dy^{3}+w_{i}(\tau ,x^{k},y^{3})dx^{i},\ \mathbf{e}%
^{4}(\tau )=dy^{4}+n_{i}(\tau ,x^{k},y^{3})dx^{i}.  \label{dmq}
\end{eqnarray}%
This d-metric possess a Killing symmetry on the time like coordinate $%
\partial _{4}=\partial _{t}$. Such d-metric and N-connection coefficients
are functions of necessary smooth class on respective coordinates. We put a
"hat" label for a family of d-metrics $\mathbf{\hat{g}}(\tau )$ in order to
emphasize that such d-metrics are with Killing symmetry on $\partial _{t}.$
It is supposed that such a parametrization can be obtained for corresponding
classes of frame/ coordinate transforms for a general family $\mathbf{\hat{g}%
}(\tau ,u)$ depending on all spacetime coordinates for other systems of
references.

Tedious computations of N-adapted coefficients of the canonical d-connection
and respective Ricci d-tensors for (\ref{dmq}) result in such formulas for
the system of nonlinear PDEs (\ref{cdeq1}): 
\begin{eqnarray}
\widehat{R}_{1}^{1}(\tau ) &=&\widehat{R}_{2}^{2}(\tau )=\frac{1}{2g_{1}g_{2}%
}[\frac{g_{1}^{\bullet }g_{2}^{\bullet }}{2g_{1}}+ \frac{(g_{2}^{%
\bullet})^{2}}{2g_{2}}-g_{2}^{\bullet \bullet } + \frac{g_{1}^{\prime
}g_{2}^{\prime}}{2g_{2}}+ \frac{\left( g_{1}^{\prime }\right) ^{2}}{2g_{1}}%
-g_{1}^{\prime \prime }]=-\ \ _{1}^{q}\mathbf{\Upsilon }(\tau ),  \notag \\
\widehat{R}_{3}^{3}(\tau ) &=&\widehat{R}_{4}^{4}(\tau )=\frac{1}{2h_{3}h_{4}%
}[\frac{\left( h_{4}^{\ast }\right) ^{2}}{2h_{4}}+\frac{h_{3}^{\ast
}h_{4}^{\ast }}{2h_{3}}-h_{4}^{\ast \ast }]=-\ \ _{2}^{q}\mathbf{\Upsilon }%
(\tau ),  \label{riccist2} \\
\widehat{R}_{3k}(\tau ) &=&\frac{\ w_{k}}{2h_{4}}[h_{4}^{\ast \ast }-\frac{%
\left( h_{4}^{\ast }\right) ^{2}}{2h_{4}}-\frac{(h_{3}^{\ast })(h_{4}^{\ast
})}{2h_{3}}]+\frac{h_{4}^{\ast }}{4h_{4}}(\frac{\partial _{k}h_{3}}{h_{3}}+%
\frac{\partial _{k}h_{4}}{h_{4}})-\frac{\partial _{k}(h_{3}^{\ast })}{2h_{3}}%
=0;  \notag \\
\widehat{R}_{4k}(\tau ) &=&\frac{h_{4}}{2h_{3}}n_{k}^{\ast \ast }+\left( 
\frac{3}{2}h_{4}^{\ast }-\frac{h_{4}}{h_{3}}h_{3}^{\ast }\right) \frac{\
n_{k}^{\ast }}{2h_{3}}=0.  \notag
\end{eqnarray}%
For simplicity, we use brief notations of partial derivatives when, for
instance, $\partial _{1}q(u^{\alpha }):=q^{\bullet }, \partial
_{2}q(u^{\alpha }):=q^{\prime }, \partial _{3}q(u^{\alpha }):=q^{\ast }$ for
an arbitrary function $q(u^{\alpha }).$ In abstract geometric form, such
formulas are written in similar forms as in various MGTs but in this work
the generating sources encode nonmetricity terms as we explained for (\ref%
{qgenersourc}).

\subsubsection{Decoupling of nonmetric geometric flow equations}

Let us express $g_{i}(\tau )=e^{\psi (\tau ,x^{k})}$ and introduce the
coefficients $\alpha _{i}(\tau )=h_{4}^{\ast }\partial _{i}[\varpi (\tau )],
\beta (\tau )=h_{4}^{\ast }(\tau )[\varpi (\tau )]^{\ast },$ $\gamma
(\tau)=(\ln |h_{4}(\tau )|^{3/2}/|h_{3}(\tau )|)^{\ast },$ for $\varpi (\tau
)=\ln |h_{4}^{\ast }(\tau )/\sqrt{|h_{3}(\tau )h_{4}(\tau )}|;$ and
considering $\ \Psi (\tau )=\exp [\varpi (\tau )]$ as a family of \textbf{%
generating functions.} The equations (\ref{riccist2}) transform into: 
\begin{eqnarray}
\psi ^{\bullet \bullet }+\psi ^{\prime \prime } &=&2\ \ _{1}^{q}\mathbf{%
\Upsilon }(\tau ),  \label{eq1} \\
(\varpi )^{\ast }h_{4}^{\ast } &=&2h_{3}h_{4}\ _{2}^{q}\mathbf{\Upsilon }%
(\tau ),  \label{e2a} \\
\beta w_{j}-\alpha _{j} &=&0,  \label{e2b} \\
\ n_{k}^{\ast \ast }+\gamma n_{k}^{\ast } &=&0,  \label{e2c}
\end{eqnarray}%
where the explicit dependence of coefficients on respective $(\tau ,x^{k})$
or $(\tau ,x^{k},y^{3})$ is omitted. This system of equations together with
the previous one possess an explicit decoupling property. In brief, this
means that $g_{i}(\tau )$ are related to a $\tau $-family of 2-d Poisson
equations (\ref{eq1}); then $h_{3}(\tau )$ and $h_{4}(\tau )$ $\ $are
related via nontrivial $\varpi (\tau )$ and $\ _{2}^{q}\mathbf{\Upsilon }%
(\tau )$ as in (\ref{e2a}). Finding any solution for $h_{a}(\tau ),$ we can
compute the families of coefficients $\beta (\tau )$ and $\alpha _{i}(\tau )$
and solve respective linear equations for $w_{j}(\tau )$ from (\ref{e2b}).
To find solutions for $n_{k}(\tau )$ we have to integrate two times on $%
y^{3} $ in (\ref{e2c}) when $\gamma (\tau )$ is determined by $h_{3}(\tau )$
and $h_{4}(\tau ).$

\subsection{Off-diagonal solutions for nonmetric quasi-stationary
configurations}

We can generate $\tau $-families of solutions of nonmetric geometric flow
equations by integrating recurrently the decoupled system of nonlinear PDEs (%
\ref{eq1})-(\ref{e2c}). Any generic off-diagonal metric (\ref{dmq}) (if the
N-coefficients vanish for certain coordinate transforms, we generate
diagonal metrics) is determined by respective families of generating
function $\varpi (\tau )$ (equivalently, $\Psi (\tau )$) and two generating
sources $\ _{1}^{q}\mathbf{\Upsilon }(\tau )$ and $\ _{2}^{q}\mathbf{%
\Upsilon }(\tau ).$ The explicit form of such solutions depends on the type
of parameterizations of generating functions and generating sources and how
such values are related to some integration functions.

\subsubsection{Generating functions and sources for nonmetric
quasi-stationary off-diagonal solutions}

By straightforward computations, we can check that exact solutions are
defined by such generic off-diagonal quasi-stationary $\tau $-families of
d-metrics,%
\begin{eqnarray}
ds^{2}(\tau ) &=&e^{\psi (\tau ,x^{k})}[(dx^{1})^{2}+(dx^{2})^{2}]+\frac{%
[\Psi ^{\ast }]^{2}}{4(\ _{2}^{q}\mathbf{\Upsilon })^{2}\{g_{4}^{[0]}-\int
dy^{3}[\Psi ^{2}]^{\ast }/4(\ \ _{2}^{q}\mathbf{\Upsilon })\}}(dy^{3}+\frac{%
\partial _{i}\Psi }{\Psi ^{\ast }}dx^{i})^{2}+  \label{qeltors} \\
&&\{g_{4}^{[0]}-\int dy^{3}\frac{[\Psi ^{2}]^{\ast }}{4(\ _{2}^{q}\mathbf{%
\Upsilon })}\}\{dt+[\ _{1}n_{k}+\ _{2}n_{k}\int dy^{3}\frac{[(\Psi
)^{2}]^{\ast }}{4(\ _{2}^{q}\mathbf{\Upsilon })^{2}|g_{4}^{[0]}-\int
dy^{3}[\Psi ^{2}]^{\ast }/4(\ _{2}^{q}\mathbf{\Upsilon })|^{5/2}}]dx^{k}\}. 
\notag
\end{eqnarray}%
If for such d-metrics there are considered parametric decompositions as in (%
\ref{qsourc}), we generate recurrently certain classes of parametric
solutions.

With respect to coordinate dual frames, the d-metrics (\ref{qeltors}) can be
represented in the form 
\begin{equation*}
\mathbf{\hat{g}}=\underline{\widehat{g}}_{\alpha \beta }(\tau ,u)du^{\alpha
}\otimes du^{\beta },
\end{equation*}
when the off-diagonal metrics are parameterized in the form 
\begin{equation*}
\underline{\widehat{g}}_{\alpha \beta }(\tau ,u)=\left[ 
\begin{array}{cccc}
e^{\psi }+(w_{1})^{2}h_{3}+(n_{1})^{2}h_{4} & w_{1}w_{2}h_{3}+n_{1}n_{2}h_{4}
& w_{1}h_{3} & n_{1}h_{4} \\ 
w_{1}w_{2}h_{3}+n_{1}n_{2}h_{4} & e^{\psi }+(w_{2})^{2}h_{3}+(n_{2})^{2}h_{4}
& w_{2}h_{3} & n_{2}h_{4} \\ 
w_{1}h_{3} & w_{2}h_{3} & h_{3} & 0 \\ 
n_{1}h_{4} & n_{2}h_{4} & 0 & h_{4}%
\end{array}%
\right] ,
\end{equation*}%
where we omit respective dependencies of coefficients on $(\tau,x^{k},y^{3}) 
$ are omitted.

Finally, we note that $\tau $-families of quasi-stationary solutions (\ref%
{qeltors}) are general in the sense that they are determined by some general
generating function $\Psi (\tau,x^{k},y^{3}),$ two generating effective
sources $\ _{1}^{q}\mathbf{\Upsilon }(\tau ,x^{k})$ (encoded as a solution $%
\psi (\tau ,x^{k})$ of 2-d Poisson equation (\ref{eq1})) and $\ _{2}^{q}%
\mathbf{\Upsilon }(\tau,x^{k},y^{3})$, and integration functions $\
_{1}n_{i}(\tau ,x^{k}),\ _{2}n_{i}(\tau ,x^{k})$ and $h_{4}^{[0]}(\tau
,x^{k}).$

\subsubsection{Nonlinear symmetries of nonmetric quasi-stationary
off-diagonal solutions}

\label{assnonlinearsym} The $\tau $-family of solutions (\ref{qeltors})
posses important nonlinear shell symmetries which allow us to define in
explicit form certain transforms of generating functions and effective
sources into other types of generating functions and effective cosmological
constants. Such formulas allow to change the generating data, $(\Psi (\tau
), \ _{2}^{q}\mathbf{\Upsilon }(\tau ))\leftrightarrow (\Phi (\tau ),\
_{2}^{q}\Lambda (\tau )=const\neq 0,$ for $\tau _{0}),$ using the formulas%
\begin{eqnarray}
\frac{\lbrack \Psi ^{2}]^{\ast }}{\ _{2}^{q}\mathbf{\Upsilon }(\tau )} &=&%
\frac{[\Phi ^{2}(\tau )]^{\ast }}{\ _{2}^{q}\Lambda (\tau )},%
\mbox{ which can be
integrated as  }  \label{ntransf1} \\
\Phi ^{2}(\tau ) &=&\ _{2}^{q}\Lambda (\tau )\int dy^{3}(\ \ _{2}^{q}\mathbf{%
\Upsilon })^{-1}[\Psi ^{2}(\tau )]^{\ast }\mbox{ and/or }\Psi ^{2}(\tau )=(\
_{2}^{q}\Lambda (\tau ))^{-1}\int dy^{3}(\ \ _{2}^{q}\mathbf{\Upsilon }%
)[\Phi ^{2}(\tau )]^{\ast }.  \label{ntransf2}
\end{eqnarray}%
Such nonlinear symmetries can be defined for other types of effective
sources. For instance, we can consider effective matter sources $\ ^{tot}%
\widehat{\mathbf{T}}_{\mu \nu }$(\ref{totem}), for (\ref{dsourcparam}) as in
(\ref{nonheinst}), when $\tau $-running effective cosmological constants are
parameterized in the form $\ _{2}^{m}\Lambda (\tau )+\ _{2}^{F}\Lambda
(\tau)+\ _{2}^{e}\Lambda (\tau ).$ Corresponding nonlinear symmetries can be
considered for generating new classes of off-diagonal solutions and in order
to compare their physical implications defined by different types of
effective cosmological constants. For nonmetric effective sources and
running cosmological constants, we use the left label $q$ when $\
_{2}^{q}\Lambda (\tau )=0$ is taken for metric compatible configurations.

We conclude that any quasi-stationary solution (\ref{qeltors}) possess
important nonlinear symmetries of type (\ref{ntransf1}) and (\ref{ntransf2}%
). As a result, the nonlinear quadratic element for quasi-stationary
solutions (\ref{qeltors}) can be written in the form%
\begin{eqnarray}
ds^{2}(\tau ) &=&g_{\alpha _{s}\beta _{s}}(\tau ,x^{k},y^{3},\Phi (\tau
),_{2}^{q}\Lambda (\tau ))du^{\alpha }du^{\beta }=e^{\psi (\tau
,x^{k})}[(dx^{1})^{2}+(dx^{2})^{2}]  \label{offdiagcosmcsh} \\
&&-\frac{\Phi ^{2}(\tau )[\Phi ^{\ast }(\tau )]^{2}}{|\ _{2}^{q}\Lambda
(\tau )\int dy^{3}\ _{2}^{q}\mathbf{\Upsilon }(\tau )[\Phi ^{2}(\tau
)]^{\ast }|[g_{4}^{[0]}(\tau )-\Phi ^{2}(\tau )/4\ _{2}^{q}\Lambda (\tau )]}%
\{dy^{3}+\frac{\partial _{i}\ \int dy^{3}\ \ _{2}^{q}\mathbf{\Upsilon }(\tau
)\ [\Phi ^{2}(\tau )]^{\ast }}{\ \ _{2}^{q}\mathbf{\Upsilon }(\tau )\ [(\
_{2}\Phi (\tau ))^{2}]^{\ast }}dx^{i}\}^{2}  \notag \\
&&-\{g_{4}^{[0]}(\tau )-\frac{\Phi ^{2}(\tau )}{4\ _{2}^{q}\Lambda (\tau )}%
\}\{dt+[\ _{1}n_{k}(\tau )\   \notag \\
&&+\ _{2}n_{k}(\tau )\int dy^{3}\frac{\Phi ^{2}(\tau )[\Phi ^{\ast }(\tau
)]^{2}}{|\ _{2}^{q}\Lambda (\tau )\int dy^{3}\ \ \ _{2}^{q}\mathbf{\Upsilon }%
(\tau )[\Phi ^{2}(\tau )]^{\ast }|[g_{4}^{[0]}(\tau )-\Phi ^{2}(\tau )/4\
_{2}^{q}\Lambda (\tau )]^{5/2}}]\},  \notag
\end{eqnarray}%
for indices: $i,j,k,...=1,2;a,b,c,...=3,4;$ generating functions $\psi (\tau
,x^{k})$ and$\ \Phi (\tau ,x^{k_{1}}y^{3});$ generating sources $\ _{2}^{q}%
\mathbf{\Upsilon }(\tau ,x^{k})$ and $\ _{2}^{q}\mathbf{\Upsilon }(\tau,
x^{k_{1}},y^{3});$ effective cosmological constants $\ _{1}^{q}\Lambda (\tau
)$ and $\ _{2}^{q}\Lambda (\tau );$ and integration functions\newline
$\ _{1}n_{k}(\tau ,x^{j}),\ _{2}n_{k}(\tau ,x^{j})$ and $g_{4}^{[0]}(\tau ,
x^{k}).$

\subsubsection{Using some d-metric coefficients as generating functions and
nonmetricity}

Formulas (\ref{offdiagcosmcsh}) allow us to write $h_{4}^{\ast }(\tau
)=-[\Psi ^{2}(\tau )]^{\ast }/4\ \ _{2}^{q}\mathbf{\Upsilon }(\tau )$ and to
compute up to certain integration functions a value of $\ \Psi (\tau )$. We
have to integrate $[\Psi ^{2}(\tau )]^{\ast }=\int dy^{3}\ _{2}^{q}\mathbf{%
\Upsilon }(\tau )h_{4}^{\ast }(\tau )$ for any prescribed $h_{4}(\tau )$ and 
$\ _{2}^{q}\mathbf{\Upsilon }(\tau ).$ So, considering generating data $%
(h_{4}(\tau ), \ _{2}^{q}\mathbf{\Upsilon }(\tau ))$, we can write the $\tau$%
-families of quasi-stationary d-metric (\ref{qeltors}) in such equivalent
forms, 
\begin{eqnarray}
d\widehat{s}^{2}(\tau ) &=&\widehat{g}_{\alpha \beta }(\tau
,x^{k},y^{3};h_{4}(\tau ),\ _{2}^{q}\mathbf{\Upsilon }(\tau ))du^{\alpha
}du^{\beta }  \label{offdsolgenfgcosmc} \\
&=&e^{\psi (\tau ,x^{k})}[(dx^{1})^{2}+(dx^{2})^{2}]-\frac{[h_{4}^{\ast
}(\tau )]^{2}}{|\int dy^{3}[\ _{2}^{q}\mathbf{\Upsilon }(\tau )h_{4}(\tau
)]^{\ast }|\ h_{4}(\tau )}\{dy^{3}+\frac{\partial _{i}[\int dy^{3}(\ \
_{2}^{q}\mathbf{\Upsilon }(\tau ))\ h_{4}^{\ast }(\tau )]}{\ \ _{2}^{q}%
\mathbf{\Upsilon }(\tau )\ h_{4}^{\ast }(\tau )}dx^{i}\}^{2}  \notag \\
&&+h_{4}(\tau )\{dt+[\ _{1}n_{k}(\tau )+\ _{2}n_{k}(\tau )\int dy^{3}\frac{%
[h_{4}^{\ast }(\tau )]^{2}}{|\int dy^{3}[\ \ _{2}^{q}\mathbf{\Upsilon }(\tau
)h_{4}(\tau )]^{\ast }|\ [h_{4}(\tau )]^{5/2}}]dx^{k}\}.  \notag
\end{eqnarray}

In a similar form using the nonlinear symmetries (\ref{ntransf1}) and (\ref%
{ntransf2}) and expressing $\Phi ^{2}(\tau )=-4\ _{2}^{q}\Lambda
(\tau)h_{4}(\tau ),$ we can eliminate $\Phi (\tau )$ from the nonlinear
quadratic element in (\ref{offdiagcosmcsh}). We construct a $\tau $-family
of solutions of type (\ref{offdsolgenfgcosmc}) determined by some generating
data $(h_{4}(\tau );\ _{2}^{q}\Lambda (\tau ),\ _{2}^{q}\mathbf{\Upsilon }%
(\tau )).$

\subsubsection{Gravitational nonmetric polarizations}

We can consider deformations of a \textbf{prime} d-metric on a metric affine
manifold 
\begin{equation}
\mathbf{\mathring{g}=}[\mathring{g}_{\alpha },\ \mathring{N}_{i}^{a}].
\label{offdiagpm}
\end{equation}%
Such a $\mathbf{\mathring{g}}$ can be an arbitrary one, a solution of some
equations in GR or a MGTs, or defined as a $\tau $-family. In this
subsection, we study transforms of a (family) primary d-metric into a a
family of \textbf{target} d-metrics $\mathbf{g}(\tau )$ defined a
quasi-stationary solutions of type (\ref{dmq}), when 
\begin{equation}
\mathbf{\mathring{g}}\rightarrow \mathbf{g}(\tau )=[g_{\alpha }(\tau )=\eta
_{\alpha }(\tau )\mathring{g}_{\alpha },N_{i}^{a}(\tau )=\eta _{i}^{a}\
(\tau )\mathring{N}_{i}^{a}].  \label{offdiagdefr}
\end{equation}%
Such transforms are defined by some values $\eta _{\alpha
}(\tau,x^{k},y^{3}) $ and $\eta _{i}^{a}(\tau ,x^{k},y^{3})$ called $\tau $%
-running gravitational polarization ($\eta $-polarization) functions.

Target d-metrics $\mathbf{g}(\tau )$ can be constructed as quasi-stationary
solutions of type (\ref{qeltors}), or (\ref{offdiagcosmcsh}), when the
nonholonomic deformations are determined by respective generating functions,
generating sources and effective cosmological constants, 
\begin{eqnarray*}
(\Psi (\tau ),\ \ _{2}^{q}\mathbf{\Upsilon }(\tau )) &\leftrightarrow &(%
\mathbf{g}(\tau ),\ _{2}^{q}\mathbf{\Upsilon }(\tau ))\leftrightarrow (\eta
_{\alpha }(\tau )\ \mathring{g}_{\alpha }\sim (\zeta _{\alpha }(\tau
)(1+\varepsilon \chi _{\alpha }(\tau ))\mathring{g}_{\alpha },\ \ _{2}^{q}%
\mathbf{\Upsilon }(\tau ))\leftrightarrow \\
(\Phi (\tau ),\ _{2}^{q}\Lambda (\tau )) &\leftrightarrow &(\mathbf{g}(\tau
),\ _{2}^{q}\Lambda (\tau ))\leftrightarrow (\eta _{\alpha }(\tau )\ 
\mathring{g}_{\alpha }\sim (\zeta _{\alpha }(\tau )(1+\varepsilon \chi
_{\alpha }(\tau ))\mathring{g}_{\alpha },\ _{2}^{q}\Lambda (\tau )),
\end{eqnarray*}%
where $\ _{2}^{q}\Lambda (\tau )$ is an effective cosmological constant in
the v-subspace, $\varepsilon $ is a small parameter $0\leq \varepsilon <1,$
with some $\zeta _{\alpha }(\tau ,x^{k},y^{3})$ and $\chi _{\alpha }(\tau
,x^{k},y^{3}).$

For families of $\eta $- and/or $\chi $-polarizations, the nonlinear
symmetries (\ref{ntransf2}) are defined in the form: 
\begin{eqnarray}
\partial _{3}[\Psi ^{2}(\tau )] &=&-\int dy^{3}\ _{2}^{q}\mathbf{\Upsilon }%
(\tau )\partial _{3}h_{4}(\tau )\simeq -\int dy^{3}\ _{2}^{q}\mathbf{%
\Upsilon }(\tau )\partial _{3}(\eta _{4}(\tau )\ \mathring{g}_{4})  \notag \\
&\simeq &-\int dy^{3}\ _{2}^{q}\mathbf{\Upsilon }(\tau )\partial _{3}[\zeta
_{4}(\tau )(1+\varepsilon \ \chi _{4}(\tau ))\ \mathring{g}_{4}],  \notag \\
\Phi ^{2}(\tau ) &=&-4\ _{2}^{q}\Lambda (\tau )h_{4}(\tau )\simeq -4\
_{2}^{q}\Lambda (\tau )\eta _{4}(\tau )\mathring{g}_{4}\simeq -4\
_{2}^{q}\Lambda (\tau )\ \zeta _{4}(\tau )(1+\varepsilon \chi _{4}(\tau ))\ 
\mathring{g}_{4}.  \label{nonlinsymrex}
\end{eqnarray}

Generating functions for families of off-diagonal $\eta $-transforms of type
(\ref{offdiagdefr}) can be parameterized for $\eta $-polarizations, 
\begin{equation}
\psi (\tau )\simeq \psi (\kappa ;\tau ,x^{k}),\eta _{4}\ (\tau )\simeq \eta
_{4}(\tau ,x^{k},y^{3}).  \label{etapolgen}
\end{equation}%
Such generating functions can be used for defining $\tau $-families of
quasi-stationary nonlinear quadratic elements of type (\ref%
{offdsolgenfgcosmc}), 
\begin{eqnarray}
d\widehat{s}^{2}(\tau ) &=&\widehat{g}_{\alpha \beta }(\tau ,x^{k},y^{3};%
\mathring{g}_{\alpha };\psi (\tau ),\eta _{4}(\tau );\ _{2}^{q}\mathbf{%
\Upsilon }(\tau ))du^{\alpha }du^{\beta }=e^{\psi (\tau
)}[(dx^{1})^{2}+(dx^{2})^{2}]  \label{offdiagpolfr} \\
&&-\frac{[\partial _{3}(\eta _{4}(\tau )\ \mathring{g}_{4})]^{2}}{|\int
dy^{3}\ _{2}^{q}\mathbf{\Upsilon }(\tau )\partial _{3}(\eta _{4}(\tau )\ 
\mathring{g}_{4})|\ \eta _{4}(\tau )\mathring{g}_{4}}\{dy^{3}+\frac{\partial
_{i}[\int dy^{3}\ _{2}^{q}\mathbf{\Upsilon }(\tau )\ \partial _{3}(\eta
_{4}(\tau )\mathring{g}_{4})]}{\ _{2}^{q}\mathbf{\Upsilon }(\tau )\partial
_{3}(\eta _{4}(\tau )\mathring{g}_{4})}dx^{i}\}^{2}  \notag \\
&&+\eta _{4}(\tau )\mathring{g}_{4}\{dt+[\ _{1}n_{k}(\tau )+\ _{2}n_{k}(\tau
)\int dy^{3}\frac{[\partial _{3}(\eta _{4}(\tau )\mathring{g}_{4})]^{2}}{%
|\int dy^{3}\ \ _{2}^{q}\mathbf{\Upsilon }(\tau )\partial _{3}(\eta
_{4}(\tau )\mathring{g}_{4})|\ (\eta _{4}(\tau )\mathring{g}_{4})^{5/2}}%
]dx^{k}\}^{2}.  \notag
\end{eqnarray}%
For $\Phi ^{2}(\tau )=-4\ _{2}^{q}\Lambda h_{4}(\tau ),$ we can transform (%
\ref{offdiagcosmcsh}) in a variant of (\ref{offdiagpolfr}) with $\eta $%
-polarizations determined by the generating data $(h_{4}(\tau );\
_{2}\Lambda ,\ (\tau )).$ For general $\eta $- and $Q$-deformations, it is
difficult to understand if such off-diagonal metrics have certain physically
important interpretation even the primary data possess certain important
physical meaning. Nevertheless, even in such cases certain quasi-periodic,
solitonic, or another type structures can be generated. It is possible
always to compute respective Perelman thermodynamic variables and solve the
issue if certain configurations with a prescribed $\ _{2}^{q}\mathbf{%
\Upsilon }(\tau ),$ or $\ _{2}^{q}\Lambda (\tau ),$ can be thermodynamically
more optimal than other ones with some prescribed effective $\
_{2}^{m}\Lambda (\tau )+\ _{2}^{F}\Lambda (\tau )+\ _{2}^{e}\Lambda (\tau ),$
see formulas (\ref{additcosm}).

\subsubsection{Generating solutions with small parametric decompositions
encoding nonmetricity}

For a small parameter $\varepsilon ,$ we can construct $\varepsilon $-linear
functions for $\eta $-polarizations in (\ref{offdiagpolfr}) and study small
nonholonomic deformations of a prime d-metric $\mathbf{\mathring{g}}$ into
so-called $\varepsilon $-parametric $\tau $-families of solutions with $%
\zeta $- and $\chi $-coefficients. Corresponding parametric decompositions
are of type 
\begin{eqnarray}
\psi (\tau ) &\simeq &\psi (\tau ,x^{k})\simeq \psi _{0}(\tau
,x^{k})(1+\varepsilon \ _{\psi }\chi (\tau ,x^{k})),\mbox{ for }\ 
\label{epsilongenfdecomp} \\
\ \eta _{2}(\tau ) &\simeq &\eta _{2}(\tau ,x^{k_{1}})\simeq \zeta _{2}(\tau
,x^{k})(1+\varepsilon \chi _{2}(\tau ,x^{k})),\mbox{ we can consider }\ \eta
_{2}(\tau )=\ \eta _{1}(\tau );  \notag \\
\eta _{4}(\tau ) &\simeq &\eta _{4}(\tau ,x^{k},y^{3})\simeq \zeta _{4}(\tau
,x^{k},y^{3})(1+\varepsilon \chi _{4}(\tau ,x^{k},y^{3})),  \notag
\end{eqnarray}%
where $\psi (\tau )$ and $\eta _{2}(\tau )=\ \eta _{1}(\tau )$ are such way
chosen to be determined by solutions of the 2-d Poisson equation $\partial
_{11}^{2}\psi (\tau )+\partial _{22}^{2}\psi (\tau )= 2\ \ _{1}^{q}\mathbf{%
\Upsilon }(\tau ,x^{k}),$ see (\ref{eq1}). For other type signatures of
d-metrics, it can be a 2-d wave equation with respective source and certain $%
\tau $-evolving scenarios.

Parameterizations of type (\ref{epsilongenfdecomp}) allow to compute $%
\varepsilon $-parametric deformations to $\tau $-families of
quasi-stationary d-metrics with $\chi $-generating functions, 
\begin{equation*}
d\ \widehat{s}^{2}(\tau )=\widehat{g}_{\alpha \beta }(\tau ,x^{k},y^{3};\psi
(\tau ),g_{4}(\tau );\ _{2}^{q}\mathbf{\Upsilon }(\tau ))du^{\alpha
}du^{\beta }=e^{\psi _{0}}(1+\varepsilon \ ^{\psi }\chi (\tau
))[(dx^{1})^{2}+(dx^{2})^{2}]
\end{equation*}%
\begin{eqnarray*}
&&-\{\frac{4[\partial _{3}(|\zeta _{4}(\tau )\mathring{g}_{4}|^{1/2})]^{2}}{%
\mathring{g}_{3}|\int dy^{3}\{\ _{2}^{q}\mathbf{\Upsilon }(\tau )\partial
_{3}(\zeta _{4}(\tau )\mathring{g}_{4})\}|}-\varepsilon \lbrack \frac{%
\partial _{3}(\chi _{4}(\tau )|\zeta _{4}(\tau )\mathring{g}_{4}|^{1/2})}{%
4\partial _{3}(|\zeta _{4}(\tau )\mathring{g}_{4}|^{1/2})}-\frac{\int
dy^{3}\{\ _{2}^{q}\mathbf{\Upsilon }\partial _{3}[(\zeta _{4}(\tau )%
\mathring{g}_{4})\chi _{4}(\tau )]\}}{\int dy^{3}\{\ _{2}^{q}\mathbf{%
\Upsilon }(\tau )\partial _{3}(\zeta _{4}(\tau )\mathring{g}_{4})\}}]\}%
\mathring{g}_{3} \\
&&\{dy^{3}+[\frac{\partial _{i}\ \int dy^{3}\ _{2}^{q}\mathbf{\Upsilon }%
(\tau )\ \partial _{3}\zeta _{4}(\tau )}{(\mathring{N}_{i}^{3})\ \ _{2}^{q}%
\mathbf{\Upsilon }(\tau )\partial _{3}\zeta _{4}(\tau )}+\varepsilon (\frac{%
\partial _{i}[\int dy^{3}\ \ _{2}^{q}\mathbf{\Upsilon }(\tau )\ \partial
_{3}(\zeta _{4}(\tau )\chi _{4}(\tau ))]}{\partial _{i}\ [\int dy^{3}\ \
_{2}^{q}\mathbf{\Upsilon }(\tau )\partial _{3}\zeta _{4}(\tau )]}-\frac{%
\partial _{3}(\zeta _{4}(\tau )\chi _{4}(\tau ))}{\partial _{3}\zeta
_{4}(\tau )})]\mathring{N}_{i}^{3}dx^{i}\}^{2}
\end{eqnarray*}%
\begin{eqnarray}
&&+\zeta _{4}(\tau )(1+\varepsilon \ \chi _{4}(\tau ))\ \mathring{g}%
_{4}\{dt+[(\mathring{N}_{k}^{4})^{-1}[\ _{1}n_{k}(\tau )+16\ _{2}n_{k}(\tau
)[\int dy^{3}\frac{\left( \partial _{3}[(\zeta _{4}(\tau )\mathring{g}%
_{4})^{-1/4}]\right) ^{2}}{|\int dy^{3}\partial _{3}[\ \ _{2}^{q}\mathbf{%
\Upsilon }(\tau )(\zeta _{4}(\tau )\mathring{g}_{4})]|}]
\label{offdncelepsilon} \\
&&+\varepsilon \frac{16\ _{2}n_{k}(\tau )\int dy^{3}\frac{\left( \partial
_{3}[(\zeta _{4}(\tau )\mathring{g}_{4})^{-1/4}]\right) ^{2}}{|\int
dy^{3}\partial _{3}[\ \ _{2}^{q}\mathbf{\Upsilon }(\tau )(\zeta _{4}(\tau )%
\mathring{g}_{4})]|}(\frac{\partial _{3}[(\zeta _{4}(\tau )\mathring{g}%
_{4})^{-1/4}\chi _{4})]}{2\partial _{3}[(\zeta _{4}(\tau )\mathring{g}%
_{4})^{-1/4}]}+\frac{\int dy^{3}\partial _{3}[\ \ _{2}^{q}\mathbf{\Upsilon }%
(\tau )(\zeta _{4}(\tau )\chi _{4}(\tau )\mathring{g}_{4})]}{\int
dy^{3}\partial _{3}[\ _{2}^{q}\mathbf{\Upsilon }(\tau )(\zeta _{4}(\tau )%
\mathring{g}_{4})]})}{\ _{1}n_{k}(\tau )+16\ _{2}n_{k}(\tau )[\int dy^{3}%
\frac{\left( \partial _{3}[(\zeta _{4}(\tau )\mathring{g}_{4})^{-1/4}]%
\right) ^{2}}{|\int dy^{3}\partial _{3}[\ \ _{2}^{q}\mathbf{\Upsilon }(\tau
)(\zeta _{4}(\tau )\mathring{g}_{4})]|}]}]\mathring{N}_{k}^{4}dx^{k}\}^{2}. 
\notag
\end{eqnarray}%
Such off-diagonal parametric solutions allow us to define, for instance,
ellipsoidal deformations of BH metrics into BE ones and to provide realistic
interpretation of nonmetric deformations under geometric flows or for
off-diagonal modifications. Quasi-stationary d-metrics of type (\ref%
{offdncelepsilon}) can be generated for by certain small parametric
deformations with generating data $(\Phi (\tau ),\ _{2}^{q}\Lambda (\tau )).$

\subsubsection{Extracting Levi-Civita configurations}

The generic off--diagonal quasi-stationary solutions considered in previous
subsections were constructed for canonical d--connections $\widehat{\mathbf{D%
}}(\tau ).$ In general, such solutions are characterized by nonholonomically
induced d--torsion coefficients $\ \widehat{\mathbf{T}}_{\ \alpha \beta
}^{\gamma}(\tau)$ (such values are completely defined by the N--connection
and d--metric structures) and contain $Q$-deformations related to nontrivial
d-torsions $\mathbf{T}_{\mu \nu \alpha }(\tau )=\mathbf{A}_{\nu }(\tau )%
\mathbf{g}_{\mu \alpha}(\tau ) - \mathbf{A}_{\alpha }(\tau )\mathbf{g}_{\mu
\nu }(\tau )$, see formulas (\ref{qsourc}). We can extract zero torsion
LC-configurations for $q$-distortions of $\nabla (\tau )$ if we impose
additionally the conditions (\ref{lccond1}). By straightforward computations
for quasi-stationary configurations, we can verify that all canonical
d-torsion coefficients $\widehat{\mathbf{T}}_{\ \alpha \beta }^{\gamma
}(\tau )$ vanish if the coefficients of N--adapted frames and $v$%
--components of $\tau $-families of d--metrics are subjected to respective
conditions, 
\begin{eqnarray}
\ w_{i}^{\ast }(\tau ) &=&\mathbf{e}_{i}(\tau )\ln \sqrt{|\ h_{3}(\tau )|},%
\mathbf{e}_{i}(\tau )\ln \sqrt{|\ h_{4}(\tau )|}=0,\partial _{i}w_{j}(\tau
)=\partial _{j}w_{i}(\tau )\mbox{ and }n_{i}^{\ast }(\tau )=0;  \notag \\
n_{k}(\tau ,x^{i}) &=&0\mbox{ and }\partial _{i}n_{j}(\tau ,x^{k})=\partial
_{j}n_{i}(\tau ,x^{k}).  \label{zerot1}
\end{eqnarray}%
The solutions for necessary type of $w$- and $n$-functions depend on the
class of vacuum, non--vacuum, $Q$-deformed and other type metrics which we
attempt to generate. We may follow such steps for finding solutions
subjected to conditions (\ref{zerot1}):

Prescribing a generating function $\Psi (\tau )=\check{\Psi}%
(\tau,x^{i_{1}},y^{3}),$ for which $[\partial _{i}(\ _{2}\check{\Psi}%
)]^{\ast }=\partial _{i}(\ _{2}\check{\Psi})^{\ast },$ we solve the
equations for $w_{j}$ from (\ref{zerot1}) in explicit form if $\ _{2}^{q}%
\mathbf{\Upsilon }=const,$ or if such an effective source can be expressed
as a functional $\ _{2}^{q}\mathbf{\Upsilon }(\tau ,x^{i},y^{3})= \ _{2}^{q}%
\mathbf{\Upsilon }[\ _{2}\check{\Psi}(\tau )].$ The conditions $\partial
_{i}w_{j}(\tau )=\partial _{j}w_{i}(\tau ),$ are solved by any generating
function $\check{A}=\check{A}(\tau ,x^{k},y^{3})$ for which 
\begin{equation*}
w_{i}(\tau )=\check{w}_{i}(\tau )=\partial _{i}\ \check{\Psi}(\tau )/(\check{%
\Psi}(\tau ))^{\ast }=\partial _{i}\check{A}(\tau ).
\end{equation*}%
The equations for $n$-functions in (\ref{zerot1}) are solved for any $%
n_{i}(\tau )=\partial _{i}[\ ^{2}n(\tau ,x^{k})].$

Putting together above formulas for respective classes of generating
functions, we construct a nonlinear quadratic element for quasi-stationary
solutions with zero canonical d-torsions, (\ref{qeltors}), 
\begin{eqnarray}
d\check{s}^{2}(\tau ) &=&\check{g}_{\alpha \beta }(\tau
,x^{k},y^{3})du^{\alpha }du^{\beta }=e^{\psi (\tau
,x^{k})}[(dx^{1})^{2}+(dx^{2})^{2}]  \label{qellc} \\
&&+\frac{[\check{\Psi}^{\ast }(\tau )]^{2}}{4(\ \ \ _{2}^{q}\mathbf{\Upsilon 
}(\tau )[\check{\Psi}(\tau )])^{2}\{h_{4}^{[0]}(\tau )-\int dy^{3}[\check{%
\Psi}(\tau )]^{\ast }/4\ _{2}^{q}\mathbf{\Upsilon }(\tau )[\check{\Psi}(\tau
)]\}}\{dy^{3}+[\partial _{i}(\check{A}(\tau ))]dx^{i}\}^{2}  \notag \\
&&+\{h_{4}^{[0]}(\tau )-\int dy^{3}\frac{[\check{\Psi}^{2}(\tau )]^{\ast }}{%
4(\ _{2}^{q}\mathbf{\Upsilon }(\tau )[\check{\Psi}(\tau )])}\}\{dt+\partial
_{i}[\ ^{2}n(\tau ,x^{k})]dx^{i}\}^{2}.  \notag
\end{eqnarray}%
Finally, we note that d-metrics (\ref{qellc}) define LC-configurations for $%
\nabla (\tau )$ that involve also nonmetricity contributions encoded into $%
_{2}^{q}\mathbf{\Upsilon }(\tau ).$ This is an example when using nonlinear
symmetries we encode nonmetricity data into, in general, generic
off-diagonal pseudo-Riemannian metric for an effective Einstein gravity with
"exotic" effective energy-momentum sources.

\setcounter{equation}{0} \renewcommand{\theequation}
{B.\arabic{equation}} \setcounter{subsection}{0} 
\renewcommand{\thesubsection}
{B.\arabic{subsection}}

\section{Generating nonmetric solitonic hierarchies via solitonic metrics
and effective sources}

\label{appendixb} Let us consider the formulas for nonmetric geometric
evolution of a d-metric $\mathbf{g}(\tau )$ constructed as a solution of (%
\ref{cdeq1}): We associate a non--stretching curve $\gamma (\tau ,\mathbf{l}%
) $ on a nonholonomic Lorentz manifold $\mathbf{V}$ and use $\tau $ both as
a curve running real parameter and a geometric flow parameter. The value $%
\mathbf{l}$ is the arclength of a curve on $\mathbf{V}$ which is defined by
an evolution d--vector $\mathbf{Y}=\varsigma _{\tau }$ and tangent d--vector 
$\mathbf{X}=\varsigma _{\mathbf{l}}$ that $\mathbf{g(X,X)=}1$. Any curve $%
\varsigma (\tau ,\mathbf{l})$ defines a two--dimensional surface in $%
T_{\varsigma (\tau ,\mathbf{l})}\mathbf{V}\subset T\mathbf{V.}$ In  \cite%
{vacaru15}, there are given details on metric
compatible curve flows. In this work, the approach is generalized for
nonmetric deformations. To any dual basis (\ref{nadif}) a coframe $\mathbf{e}%
\in T_{\varsigma }^{\ast }\mathbf{V}_{\mathbf{N}}\otimes (h\mathfrak{p\oplus 
}v\mathfrak{p})$ can be associated. It is a N--adapted $\left( SO(n)%
\mathfrak{\oplus }SO(m)\right) $--parallel basis along $\varsigma .$

\subsection{Preliminaries on geometric models and solitons}

\label{ssb1}We can associate a canonical d-connection $\widehat{\mathbf{D}}$
(\ref{cdist}) with a linear connection 1--form parameterized as $\widehat{%
\mathbf{\Gamma }}\in T_{\varsigma }^{\ast }\mathbf{V}_{\mathbf{N}}\otimes (%
\mathfrak{so}(n)\mathfrak{\oplus so}(m)).$ The frame bases are 1-forms $%
\mathbf{e}_{\mathbf{X}}=\mathbf{e}_{h\mathbf{X}}+\mathbf{e}_{v\mathbf{X}}$
defined by N-adapted frames (\ref{nader}), which (for $(1,\overrightarrow{0}%
)\in \mathbb{R}^{n},\overrightarrow{0}\in \mathbb{R}^{n-1}$ and $(1,%
\overleftarrow{0})\in \mathbb{R}^{m}, \overleftarrow{0}\in \mathbb{R}%
^{m-1}), $ can be parameterized by respective matrices, 
\begin{equation*}
\mathbf{e}_{h\mathbf{X}}=\varsigma _{h\mathbf{X}}\rfloor h\mathbf{e=}\left[ 
\begin{array}{cc}
0 & (1,\overrightarrow{0}) \\ 
-(1,\overrightarrow{0})^{T} & h\mathbf{0}%
\end{array}%
\right] ,\mathbf{e}_{v\mathbf{X}}=\varsigma _{v\mathbf{X}}\rfloor v\mathbf{e=%
}\left[ 
\begin{array}{cc}
0 & (1,\overleftarrow{0}) \\ 
-(1,\overleftarrow{0})^{T} & v\mathbf{0}%
\end{array}%
\right] .
\end{equation*}%
Such d-operators act on the spaces of curves on $\mathbf{V}$.

For a N-connection (\ref{ncon}), we can construct a corresponding $h$- and $%
v $-splitting of canonical d-connections 1-forms, $\widehat{\mathbf{\Gamma }}%
=\left[ \widehat{\mathbf{\Gamma}}_{h\mathbf{X}},\widehat{\mathbf{\Gamma }}_{v%
\mathbf{X}}\right] ,$ parameterizing 
\begin{eqnarray*}
\widehat{\mathbf{\Gamma }}_{h\mathbf{X}} &=&\varsigma _{h\mathbf{X}}\rfloor 
\widehat{\mathbf{L}}=\left[ 
\begin{array}{cc}
0 & (0,\overrightarrow{0}) \\ 
-(0,\overrightarrow{0})^{T} & \widehat{\mathbf{L}}%
\end{array}%
\right] \in \mathfrak{so}(n+1),\mbox{ where } \\
\widehat{\mathbf{L}} &=&\left[ 
\begin{array}{cc}
0 & \overrightarrow{v} \\ 
-\overrightarrow{v}^{T} & h\mathbf{0}%
\end{array}%
\right] \in \mathfrak{so}(n),~\overrightarrow{v}\in \mathbb{R}^{n-1},~h%
\mathbf{0\in }\mathfrak{so}(n-1),\mbox{ and } \\
\widehat{\mathbf{\Gamma }}_{v\mathbf{X}} &=&\varsigma _{v\mathbf{X}}\rfloor 
\widehat{\mathbf{C}}=\left[ 
\begin{array}{cc}
0 & (0,\overleftarrow{0}) \\ 
-(0,\overleftarrow{0})^{T} & \widehat{\mathbf{C}}%
\end{array}%
\right] \in \mathfrak{so}(m+1),\mbox{ where } \\
\widehat{\mathbf{C}} &=&\left[ 
\begin{array}{cc}
0 & \overleftarrow{v} \\ 
-\overleftarrow{v}^{T} & v\mathbf{0}%
\end{array}%
\right] \in \mathfrak{so}(m),~\overleftarrow{v}\in \mathbb{R}^{m-1},~v%
\mathbf{0\in }\mathfrak{so}(m-1).
\end{eqnarray*}

Using the canonical d--connection $\widehat{\mathbf{D}},$ we can define some
d-matrices being decomposed with respect to the flow direction:\ in the
h--direction, 
\begin{eqnarray*}
\mathbf{e}_{h\mathbf{Y}} &=&\varsigma _{\tau }\rfloor h\mathbf{e}=\left[ 
\begin{array}{cc}
0 & \left( h\mathbf{e}_{\parallel },h\overrightarrow{\mathbf{e}}_{\perp
}\right) \\ 
-\left( h\mathbf{e}_{\parallel },h\overrightarrow{\mathbf{e}}_{\perp
}\right) ^{T} & h\mathbf{0}%
\end{array}%
\right] ,\mbox{ when } \\
\mathbf{e}_{h\mathbf{Y}} &\in &h\mathfrak{p,}\left( h\mathbf{e}_{\parallel
},h\overrightarrow{\mathbf{e}}_{\perp }\right) \in \mathbb{R}^{n}\mbox{ and }%
h\overrightarrow{\mathbf{e}}_{\perp }\in \mathbb{R}^{n-1},
\end{eqnarray*}%
and 
\begin{eqnarray*}
\widehat{\mathbf{\Gamma }}_{h\mathbf{Y}} &=&\varsigma _{h\mathbf{Y}}\rfloor 
\widehat{\mathbf{L}} =\left[%
\begin{array}{cc}
0 & (0,\overrightarrow{0}) \\ 
-(0,\overrightarrow{0})^{T} & h\mathbf{\varpi }_{\tau }%
\end{array}%
\right] \in \mathfrak{so}(n+1),\mbox{ where } \\
\ h\mathbf{\varpi }_{\tau } &=&\left[ 
\begin{array}{cc}
0 & \overrightarrow{\varpi } \\ 
-\overrightarrow{\varpi }^{T} & h\widehat{\mathbf{\Theta }}%
\end{array}%
\right] \in \mathfrak{so}(n),~\overrightarrow{\varpi }\in \mathbb{R}^{n-1},~h%
\widehat{\mathbf{\Theta }}\mathbf{\in }\mathfrak{so}(n-1).
\end{eqnarray*}
Then, we introduce similar parameterizations for the v--direction, 
\begin{eqnarray*}
\mathbf{e}_{v\mathbf{Y}} &=&\varsigma _{\tau }\rfloor v\mathbf{e}=\left[ 
\begin{array}{cc}
0 & \left( v\mathbf{e}_{\parallel },v\overleftarrow{\mathbf{e}}_{\perp
}\right) \\ 
-\left( v\mathbf{e}_{\parallel },v\overleftarrow{\mathbf{e}}_{\perp }\right)
^{T} & v\mathbf{0}%
\end{array}%
\right] ,\mbox{ when } \\
\mathbf{e}_{v\mathbf{Y}} &\in &v\mathfrak{p,}\left( v\mathbf{e}_{\parallel
},v\overleftarrow{\mathbf{e}}_{\perp }\right) \in \mathbb{R}^{m}\mbox{ and }v%
\overleftarrow{\mathbf{e}}_{\perp }\in \mathbb{R}^{m-1};
\end{eqnarray*}
and 
\begin{eqnarray*}
\widehat{{\mathbf{\Gamma }}}_{v\mathbf{Y}} &\mathbf{=}&\varsigma _{v\mathbf{Y%
}}\rfloor \widehat{\mathbf{C}}\mathbf{=}\left[ 
\begin{array}{cc}
0 & (0,\overleftarrow{0}) \\ 
-(0,\overleftarrow{0})^{T} & v\widehat{\mathbf{\varpi }}_{\tau }%
\end{array}%
\right] \in \mathfrak{so}(m+1), \\
v\mathbf{\varpi }_{\tau } &\mathbf{=}&\left[ 
\begin{array}{cc}
0 & \overleftarrow{\varpi } \\ 
-\overleftarrow{\varpi }^{T} & v\widehat{\mathbf{\Theta }}%
\end{array}%
\right] \in \mathfrak{so}(m),~\overleftarrow{\varpi }\in \mathbb{R}^{m-1},~v%
\widehat{\mathbf{\Theta }}\mathbf{\in }\mathfrak{so}(m-1).
\end{eqnarray*}

We summarize and adapt for nonmetric geometric  flows and gravity three important results proven in \cite{vacaru15} for
parameterizations related to nonholonomic canonical geometric flows of 4-d Lorentzian metrics:

For any solution of N-adapted Hamilton-Friedan equations in canonical
variables (\ref{ricciflowr2}), or for relativistic nonholonomic Ricci
soliton equations (\ref{cdeq1}), there is a canonical hierarchy of
N--adapted flows of curves $\varsigma (\tau ,\mathbf{l})=h\varsigma (\tau ,%
\mathbf{l})+ v\varsigma (\tau ,\mathbf{l})$ described by nonholonomic
geometric map equations encoding nonmetric sources:

\begin{itemize}
\item The $0$ flows are convective (travelling wave) maps $\varsigma _{\tau
}=\varsigma _{\mathbf{l}}$ distinguished as $\left( h\varsigma \right)
_{\tau }=\left( h\varsigma \right) _{h\mathbf{X}}$ and $\left( v\varsigma
\right) _{\tau }=\left( v\varsigma \right) _{v\mathbf{X}}$. The
classification of such maps depend on the type of d-connection structure.

\item There are +1 flows defined as non--stretching mKdV maps 
\begin{eqnarray}
-\left( h\varsigma \right) _{\tau }&=&\widehat{\mathbf{D}}_{h\mathbf{X}%
}^{2}\left( h\varsigma \right) _{h\mathbf{X}} +\frac{3}{2}|\widehat{\mathbf{D%
}}_{h\mathbf{X}}\left( h\varsigma \right) _{h\mathbf{X}}|_{h\mathbf{g}}^{2}
\left( h\varsigma \right) _{h\mathbf{X}},  \notag \\
-\left( v\varsigma \right) _{\tau }&=&\widehat{\mathbf{D}}_{v\mathbf{X}%
}^{2}\left( v\varsigma \right) _{v\mathbf{X}}+\frac{3}{2}|\widehat{\mathbf{D}%
}_{v\mathbf{X}}\left( v\varsigma \right) _{v\mathbf{X}}|_{v\mathbf{g}%
}^{2}~\left( v\varsigma \right) _{v\mathbf{X}},  \label{mkdv}
\end{eqnarray}%
and the +2,... flows as higher order analogs.

\item Finally, the -1 flows are defined by the kernels of the canonical
recursion h--operator, 
\begin{equation*}
h\widehat{{\mathfrak{R}}}=\widehat{\mathbf{D}}_{h\mathbf{X}}\left( \widehat{%
\mathbf{D}}_{h\mathbf{X}}+\widehat{\mathbf{D}}_{h\mathbf{X}}^{-1}\left( 
\overrightarrow{v}\cdot \right) \overrightarrow{v}\right) +\overrightarrow{v}%
\rfloor \widehat{\mathbf{D}}_{h\mathbf{X}}^{-1}\left( \overrightarrow{v}%
\wedge \widehat{\mathbf{D}}_{h\mathbf{X}}\right) ,
\end{equation*}%
and of the canonical recursion v--operator, 
\begin{equation*}
v\widehat{{\mathfrak{R}}}=\widehat{\mathbf{D}}_{v\mathbf{X}}\left( \widehat{%
\mathbf{D}}_{v\mathbf{X}}+\widehat{\mathbf{D}}_{v\mathbf{X}}^{-1}\left( 
\overleftarrow{v}\cdot \right) \overleftarrow{v}\right) +\overleftarrow{v}%
\rfloor \widehat{\mathbf{D}}_{v\mathbf{X}}^{-1}\left( \overleftarrow{v}%
\wedge \widehat{\mathbf{D}}_{v\mathbf{X}}\right) ,
\end{equation*}
inducing non--stretching maps $\widehat{\mathbf{D}}_{h\mathbf{Y}%
}\left(h\varsigma \right) _{h\mathbf{X}}=0$ and $\widehat{\mathbf{D}}_{v%
\mathbf{Y}}\left( v\varsigma \right) _{v\mathbf{X}}=0$.
\end{itemize}

The canonical recursion d-operator $\widehat{{\mathfrak{R}}}=(h\widehat{{%
\mathfrak{R}}},v\widehat{{\mathfrak{R}}})$ is related to respective
bi-Hamiltonian structures for curve flows (in our case, determined by
geometric flows and respective solitonic models).

\subsection{Examples of solitonic space like stationary distributions and
nonlinear waves}

To generate quasi-stationary solutions of nonmetric geometric flow and
gravity equations we can consider $\tau $-running of fixed $\tau _{0}$ space
distributions which can anisotropic on certain angular type coordinates $%
(\vartheta ,\varphi ).$

\subsubsection{Quasi-stationary solitonic distributions}

We shall use distributions $\ \wp =\wp (r,\vartheta ,\varphi )$ as solutions
of a respective six classes of solitonic 3-d equations 
\begin{eqnarray}
\partial _{rr}^{2}\wp +\epsilon \partial _{\varphi }(\partial _{\vartheta
}\wp +6\wp \partial _{\varphi }\wp +\partial _{\varphi \varphi \varphi
}^{3}\wp ) &=&0,  \label{solitdistr} \\
\ \partial _{rr}^{2}\wp +\epsilon \partial _{\vartheta }(\partial _{\varphi
}\wp +\wp \partial _{\vartheta }\wp +\partial _{\vartheta \vartheta
\vartheta }^{3}\wp ) &=&0,  \notag \\
\partial _{\vartheta \vartheta }^{2}\wp +\epsilon \partial _{\varphi
}(\partial _{r}\wp +6\wp \partial _{\varphi }\wp +\partial _{\varphi \varphi
\varphi }^{3}\wp ) &=&0,  \notag \\
\partial _{\vartheta \vartheta }^{2}\wp +\epsilon \partial _{r}(\partial
_{\varphi }\wp +6\wp \partial _{r}\wp +\partial _{rrr}^{3}\wp ) &=&0,  \notag
\\
\partial _{\varphi \varphi }^{2}\wp +\epsilon \partial _{r}(\partial
_{\vartheta }\wp +6\wp \partial _{r}\wp +\partial _{rrr}^{3}\wp ) &=&0, 
\notag \\
\ \partial _{\varphi \varphi }^{2}\wp +\epsilon \partial _{\vartheta
}(\partial _{r}\wp +6\wp \partial _{\vartheta }\wp +\partial _{\vartheta
\vartheta \vartheta }^{3}\wp ) &=&0,  \notag
\end{eqnarray}%
for $\epsilon =\pm 1$. To construct in explicit form solutions of such
nonlinear PDEs is a very difficult task. Nevertheless, their physical
properties are well known from the theory of solitonic hierarchies. We can
take any $\wp (u)$ as a parametric, or exact solution of an equation (\ref%
{solitdistr}) and consider as a generating function and/or generating source
which does not depend on the time coordinate. These equations and their
solutions can be redefined via frame/coordinate transforms for stationary
generating functions parameterized in non-spherical coordinates and labeled
in the form $\wp =\wp (x^{i},y^{3})$. We can use such functions as
generating functions and/or generating sources for nonmetric Ricci solitons (%
\ref{nonheinst}) when the constructions can be extended for quasi-stationary
geometric flows.

\subsubsection{Generating nonlinear solitonic waves on temperature like
parameter}

Stationary geometric flow evolution on a Lorentz manifold can be
characterized by 3-d solitonic waves with explicit dependence flow parameter 
$\tau $ defined by functions $\wp (\tau ,u)$ as solutions of such nonlinear
PDEs: 
\begin{equation}
\ \ \wp =\left\{ 
\begin{array}{ccc}
\wp (\tau ,\vartheta ,\varphi ) & \mbox{ as a solution of } & \partial
_{\tau \tau }^{2}\wp +\epsilon \frac{\partial }{\partial \varphi }[\partial
_{\vartheta }\wp +6\wp \frac{\partial }{\partial \varphi }\wp +\frac{%
\partial ^{3}}{(\partial \varphi )^{3}}\wp ]=0; \\ 
\wp (\vartheta ,\tau ,\varphi ) & \mbox{ as a solution of } & \partial
_{\vartheta \vartheta }^{2}\wp +\epsilon \frac{\partial }{\partial \varphi }%
[\partial _{t}\wp +6\wp \frac{\partial }{\partial \varphi }\wp +\frac{%
\partial ^{3}}{(\partial \varphi )^{3}}\wp ]=0; \\ 
\wp (\tau ,r,\varphi ) & \mbox{ as a solution of } & \partial _{\tau \tau
}^{2}\wp +\epsilon \frac{\partial }{\partial \varphi }[\partial _{r}\wp
+6\wp \frac{\partial }{\partial \varphi }\wp +\frac{\partial ^{3}}{(\partial
\varphi )^{3}}\wp ]=0; \\ 
\wp (r,\tau ,\varphi ) & \mbox{ as a solution of } & \partial _{rr}^{2}\wp
+\epsilon \frac{\partial }{\partial \varphi }[\partial _{\tau }\wp +6\wp 
\frac{\partial }{\partial \varphi }\wp +\frac{\partial ^{3}}{(\partial
\varphi )^{3}}\wp ]=0; \\ 
\wp (\tau ,\varphi ,\vartheta ) & \mbox{ as a solution of } & \partial
_{\tau \tau }^{2}\wp +\epsilon \frac{\partial }{\partial \vartheta }%
[\partial _{\varphi }\wp +6\wp \frac{\partial }{\partial \vartheta }\wp +%
\frac{\partial ^{3}}{(\partial \vartheta )^{3}}\wp ]=0; \\ 
\ \wp (\varphi ,\tau ,\vartheta ) & \mbox{ as a solution of } & \partial
_{\varphi \varphi }^{2}\wp +\epsilon \frac{\partial }{\partial \vartheta }%
[\partial _{\tau }\wp +6\iota \frac{\partial }{\partial \vartheta }\wp +%
\frac{\partial ^{3}}{(\partial \vartheta )^{3}}\ \wp ]=0.%
\end{array}%
\right.  \label{swaves}
\end{equation}%
Applying general frame/coordinate transforms on respective solutions (\ref%
{swaves}), we construct solitonic waves parameterized by functions labeled
in the form $\ \wp =\wp (\tau ,x^{i}),$ $=\wp (\tau ,x^{1},y^{3}),$ or $=\wp
(\tau ,x^{2},y^{3}).$

\subsubsection{Ansatz for quasi-stationary geometric flows and solitonic
hierarchies}

We can consider different types of solitonic stationary configurations
determined, for instance, by sine-Gordon (using $\tau $-derivatives) and
various types of nonlinear temperature like wave configurations
characterized by nonholonomic geometric curve flows. Any such solitonic
hierarchy configuration, nonlinear wave and solitonic distribution of type $%
\wp (\tau ,u)$ (\ref{swaves}) or (\ref{solitdistr}) \ can be can be used as
generating functions for quasi-stationary d-metrics of type(\ref{dmq}), 
\begin{equation}
\mathbf{g}(\tau )=\mathbf{g[}\wp (\tau ,u\mathbf{)]=g[\wp ]=}(g_{i}[\mathbf{%
\wp }],g_{a}[\mathbf{\wp }]).  \label{solitondm}
\end{equation}%
In terms of polarization functions (\ref{offdiagpolfr}) determined by
solitonic hierarchies, we write $\eta _{i}(\tau )=\eta _{i}(\tau
,x^{k})=\eta _{i}[\mathbf{\wp }],\eta _{a}(\tau )=\eta _{a}(\tau
,x^{k},y^{b})=\eta _{a}[\mathbf{\wp }]$ and $\eta _{i}^{a}(\tau )=\eta
_{i}^{a}(\tau ,x^{k},y^{b})=\eta _{i}^{a}[\mathbf{\wp }].$ In general, a
functional dependence $\mathbf{[\wp ]}$ can be defined by a superpositions
of some solitonic hierarchies of type (\ref{mkdv}) (we can mix also
configurations of type (\ref{swaves}) and/or (\ref{solitdistr})). This can
be written, for instance, in the form $[\mathbf{\wp }]=[\ _{1}\mathbf{\wp }%
,\ _{2}\mathbf{\wp },...]$ where the left label is for numbering the
solitonic hierarchies.

Solitonic hierarchies can be used for modeling nonholonomic flow evolution
of a $Q$-source (\ref{qsourc}), when the generating sources (\ref%
{qgenersourc}) are determined by some functionals 
\begin{equation}
\ ^{q}\mathbf{\Upsilon }_{\ \nu }^{\mu }(\wp (\tau ,u\mathbf{)})=[~\
_{h}^{q}\Upsilon \lbrack \mathbf{\wp }]\delta _{j}^{i},~\ ^{q}\Upsilon
\lbrack \mathbf{\wp }]\delta _{b}^{a}].  \label{qgenersourc1}
\end{equation}%
The $\tau $-modified Einstein equations (\ref{cdeq1}) with such nonmetricity
sources and written in canonical nonholonomic variables transform into
functional equations $\widehat{\mathbf{R}}_{\ \ \beta }^{\alpha }(\tau )=\
^{q}\mathbf{\Upsilon }_{\ \ \beta }^{\alpha }(\wp ).$ We can extract
LC-configurations by imposing conditions (\ref{lccond1}), when $\widehat{%
\mathbf{T}}_{\ \alpha \beta }^{\gamma }=0,$for $\nabla .$ For such nonmetric
geometric flow configurations, the nonmetricity field $\mathbf{Q}%
_{\alpha\beta \gamma }(\wp )$ and d-torsion $\mathbf{T}_{\mu \nu \alpha
}(\wp )=\mathbf{A}_{\nu }\mathbf{g}_{\mu \alpha }(\wp )-\mathbf{A}_{\alpha }%
\mathbf{g}_{\mu \nu }(\wp )$ are determined by solitonic hierarchies.

\end{document}